\documentclass{jpsj3}

\def\gsim{\mathop {\vtop {\ialign {##\crcr 
$\hfil \displaystyle {>}\hfil $\crcr \noalign {\kern1pt \nointerlineskip } 
$\,\sim$ \crcr \noalign {\kern1pt}}}}\limits}
\def\lsim{\mathop {\vtop {\ialign {##\crcr 
$\hfil \displaystyle {<}\hfil $\crcr \noalign {\kern1pt \nointerlineskip } 
$\,\,\sim$ \crcr \noalign {\kern1pt}}}}\limits}

\def\runtitle{
Ginzburg-Landau Formalism for $2n$-Body Condensation
}
\def\runauthor{Atsuhi {Tsuruta}, Shinya {Imamura}, Kazumasa {Miyake}}

\title{
Ginzburg-Landau Formalism for {$2n$-Body} Condensation
}          
\author{Atsuhi {Tsuruta}$^{1}$, Shinya {Imamura}$^{1}$, 
and Kazumasa {Miyake}$^{1,2}$}
\inst
{
$^{1}$Division of Materials Physics, Department of Materials Engineering Science, \\
Graduate School of Engineering Science, Osaka University, Toyonaka, 
Osaka 560-8531, Japan\\
$^{2}$
Toyota Physical and Chemical Research Institute, Nagakute, Aichi 480-1192, Japan
}

\recdate
{April 1, 2014}

\abst
{
The Ginzburg-Landau formalism is constructed for Fermi superfluidity based on the 2$n$-body 
condensation in parallel with the usual Ginzburg-Landau formalism 
for the Cooper pair condensation.  By this formalism, the transition temperature of the 
2$n$-body condensation is given compactly by the zero of coefficient of the quadratic term 
in the 2$n$-body condensed order parameter without counting complicated Feynman diagrams 
for the 2$n$-body condensation susceptibility.  It is shown that the 2$n$-body condensed state is 
stabilized in the intermediate- or strong-coupling regime against the BCS state based on the 
Cooper pair condensation.  This theory is applied to the case of the possible superfluid state 
of cold atomic gases such as $^9$Be and $^{173}$Yb; the quartet state is possible in the 
former case, while the sextet state is expected in the latter case. It is predicted that the $^{173}$Yb 
atomic gas so far attained satisfies the condition for the sextet condensed state to be 
{realized}.   
}

\begin{document}
\sloppy
\maketitle
\section{Introduction}
The superfluidity known so far is sustained by the Bose-Einstein or 
Cooper pair condensate.  The former is realized in liquid $^4$He and some 
atomic gases of alkali metal {elements}~\cite{BEC1,BEC2,BEC3,BEC4}, 
while the latter is realized in liquid $^3$He and a variety of superconductors.  
Over the past decade, it has been found that the superfluidity based on the Cooper pair 
condensation is also realized in fermionic atomic gases of alkali metal {elements}~\cite{Regal}.
A new aspect of the latter case is that the crossover to the Bose-Einstein 
condensation of diatomic molecules is possible with the help of the so-called effect of 
Feshbach resonance~\cite{Ohashi,Review}.  

In principle, there exists another possibility that superfluidity is sustained by 
a condensate based on four fermions (quartet) as in an $\alpha$-particle 
correlation in light nucleus~\cite{Ropke}.  The $\alpha$-particle consists of 
two protons and two neutrons that have approximate quadruple degeneracy corresponding 
to the $2\times2$ degeneracy of the real spin and isotopic spin states.  
In this context, the problem of the quartet condensation has been discussed from time to 
time over the past decade or so~\cite{Ropke,Tohsaki,Funaki,Sogo1,Sogo2,Schuk}.  
The problem of the quartet condensation has also been addressed in the context of 
a fermionic atomic gas with fourfold degeneracy 
in internal degrees of freedom such as the $^{9}$Be atom, which has a nuclear spin $I=3/2$ 
with an electron spin $S=0$~\cite{Kamei}.  

{
The ground state of the four-particle system of such {a}
particle is known to be fully antisymmetric with respect to spin coordinates $I_{z}$ 
and fully symmetric with respect 
to space (or wavenumber) coordinates\cite{Nagaoka}.  It has been shown, by 
solving the so-called {``}Cooper problem", that the quartet state 
can be stabilized against the Cooper pairing state when four particles move outside  
a rigid Fermi surface in a moderately strong or strong-coupling region of dilute 
systems~\cite{Kamei}. It is expected that the quartet superfluid state is possible, 
in principle, in fermionic atomic gases with a nuclear spin $I=3/2$ {and an electron spin 
$S=0$}, such as $^{9}$Be.  
}

Such a superfluid state with the 2$n$-body condensation beyond the Cooper pair 
condensation ($n=1$) may also be possible for $n\ge 2$.  
{The possible $n$ is restricted by the condition that $2n\le 2I+1$ with 
$I$ being the nuclear spin.}  For example, the ground state of the 
$^{173}$Yb atom is sextuply degenerate, i.e., nuclear spin $I=5/2$ and electron 
spin $S=0$, so that a sextet condensed state ($n=3$) is possible in principle.  
Indeed, the scattering length analysis (within $s$-wave scattering) of $^{173}$Yb 
shows that it is located in a rather strong-coupling region with a scattering 
length $a_{s}\simeq 11$ nm~\cite{Kitagawa}. 
This implies that the shallow two-body s-wave bound state 
exists, which guarantee{s} the existence of a 
{6-body bound state because the ground state of a 6-particle system with 
$I=5/2$ is fully symmetric in space coordinates and fully antisymmetric in spin coordinates 
according to the theorem by Nagaoka and Usui~\cite{Nagaoka}, and has a lower energy than 
three 2-body bound states.  Namely, in the dilute limit, the 6-body correlation 
is expected to dominate the 2-body correlation, promoting the sextet condensed state of 
fermionic atomic gas of $^{173}$Yb compared with the Cooper pairing state. 
However, in the case of an intermediate or high density of atoms, these two condensed states compete 
with each other, as discussed in the ``Cooper problem" of quartet condensation.~\cite{Kamei}  
Therefore, we need to investigate the relative stability of these two states.  
}
It was reported that the atomic gas of  $^{173}$Yb is 
cooled to $T/T_{\rm F}=0.37$ in an optical trap~\cite{Fukuhara}.  
In th{ese} situation{s, a} theory for discussing the 2$n$-body condensation 
with $n\ge 2$ is desired.  

For the quartet condensation, considerable theoretical research studies have been accumulated 
over the past decade or so, not only as a problem of nuclear 
physics~\cite{Ropke,Tohsaki,Funaki,Sogo1,Sogo2,Schuk} but also as a subject of 
fundamental interest in materials physics.  
The quartet condensation was shown to {be} possible in one-dimensional models
with quadruply degenerate internal degrees of freedom~\cite{Schlottman,Schlottman2,Wu1}.  
Possible phases of cold atomic systems with a spin $I=3/2$ were reviewed from 
a wide theoretical point of view.~\cite{Wu2}  A possibility of four-electron 
attractive interaction in electron-phonon coupled systems was also discussed~\cite{Tarasewicz}.  
However, a concise formalism that enables the estimation of the transition temperature 
for the 2$n$-body condensation with $n\ge2$, including the quartet condensation ($n=2$){,} 
is expected, while some trials have been reported in a community of nuclear physics for the 
quartet condensation~\cite{Ropke,Sogo1,Schuk} . 

The purpose of this paper is to construct a Ginzburg-Landau-type formalism for the 2$n$-body 
condensation with $n\ge 2$ in general.  It will turn out that this is possible by using 
numerical calculations at  
{a realistic cost} for any $n$ as far as the 2$n$-body 
condensation with a zero 
center-of-mass momentum is concerned.  Namely, a theoretical treatment of {\it \`a  la} 
Nozi\`eres and Schmitt-Rink is beyond the scope of the present paper.  
Nevertheless, we can give a physical picture of the 2$n$-body condensed states.  In particular, 
our result is applicable for discussing the possibility of observing the sextet superfluidity in a 
cold atomic gas of $^{173}$Yb in an optical trap~\cite{Fukuhara}.  

The organization of the paper is as follows.  
In Sect. 2, the idea of the Ginzburg-Landau ({GL}) theory for the Cooper pair condensation is 
extended to the case of the quartet condensation following the idea of the variational 
principles of the mean-field approximation based on the Feynman inequality for the thermodynamic 
potential.  The {GL} thermodynamic potential (up to quadratic terms) is given in compact 
form, which is tractable with reasonable computation time.  
In Sect. 3, it is shown that the {GL} formalism is extended to the 
case of the 2$n$-body condensation without any essential difficulties.  Explicit forms of {GL} 
thermodynamic potential in general form for any $n$ are obtained.  
In Sect. 4, the transition temperature 
$T_{\rm c}$ for the three-dimensional free space is calculated for $n=2\,\sim\,5$, 
namely, from the quartet condensation to the dectet condensation, together with the case of 
the Cooper pair ($n=1$) condensation.   
In Sect. 5, the case of a two-dimensional square lattice, to which the fast-Fourier-transformation (FTT) 
technique is applicable, is discussed and $T_{\rm c}$ is calculated for {any filling 
of particles}.  
In Sect. 6, the filling dependence of $T_{\rm c}$ and the quartic terms of the {GL} thermodynamic 
potential on the square lattice are discussed for the quartet condensation.   
In Sect. 7, a possibility of the sextet condensation in $^{173}$Yb atomic gas is discussed.  
In Appendix A, the {GL} theory is reformulated on the basis of the idea of the variational 
principles of the mean-field approximation based on the Feynman inequality.  
In Appendix B, single-particle Green's function in real- and imaginary-time spaces is given.  
In Appendix C, the expressions of quartic terms in {GL}  expansion for the 
quartet condensation are derived.

\section{Ginzburg-Landau Theory for Quartet Condensation}
{In this section}, we consider {a} many-particle system of fermions with fourfold-degenerate 
internal degrees of freedom.  For example, $^9$Be has a nuclear spin $I=3/2$ and the states 
with $I_{z}=\pm 3/2,\,\pm 1/3$ are degenerate.  The Hamiltonian of such a system is 
expressed as 
\begin{equation}
H=\sum_{{\bf k},\sigma} \xi_k a^{\dagger}_{{\bf k}\sigma} a_{{\bf k} \sigma}
+{1\over 2}\sum_{\bf q}\sum_{{\bf k},{\bf k}^{\prime}}
V_{{\bf k},{\bf k}^{\prime}}\sum_{\sigma,\sigma^{\prime}}
a_{{\bf k}+{\bf q}/2, \sigma}^{\dagger}a_{-{\bf k}+{\bf q}/2, \sigma^{\prime}}^{\dagger}
a_{-{\bf k}^{\prime}+{\bf q}/2, \sigma^{\prime}}a_{{\bf k}^{\prime}+{\bf q}/2, \sigma},
\label{Q:1} 
\end{equation}
where the summation with respect to the {spin variables} $\sigma$ and $\sigma^{\prime}$ is 
taken over {$\alpha=3/2$, $\beta=1/2$, $\gamma=-1/2$, and {$\delta=-3/2$}}, which represent 
the internal degrees of freedom, e.g., $I_{z}$.  
Here, $\xi_{k}\equiv \varepsilon_{k}-\mu$, $\varepsilon=k^{2}/2m$ b{e}ing the kinetic energy of particles 
and $\mu$ being the chemical potential, 
and the two particle interaction $V_{{\bf k},{\bf k}^{\prime}}$ is assumed to be 
independent of the internal degrees of freedom.  This Hamiltonian is re{garded} as a generalization 
of that used in the Cooper pair condensat{ion}.  However, it is more convenient for 
discussing the quartet 
condensation or the 2$n$-body condensation with {$n\ge 3$} to represent Eq.\ (\ref{Q:1}) 
in the form 
\begin{eqnarray}
& &H=\sum_{{\bf p},\sigma} \xi_p a^{\dagger}_{{\bf p}\sigma} a_{{\bf p} \sigma}
\nonumber
\\
& &\qquad\quad
+{1\over 2}\sum_{{\bf p}_{1},\cdots,{\bf p}_{4}}\sum_{\sigma,\sigma^{\prime}}V_{{\bf p}_{1}-{\bf p}_{4}}
\delta\left({\bf p}_{1}+{\bf p}_{2}-{\bf p}_{3}-{\bf p}_{4}\right)
a_{{\bf p}_{1}, \sigma}^{\dagger}a_{{\bf p}_{2}, \sigma^{\prime}}^{\dagger}
a_{{\bf p}_{3}, \sigma^{\prime}}a_{{\bf k}_{4}, \sigma}.
\label{Q:2} 
\end{eqnarray}

Similarly to the case of the Cooper pair condensation (discussed in Appendix A), the 
mean-field Hamiltonian can be represented as 
\begin{equation}
H_{\rm mf}=\sum_{{\bf p},\sigma} \xi_p a^{\dagger}_{{\bf p}\sigma} a_{{\bf p} \sigma}
-\sum_{{\bf p}_{1},\cdots,{\bf p}_{4}}\Delta({\bf p}_{1},{\bf p}_{2},{\bf p}_{3},{\bf p}_{4})\,
\delta\left(\sum_{i=1}^{4}{\bf p}_{i}\right)
a^{\dagger}_{{\bf p}_{1}\alpha}a^{\dagger}_{{\bf p}_{2}\beta}
a^{\dagger}_{{\bf p}_{3} \gamma}a^{\dagger}_{{\bf p}_{4} \delta}+{\rm h.c.},  
\label{Q:3}
\end{equation} 
where $\Delta({\bf p}_{1},{\bf p}_{2},{\bf p}_{3},{\bf p}_{4})$ is the mean field in the present 
case and assumed to be independent of the internal degrees of freedom, $\alpha$, $\beta$, $\gamma$, 
and $\delta$, as in the case of the ``Cooper problem" discussed in Ref.\ \citen{Kamei}, where 
it was assumed that the wave function for the spin state is fully antisymmetric as in the case of 
the four-particle state.  
As discussed in Appendix A, the {GL} thermodynamic potential $\Omega_{\rm {GL}}$ is 
given {explicitly} by
\begin{equation}
\Omega_{\rm {GL}}{=} \Omega_{\rm mf}+\langle H-H_{\rm mf} \rangle_{\rm mf}.
\label{Q:4}
\end{equation}
The operator corresponding to the second term in Eq.\ (\ref{Q:4}) is given by 
\begin{eqnarray}
& &H-H_{\rm mf}=
{1\over 2}\sum_{{\bf p}_{1},\cdots,{\bf p}_{4}}\sum_{\sigma,\sigma^{\prime}}V_{{\bf p}_{1}-{\bf p}_{4}}
\delta\left({\bf p}_{1}+{\bf p}_{2}-{\bf p}_{3}-{\bf p}_{4}\right)
a_{{\bf p}_{1}, \sigma}^{\dagger}a_{{\bf p}_{2}, \sigma^{\prime}}^{\dagger}
a_{{\bf p}_{3}, \sigma^{\prime}}a_{{\bf k}_{4}, \sigma}
\nonumber
\\
& &\qquad\qquad\qquad
+\sum_{{\bf p}_{1},\cdots,{\bf p}_{4}}\Delta({\bf p}_{1},{\bf p}_{2},{\bf p}_{3},{\bf p}_{4})\,
\delta\left(\sum_{i=1}^{4}{\bf p}_{i}\right)
a^{\dagger}_{{\bf p}_{1}\alpha}a^{\dagger}_{{\bf p}_{2}\beta}
a^{\dagger}_{{\bf p}_{3} \gamma}a^{\dagger}_{{\bf p}_{4} \delta}+{\rm h.c.}.
\label{Q:5}
\end{eqnarray} 

First, we calculate $\Omega_{\rm mf}$ by perturbation expansion with respect to the mean field  
$\Delta$ in the mean-field Hamiltonian ({\ref{Q:3}}) up to the quadratic term 
in $\Delta({\bf p}_{1},{\bf p}_{2},{\bf p}_{3},{\bf p}_{4})$ because we are interested in 
obtaining the transition temperature for the moment.  The quartic term will be 
discussed later.  Hereafter, we assume that the wavenumber 
dependence of $\Delta$ is fully symmetric with respect to ${\bf p}_{i}$ ($i=1\sim 4$) as in the 
case of the four-particle ground state~\cite{Nagaoka}, and is given with a variational function 
$f({\bf p})$ as 
\begin{equation}
\Delta({\bf p}_{1},{\bf p}_{2},{\bf p}_{3},{\bf p}_{4})=\Delta\prod_{i=1}^{4}f({\bf p}_{i}). 
\label{Q:6}
\end{equation}
The wave function $f({\bf p})$ is a generalization of that introduced in Ref.\ \citen{Kamei} for 
the ``Cooper problem" of the quartet bound state.  

\begin{figure}[h]
\begin{center}
\rotatebox{0}{\includegraphics[width=0.5\linewidth]{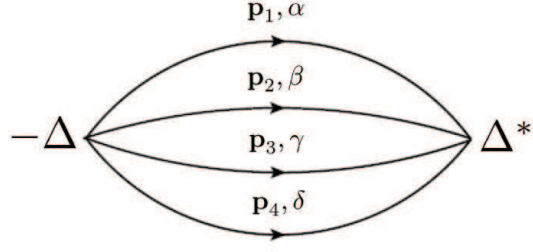}}
\caption{
Feynman diagram for  $\Omega_{\rm mf}$ of the quadratic term with respect to $\Delta$ and $\Delta^{*}$.  
}
\label{Fig:1}
\end{center}
\end{figure}

The result for $\Omega_{\rm mf}$ is given as  
\begin{equation}
\Omega_{\rm mf}\simeq \Omega_{0}-A_{4}(T)|\Delta|^{2}+{\cal O}(|\Delta|^{4}), 
\label{Q:7}
\end{equation} 
where $\Omega_{0}$ is the thermodynamic potential in the normal state, and the coefficient 
$A_{4}(T)$ is given by the Feynman diagram shown in Fig.\ \ref{Fig:1}, and its  
analytical expression is given as  
\begin{eqnarray}
& &A_{4}(T)=T^3 
\prod_{i=1}^{4}
\displaystyle \int \displaystyle \frac{d{\bf p}_i}{(2\pi)^3}
\displaystyle \sum \limits _{\varepsilon_{ni}} 
|f({\bf p}_{i})|^{2}G({\bf p}_{i},{\rm i}\epsilon_{n_i})
\nonumber \\
 & &\qquad\qquad\quad
  \times \delta\left({\bf p}_1+{\bf p}_2+{\bf p}_3+{\bf p}_4\right)  
            \times  \delta_{\epsilon_{n_1}+\epsilon_{n_2}+\epsilon_{n_3}+\epsilon_{n_4},0},
\label{Q:8}
\end{eqnarray}
where $G$ is the Matsubara Green function of quasiparticles in the normal state and is 
assumed to be independent of the four spin variables $\alpha$, $\beta$, $\gamma$, and $\delta$.  
Hereafter, $\epsilon_{n}\equiv (2n+1)\pi T$ is the fermionic Matsubara frequency.  
By using the identities  
\begin{equation}
\delta\left({\bf p}_1+{\bf p}_2+{\bf p}_3+{\bf p}_4\right)
=\int \displaystyle \frac{d {\bf r}}{(2\pi)^3} 
e^{{\rm i}({\bf p}_1+{\bf p}_2+{\bf p}_3+{\bf p}_4) \cdot {\bf r}},
\label{Q:9}
\end{equation}
and 
\begin{equation}
\delta_{\epsilon_{n_1}+\epsilon_{n_2}+\epsilon_{n_3}+\epsilon_{n_4},0}
=T\int^\beta _0 d\tau e^{-{\rm i}
(\epsilon_{n_1}+\epsilon_{n_2}+\epsilon_{n_3}+\epsilon_{n_4})\tau},
\label{Q:10}
\end{equation}
the coefficient $A_{4}(T)$, given in Eq.\ (\ref{Q:8}), is reduced to a {compact} form as 
\begin{equation}
A_{4}(T)=
\int \displaystyle \frac{d {\bf r}}{(2\pi)^3} \int^\beta _0 d\tau \, 
\left[\displaystyle \int
\frac{d{\bf p}}{(2\pi)^3}\,T\sum \limits _{\epsilon_n} |f({\bf p})|^{2}G({\bf p},{\rm i}\epsilon_n)
 e^{{\rm i}({\bf p}\cdot {\bf r}-\epsilon_n \tau)}\right]^4. 
\label{Q:11}
\end{equation}
Note here that the numbers of integration and summation variables are greatly {reduced}. 
This point is much more crucial for extending the discussion to the cases of the sextet{,   
octet, and dectet} condensations.  

\begin{figure}[h]
\begin{center}
\rotatebox{0}{\includegraphics[width=0.5\linewidth]{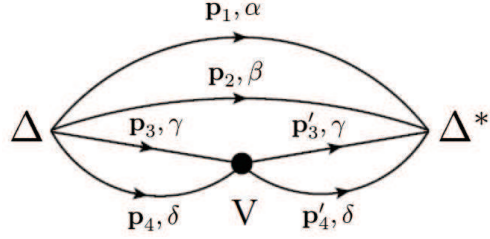}}
\caption{
Feynman diagram for $\langle H-H_{\rm mf} \rangle_{\rm mf}$ including the interaction $V$ 
of the quadratic term with respect to $\Delta$ and $\Delta^{*}$ .  
}
\label{Fig:2}
\end{center}
\end{figure}

Next, we calculate the grand canonical average of Eq.\ (\ref{Q:5}) with the mean-field 
Hamiltonian ({\ref{Q:3}}) up to quadratic terms in the gap $\Delta$.  
These terms are given by the Feynman diagrams shown in two terms of Fig.\ \ref{Fig:1} 
(with a positive sign) and 
Fig.\ \ref{Fig:2}, and their analytical expressions are given as
\begin{equation}
\langle H-H_{\rm mf} \rangle_{\rm mf}\simeq
2A_{4}(T)|\Delta|^{2}+VB_{4}(T)|\Delta|^{2}+{\cal O}(|\Delta|^{4}),
\label{Q:12}
\end{equation}
where we have assumed that the two-particle interaction $V_{\bf q}$ is wave-vector-independent 
with the energy cutoff $\varepsilon_{\rm c}$ (on the order of the Fermi energy 
$\varepsilon_{\rm F}$) considering the case of a 
dilute atomic gas with an $s$-wave attractive interaction, or a model case of fermion with 
multi-internal degrees of freedom moving on a lattice.  
The expression for $B_{4}(T)$ in Eq.\ (\ref{Q:12}) is given as 
\begin{eqnarray}
& &B_{4}(T)=\,_{4}C_{2}\,T^4 
\prod_{i=1}^{2}
\displaystyle \int \displaystyle \frac{d{\bf p}_i}{(2\pi)^3}
\displaystyle \sum \limits _{\epsilon_{n_i}} 
|f({\bf p}_{i})|^{2}G({\bf p}_{i},{\rm i}\epsilon_{n_i})
\nonumber \\
 & &\qquad\qquad\quad
\times\prod_{j=3}^{4}
\displaystyle \int \displaystyle \frac{d{\bf p}_j}{(2\pi)^3}
\displaystyle \sum \limits _{\epsilon_{n_j}} 
\displaystyle \int \displaystyle \frac{d{\bf p}^{\prime}_j}{(2\pi)^3}
\displaystyle \sum \limits _{\epsilon_{n'_j}} 
[f({\bf p}_j)]^{*}f({\bf p}^{\prime}_j)
G({\bf p}_{j},{\rm i}\epsilon_{n_j})
G({\bf p}^{\prime}_{j},{\rm i}\epsilon_{n'_j})
\nonumber \\
 & &\qquad\qquad\qquad
  \times \delta\left({\bf p}_1+{\bf p}_2+{\bf p}_3+{\bf p}_4\right)  
            \times  \delta_{\epsilon_{n_1}+\epsilon_{n_2}+\epsilon_{n_3}+\epsilon_{n_4},0}
\nonumber \\
 & &\qquad\qquad\qquad
  \times \delta\left({\bf p}_1+{\bf p}_2+{\bf p}^{\prime}_3+{\bf p}^{\prime}_4\right)  
            \times  \delta_{\epsilon_{n_1}+\epsilon_{n_2}+\epsilon_{n'_3}+\epsilon_{n'_4},0}.
\label{Q:13}
\end{eqnarray}
Here, the combination factor $_4C_2$ represents the number of ways of choosing two 
(connected to the interaction $V$) of four Green functions.   
By using Eqs.\ (\ref{Q:9}) and (\ref{Q:10}) and similar ones, the coefficient $B_{4}(T)$ 
is reduced to 
\begin{eqnarray}
& &B_{4}(T)= \displaystyle \frac{_{4}C_{2}}{(2\pi)^6}
\prod_{i=1}^{2}\displaystyle \int^\beta _0 d\tau_{i}
\displaystyle \int d{\bf r}_{i}
\left[\int \displaystyle \frac{d{\bf q}_1}{(2\pi)^3}
T\sum \limits_{\epsilon_{m_1}} [f({\bf q}_1)]^{*}G({\bf q}_1,{\rm i}\epsilon_{m_1}) 
e^{{\rm i}({\bf q}_1\cdot{\bf r}_1-\epsilon_{m_1} \tau_1)}\right]^2 
\nonumber 
\\
& &\qquad\qquad\qquad
\times 
\left[\int \displaystyle \frac{d{\bf q}_2}{(2\pi)^3}
T\sum \limits_{\epsilon_{m_2}} f({\bf q}_2)G({\bf q}_2,{\rm i}\epsilon_{m_2}) 
e^{{\rm i}({\bf q}_2\cdot{\bf r}_2-\epsilon_{m_2} \tau_2)}\right]^2
\nonumber 
\\
& &\qquad\qquad\qquad
\times 
\left[\int \displaystyle \frac{d{\bf q}_3}{(2\pi)^3}
T\sum \limits _{\epsilon_{m_3}} |f({\bf q}_3)|^{2}
G({\bf q}_3,{\rm i}\epsilon_{m_3}) e^{{\rm i}({\bf q}_3\cdot ({\bf r}_1+{\bf r}_2)
-\epsilon_{m_3} (\tau_1+\tau_2))}\right]^2.
\label{Q:14}
\end{eqnarray}
This expression is also numerically tractable as that for $A_{{2n}}(T)$, given in Eq.\ (\ref{Q:11}).  
This is also the case for the sextet{, octet and dectet} condensation{s}, 
as discussed in the next section.    

Adding Eqs.\ (\ref{Q:7}) and (\ref{Q:12}), the {GL} thermodynamic 
potential ${\Omega}_{\rm {GL}}(\Delta)$ is expressed as  
\begin{equation}
{\Omega}_{\rm {GL}}(\Delta)\simeq
\Omega_{0}+\left[A_{4}(T)+VB_{4}(T)\right]|\Delta|^{2}+{\cal O}(|\Delta|^{4}).
\label{Q:15}
\end{equation}
Then, the transition temperature $T_{\rm c}$ for the quartet condensation 
is determined by the relation 
\begin{equation}
A_{4}(T_{\rm c})=|V|B_{4}(T_{\rm c}).
\label{Q:16}
\end{equation}
This is a natural extension of that for the Cooper pair condensation, i.e., Eq.\ (\ref{A:11}), 
leading to the BCS formula Eq.\ (\ref{A:12}). 

The quartic terms in $\Delta$ and $\Delta^{*}$ include the integration with 
respect to ${\bf r}_{1}$, ${\bf r}_{2}$, and ${\bf r}_{3}$, and $\tau_{1}$, $\tau_{2}$, and 
$\tau_{3}$, respectively, as discussed in Appendix C 
[See, {e.g.}, Eqs.\ (\ref{Q:18}) and (\ref{Q:18bc})].    
Therefore, integrations similar to 
Eqs.\ (\ref{Q:11}) and (\ref{Q:14}) are technically impossible to perform within a reasonable 
computation time in the case of three-dimensional space, which will be discussed in Sect.4.  

On the other hand, in the case of a two-dimensional square lattice, it is possible 
to perform the calculations by exploiting the technique of 
fast Fourier transformation (FFT), as will be discussed in Sect.6, in which the quartic term will 
be shown to be positive {for relevant parameter sets}.  Therefore, it is reasonable to assume 
that the quartic term with respect to {$\Delta$ and $\Delta^{*}$} has a positive finite value 
also {in} the case of three-dimensional free space
{, making the transition a second-order one}.

\section{Generalization to 2$n$-Body Condensation}
The formalism {determining} the transition temperature $T_{\rm c}$ developed in the previous
section for the quartet condensation is easily generalized to the sextet{, octet, and dectet 
condensations}.  

\begin{figure}[h]
\begin{center}
\rotatebox{0}{\includegraphics[width=0.5\linewidth]{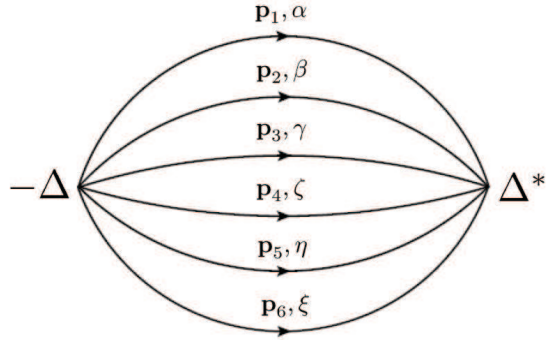}}
\caption{
Feynman diagram for $\Omega_{\rm mf}$ of the quadratic term with respect to $\Delta$ and $\Delta^{*}$.  
}
\label{Fig:4}
\end{center}
\end{figure}

For the sextet condensation, the coefficient $A_{6}(T)$ is given by the Feynman diagram shown in 
Fig.\ \ref{Fig:4}, 
and its analytical expression is given in parallel with Eq.\ (\ref{Q:8}) as follows: 
\begin{eqnarray}
& &A_{6}(T)=T^5 
\prod_{i=1}^{6}
\displaystyle \int \displaystyle \frac{d{\bf p}_i}{(2\pi)^3}
\displaystyle \sum \limits _{\varepsilon_{ni}} 
|f({\bf p}_{i})|^{2}G({\bf p}_{i},{\rm i}\epsilon_{n_i})
\nonumber \\
 & &\qquad\qquad\quad
  \times \delta\left({\bf p}_1+{\bf p}_2+{\bf p}_3+{\bf p}_4+{\bf p}_5+{\bf p}_6\right)  
            \times 
\delta_{\epsilon_{n_1}+\epsilon_{n_2}+\epsilon_{n_3}+\epsilon_{n_4}+\epsilon_{n_5}+\epsilon_{n_6},0},
\label{Q:19}
\end{eqnarray}
where $G$ is the Matsubara Green function of quasiparticles in the normal state and 
assumed to be independent of the six spin variables $\alpha$, $\beta$, $\gamma$, $\zeta$, $\eta$, and 
$\xi$.  By using the identities, similar to Eqs.\ (\ref{Q:9}) and (\ref{Q:10}), 
\begin{equation}
\delta\left({\bf p}_1+{\bf p}_2+{\bf p}_3+{\bf p}_4+{\bf p}_5+{\bf p}_6\right)
=\int \displaystyle \frac{d {\bf r}}{(2\pi)^3} 
e^{{\rm i}({\bf p}_1+{\bf p}_2+{\bf p}_3+{\bf p}_4+{\bf p}_5+{\bf p}_6) \cdot {\bf r}},
\label{Q:20}
\end{equation}
and 
\begin{equation}
\delta_{\epsilon_{n_1}+\epsilon_{n_2}+\epsilon_{n_3}+\epsilon_{n_4}+\epsilon_{n_5}+\epsilon_{n_6},0}
=T\int^\beta _0 d\tau e^{-{\rm i}
(\epsilon_{n_1}+\epsilon_{n_2}+\epsilon_{n_3}+\epsilon_{n_4}+\epsilon_{n_5}+\epsilon_{n_6})\tau},
\label{Q:21}
\end{equation}
the coefficient $A_{6}(T)$, given in Eq.\ (\ref{Q:19}), is {again} reduced to a 
{compact} form as 
\begin{equation}
A_{6}(T)=
\int \displaystyle \frac{d {\bf r}}{(2\pi)^3} \int^\beta _0 d\tau \, 
\left[\displaystyle \int
\frac{d{\bf p}}{(2\pi)^3}\,T\sum \limits _{\epsilon_n} |f({\bf p})|^{2}G({\bf p},{\rm i}\epsilon_n)
 e^{{\rm i}({\bf p}\cdot {\bf r}-\epsilon_n \tau)}\right]^{6}. 
\label{Q:22}
\end{equation}
The numerical calculation of Eq.\ (\ref{Q:22}) can be performed {at} the same 
{computational cost} as Eq.\ (\ref{Q:11}).  
Namely, the increase in the integral or summation variables in Eq.\ (\ref{Q:19}), compared with that in 
the case of quartet condensation, is absorbed by the identities Eqs.\ (\ref{Q:20}) and (\ref{Q:21}). 

\begin{figure}[h]
\begin{center}
\rotatebox{0}{\includegraphics[width=0.6\linewidth]{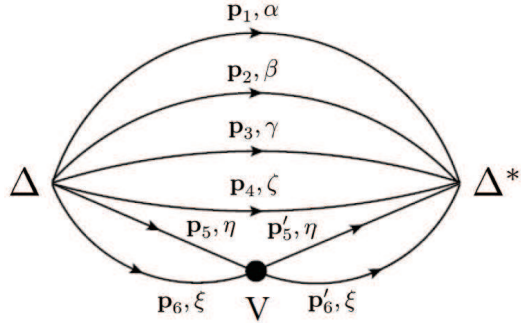}}
\caption{
Feynman diagram for $\langle H-H_{\rm mf} \rangle_{\rm mf}$ including the interaction $V$ of 
the quadratic term with respect to $\Delta$ {and $\Delta^{*}$}.  
}
\label{Fig:5}
\end{center}
\end{figure}

Similarly, the coefficient $B_{6}(T)$, whose Feynman diagram is given by Fig.\ \ref{Fig:5}, 
is calculated in parallel with  Eq.\ (\ref{Q:13}) as follows: 
\begin{eqnarray}
& &B_{6}(T)=\,_{6}C_{2}\,T^6 
\prod_{i=1}^{4}
\displaystyle \int \displaystyle \frac{d{\bf p}_i}{(2\pi)^3}
\displaystyle \sum \limits _{\epsilon_{n_i}} 
|f({\bf p}_{i})|^{2}G({\bf p}_{i},{\rm i}\epsilon_{n_i})
\nonumber \\
 & &\qquad\qquad\quad
\times\prod_{j=5}^{6}
\displaystyle \int \displaystyle \frac{d{\bf p}_j}{(2\pi)^3}
\displaystyle \sum \limits _{\epsilon_{n_j}} 
\displaystyle \int \displaystyle \frac{d{\bf p}^{\prime}_j}{(2\pi)^3}
\displaystyle \sum \limits _{\epsilon_{n'_j}} 
[f({\bf p}_j)]^{*}f({\bf p}^{\prime}_j)
G({\bf p}_{j},{\rm i}\epsilon_{n_j})
G({\bf p}^{\prime}_{j},{\rm i}\epsilon_{n'_j})
\nonumber \\
 & &\qquad\qquad\qquad
  \times \delta\left({\bf p}_1+{\bf p}_2+{\bf p}_3+{\bf p}_4+{\bf p}_5+{\bf p}_6\right)  
  \times \delta_{\epsilon_{n_1}+\epsilon_{n_2}+\epsilon_{n_3}+\epsilon_{n_4}
   +\epsilon_{n_5}+\epsilon_{n_6},0}
\nonumber \\
 & &\qquad\qquad\qquad
  \times 
\delta\left({\bf p}_1+{\bf p}_2+{\bf p}_3+{\bf p}_4+{\bf p}^{\prime}_5+{\bf p}^{\prime}_6\right)  
  \times  \delta_{\epsilon_{n_1}+\epsilon_{n_2}+\epsilon_{n_3}+\epsilon_{n_4}
            +\epsilon_{n'_5}+\epsilon_{n'_6},0}.
\label{Q:23}
\end{eqnarray}
Here, the combination factor $_6C_2$ represents the number of ways fo choosing two  
(connected to the interaction $V$) of six Green functions.   
By using Eqs.\ (\ref{Q:20}) and (\ref{Q:21}) and similar ones, the coefficient $B_{6}(T)$ 
is reduced to 
\begin{eqnarray}
& &B_{6}(T)= \displaystyle \frac{_{6}C_{2}}{(2\pi)^6}
\prod_{i=1}^{2}\displaystyle \int^\beta _0 d\tau_{i}
\displaystyle \int d{\bf r}_{i}
\left[\int \displaystyle \frac{d{\bf q}_1}{(2\pi)^3}
T\sum \limits_{\epsilon_{m_1}} [f({\bf q}_1)]^{*}G({\bf q}_1,{\rm i}\epsilon_{m_1}) 
e^{{\rm i}({\bf q}_1\cdot{\bf r}_1-\epsilon_{m_1} \tau_1)}\right]^2 
\nonumber 
\\
& &\qquad\qquad\qquad
\times 
\left[\int \displaystyle \frac{d{\bf q}_2}{(2\pi)^3}
T\sum \limits_{\epsilon_{m_2}} f({\bf q}_2)G({\bf q}_2,{\rm i}\epsilon_{m_2}) 
e^{{\rm i}({\bf q}_2\cdot{\bf r}_2-\epsilon_{m_2} \tau_2)}\right]^2
\nonumber 
\\
& &\qquad\qquad\qquad
\times 
\left[\int \displaystyle \frac{d{\bf q}_3}{(2\pi)^3}
T\sum \limits _{\epsilon_{m_3}} |f({\bf q}_3)|^{2}
G({\bf q}_3,{\rm i}\epsilon_{m_3}) e^{{\rm i}({\bf q}_3\cdot ({\bf r}_1+{\bf r}_2)
-\epsilon_{m_3} (\tau_1+\tau_2))}\right]^{4}.
\label{Q:24}
\end{eqnarray}
The calculation of Eq.\ (\ref{Q:24}) is performed at the same numerical cost as Eq.\ (\ref{Q:14}) 
for the quartet condensation.  

Then, the transition temperature $T_{\rm c}$ of the sextet condensation is also given by 
Eq.\ (\ref{Q:16}) with $A_{6}(T)$, given by Eq.\ (\ref{Q:22}), and $B_{6}(T)$, given by 
Eq.\ (\ref{Q:24}), as in the 
case of the quartet condensation.   

As one can see from the derivation of the coefficients $A_{6}(T)$ and $B_{6}(T)$ above, 
one can infer a general expression for these coefficients.  Namely, for the octet condensation, 
the exponent of $[\int d{\bf p}/(2\pi)^{3}\,T\sum_{\epsilon_n}\,\cdots]$ in the expression of $A_{6}(T)$, 
i.e., Eq.\ (\ref{Q:22}), is only replaced by 8, and the exponent of the last factor 
$[\int d{\bf q}_{3}/(2\pi)^{3}\,T\sum_{\epsilon_{m_3}}\,\cdots]$ in the expression of $B_{6}(T)$, 
i.e., Eq.\ (\ref{Q:24}), is only replaced by 6.  This is easily generalized to the case of higher number of 
condensation unit, say the {octet or} dectet condensation.    
For $2n$-body condensation, the exponent of 
$[\int d{\bf p}/(2\pi)^{3}\,T\sum_{\epsilon_n}\,\cdots]$ in the expression of $A_{2n}(T)$, 
i.e., Eq.\ (\ref{Q:11}), is given by $2n$, and the exponent of the last factor 
$[\int d{\bf q}_{3}/(2\pi)^{3}\,T\sum_{\epsilon_{m_3}}\,\cdots]$ in the expression of $B_{2n}(T)$, 
i.e., Eq.\ (\ref{Q:14}), is given by $2(n-1)$.  A combination factor of $B_{2n}(T)$ is given by 
the number of ways of choosing 2 lines from $2n$ lines, i.e., $_{2n}C_{2}$ in general.  
Namely, $A_{2n}(T)$ and $B_{2n}(T)$ are given by the following expressions: 
\begin{equation}
A_{2n}(T)=
\int \displaystyle \frac{d {\bf r}}{(2\pi)^3} \int^\beta _0 d\tau \, 
\left[\displaystyle \int
\frac{d{\bf p}}{(2\pi)^3}\,T\sum \limits _{\epsilon_n} |f({\bf p})|^{2}G({\bf p},{\rm i}\epsilon_n)
 e^{{\rm i}({\bf p}\cdot {\bf r}-\epsilon_n \tau)}\right]^{2n},
\label{Q:22G}
\end{equation}
and
\begin{eqnarray}
& &B_{2n}(T)= \displaystyle \frac{_{2n}C_{2}}{(2\pi)^6}
\prod_{i=1}^{2}\displaystyle \int^\beta _0 d\tau_{i}
\displaystyle \int d{\bf r}_{i}
\left[\int \displaystyle \frac{d{\bf q}_1}{(2\pi)^3}
T\sum \limits_{\epsilon_{m_1}} [f({\bf q}_1)]^{*}G({\bf q}_1,{\rm i}\epsilon_{m_1}) 
e^{{\rm i}({\bf q}_1\cdot{\bf r}_1-\epsilon_{m_1} \tau_1)}\right]^2 
\nonumber 
\\
& &\qquad\qquad\qquad
\times 
\left[\int \displaystyle \frac{d{\bf q}_2}{(2\pi)^3}
T\sum \limits_{\epsilon_{m_2}} f({\bf q}_2)G({\bf q}_2,{\rm i}\epsilon_{m_2}) 
e^{{\rm i}({\bf q}_2\cdot{\bf r}_2-\epsilon_{m_2} \tau_2)}\right]^2
\nonumber 
\\
& &\qquad\qquad\qquad
\times 
\left[\int \displaystyle \frac{d{\bf q}_3}{(2\pi)^3}
T\sum \limits _{\epsilon_{m_3}} |f({\bf q}_3)|^{2}
G({\bf q}_3,{\rm i}\epsilon_{m_3}) e^{{\rm i}({\bf q}_3\cdot ({\bf r}_1+{\bf r}_2)
-\epsilon_{m_3} (\tau_1+\tau_2))}\right]^{2(n-1)}.
\label{Q:24G}
\end{eqnarray}

Then, the transition temperature of $2n$-body condensation 
is determined by Eq.\ (\ref{Q:16}) by using the coefficients $A_{2n}(T)$ and $B_{2n}(T)$ 
{instead of $A_{4}(T)$ and $B_{4}(T)$}.  
It is remarkable that the numerical cost for the transition temperature $T_{\rm c}$ does not increase 
with increasing number $n$ for $2n$-body condensation.  This is a secret of attacking the 
problem using the generalized Ginzburg-Landau formalism.  

\section{Three-Dimensional Free Space}
In this section, we {calculate} the transition temperatures for $2n$-body ($n=2\,\sim\,5$) 
condensation and compare them with that for the Cooper pair condensation in three-dimensional 
free space.  Precisely speaking, 
the variational wave function $f({\bf p})$ in Eq.\ (\ref{Q:6}) should {be} determined so as to 
minimize the thermodynamic potential or free energy.  However, since such a calculation 
needs a much longer time, we here adopt an approximate solution by assuming 
\begin{equation}
f({\bf p})=\begin{cases}
1,&0<\varepsilon_{p}<\varepsilon_{\rm c};\\
                  0,&\varepsilon_{\rm c}<\varepsilon_{p}.\\
                  \end{cases}
\label{fp}
\end{equation}
Nevertheless, a fundamental aspect of $2n$-body condensation {is expected to be captured}.  

Let us define the quantity in the square brackets of Eq.\ (\ref{Q:11}) by 
${\tilde G}({\bf r},\tau)$ which is given explicitly as follows 
(see Appendix B for its derivation): 
\begin{eqnarray}
& &{\tilde G}({\bf r},\tau)=\displaystyle  \int
\frac{d{\bf p}}{(2\pi)^3}\,T\sum
\limits _{\epsilon_n} |f({\bf p})|^{2}G({\bf p},{\rm i}\epsilon_n)
e^{{\rm i}({\bf p}\cdot {\bf r}-\epsilon_n \tau)}
\nonumber 
\\
&&\qquad\qquad
=-\displaystyle \frac{m}{2\pi ^2}  \quad \displaystyle
\frac{\varepsilon_{\rm F}}{r} \displaystyle \int ^{x_{\rm c}}_{-1} dx
\sin\left[\sqrt{x+1}(k_{\rm F} r)\right]\, \displaystyle
\frac{e^{(\beta -\tau) \varepsilon_{\rm F} x}}{e^{\beta \varepsilon_{\rm F} x}+1}, 
\label{R:1}
\end{eqnarray}
where $\varepsilon_{\rm F}$ and $k_{\rm F}$ are the Fermi energy and Fermi wave number, 
respectively, and $x_{\rm c}\equiv (\varepsilon_{\rm c}/\varepsilon_{\rm F})-1$.  
The $x$-integration in Eq.\ (\ref{R:1}) well converges as $x_{\rm c}$ increases.  Then, 
we put $\varepsilon_{\rm c}=2\varepsilon_{\rm F}$, i.e., $x_{\rm c}=1$.   
It turns out by explicit numerical calculations that the integrations with respect to ${\bf r}$ 
in the expressions of $A_{2n}(T)$, i.e., Eqs.\ (\ref{Q:11}) and (\ref{Q:22}), and $B_{2n}(T)$, 
i.e., Eqs.\ (\ref{Q:14}) and (\ref{Q:24}), should be taken over a sufficiently wide $r$-region. 
{On the other hand, angular integration with respect to the direction 
${\hat {\bf r}}={\bf r}/r$ is easily performed, giving only the factor $4\pi$. } 
Therefore, proper $r$-integration remains {to be performed}.  
{In order for $A_{2}(T)$ for the Cooper pair condensation to exhibit a 
logarithmic $T$ dependence down to $T=10^{-3}\ \varepsilon_{\rm F}$, we have to take 
the integration over $0<rk_{\rm F}<1000/\sqrt{2}$.  Moreover, the contribution from 
the region $rk_{\rm F}\ll 1$ should also be calculated properly so that we have to take finer 
meshes there.  Therefore, we choose the following points on the $r$-axis: 
\begin{equation}
r_{n}=r^{*}\frac{\rho_{r}^{n}-1}{\rho_{r}-1},
\label{eq:mesh1}
\end{equation}
and take the summation from $n=1$ to $n=M_{r}$ by multiplying the width of each mesh, 
$\Delta r_{1}=r^{*}(1+\rho_{r}/2)$ for $n=1$, and  
\begin{equation}
\Delta r_{n}=\frac{r_{n}+r_{n+1}}{2}-\frac{r_{n-1}+r_{n}}{2}
=\frac{r^{*}}{2}\rho_{r}^{n-1}(\rho_{r}+1),
\label{eq:mesh2}
\end{equation}
for $2\le n\le M_{r}$.   
Namely, we use a modified trapezoidal rule. 
Explicitly, we take $r^{*}=10^{-9}/(\sqrt{2}k_{\rm F})$, 
$\rho_{r}=1.02419764544894$, and $M_{r}=1000$, which yields $r_{M_{r}}$, given by Eq.\ (\ref{eq:mesh1}), 
$r_{M_{r}}\simeq 1000.0000000/(\sqrt{2}k_{\rm F})$.    
}

On the other hand, the $\tau$ dependence of ${\tilde G}({\bf r},\tau)$, given by 
{Eq.}\ (\ref{R:1}), 
near $\tau=0$ and $\beta$ is 
very sharp in the limit $\beta\varepsilon_{\rm F}\gg1$ because the factor 
$e^{(\beta -\tau) \varepsilon_{\rm F} x}/[e^{\beta \varepsilon_{\rm F} x}+1]$ is exponentially 
small for $-1<x<x_{\rm c}$ in the intermediate region $0<\tau<\beta$, 
while it is nearly equal to 1 for $0<x<x_{\rm c}$ and $-1<x<0$ at $\tau=0$ and $\tau=\beta$, 
respectively.  Therefore, it is crucial to properly take into 
account the sharp variation of ${\tilde G}({\bf r},\tau)$ near $\tau=0$ and $\beta$ in numerical 
integrations in Eqs.\ (\ref{Q:11}), (\ref{Q:14}), (\ref{Q:22}), and (\ref{Q:24}).  
To this end, we take meshes of the $\tau$-integration as follows.  
{
Similarly to the case of $r$-integration, we choose the following points in $0\le\tau\le\beta/2$ on 
the $\tau$-axis
\begin{equation}
\tau_{n}=\tau_{\delta}+\tau^{*}\frac{\rho_{\tau}^{n}-1}{\rho_{\tau}-1},
\label{eq:mesh4a}
\end{equation}
and in $\beta/2\le\tau\le\beta$ 
\begin{equation}
\tau_{n}=\beta-\tau_{\delta}-\tau^{*}\frac{\rho_{\tau}^{n}-1}{\rho_{\tau}-1},
\label{eq:mesh4b}
\end{equation}
and take the summation from $n=0$ to $n=M_{\tau}/2$ ($M_{\tau}$ being chosen as an even natural 
integer) by multiplying the width of each mesh:  
$\Delta \tau_{0}=\tau_{\delta}+\tau^{*}/2$ for $n=0$, and 
\begin{equation}
\Delta \tau_{n}=\frac{\tau_{n}+\tau_{n+1}}{2}-\frac{\tau_{n-1}+\tau_{n}}{2}
=\frac{\tau^{*}}{2}\rho_{\tau}^{n-1}(\rho_{\tau}+1),
\label{eq:mesh5}
\end{equation}
for $1\le n<M_{\tau}/2$, and 
\begin{equation}
\Delta \tau_{(M_{\tau}/2)}=\tau_{(M_{\tau}/2)}-\frac{\tau_{(M_{\tau}/2)-1}+\tau_{(M_{\tau}/2)}}{2}
=\frac{\tau^{*}}{2}\rho_{\tau}^{(M_{\tau}/2)-1},
\label{eq:mesh6}
\end{equation}
for $n=M_{\tau}/2$.  Here, we have introduced a small $\tau_{\delta}$ in order to avoid 
singular behaviors at $\tau=0$ and $\tau=\beta$.  
Explicitly, we take $\tau^{*}=10^{-5}\,\beta$, $\tau_{\delta}=10^{-8}\beta$, 
$\rho_{\tau}=1.49932125806831$,  
and $M_{\tau}=50$, which yields $\tau_{(M_{\tau}/2)}$, given by Eqs.\ (\ref{eq:mesh4a}) and 
(\ref{eq:mesh4b}), $\tau_{(M_{\tau}/2)}\simeq 0.5000000\,\beta$.  
}

The relation determining the transition temperature $T_{\rm c}$, i.e., Eq.\ (\ref{Q:16}), is 
transformed to  
\begin{equation}
1=|V|\chi_{2n}(T_{\rm c}),
\label{R:3}
\end{equation}
where the ``$2n$-body condensation susceptibility" $\chi_{2n}(T)$ is defined by 
\begin{equation}
\chi_{2n}(T)\equiv{B_{2n}(T)\over A_{2n}(T)},
\label{R:4}
\end{equation}
where $A_{2n}(T)$ and $B_{2n}(T)$ are the expressions for $2n$-body condensation, respectively: 
e.g., $A_{2n}(T)$ and $B_{2n}(T)$ are given by Eqs.\ (\ref{Q:11}) and (\ref{Q:22}), and 
Eqs.\ (\ref{Q:14}) and (\ref{Q:24}) in the cases of the quartet ($n=2$) and sextet ($n=3$) condensations, 
respectively.  
{However, it should be noted that $\chi_{2n}(T)$ cannot be represented by a canonical 
correlation function of any quantities. This is in marked contrast with} the case of the Cooper pair 
condensation ($n=1$), {in which $\chi_{2}(T)$ is given by 
$A_{2}(T)$ whose explicit form is given by} 
\begin{equation}
A_{2}(T)=
\int \displaystyle \frac{d {\bf r}}{(2\pi)^3} \int^\beta _0 d\tau \, 
\left[\displaystyle \int
\frac{d{\bf p}}{(2\pi)^3}\,T\sum \limits _{\epsilon_n} |f({\bf p})|^{2}G({\bf p},{\rm i}\epsilon_n)
 e^{{\rm i}({\bf p}\cdot {\bf r}-\epsilon_n \tau)}\right]^{2}, 
\label{Q:25}
\end{equation}
with the same energy cutoff 
$\varepsilon_{\rm c}=2\varepsilon_{\rm F}$ as that in the case of $n\ge 2$.  This $A_{2}(T)$ is 
simply $K_{1}(T)$, given by Eq.\ (\ref{A:7}), {which is the canonical correlation of 
the pair operator, as} discussed in Appendix A.   

Figure \ref{Fig:6} shows the temperature dependence of the 
``$2n$-body condensation susceptibility" $\chi_{2n}(T)$ for 
$n=2\,\sim\,5$, i.e., from the quartet condensation to the dectet condensation, together with the Cooper 
pair susceptibility ${\chi_{2}(T)=A_{2}(T)\equiv}\, K_{1}(T)$.  
The unit of $\chi_{2n}$ is $N(\varepsilon_{\rm F})$, the density of states at the Fermi level per 
spin component.  This result implies that the $2n$-body condensation (with $n\ge 2$) 
has a larger ``susceptibility" than the Cooper pair condensation in the 
high-temperature region $T\gsim 10^{-1}\varepsilon_{\rm F}$ and vice versa.  
Another intriguing aspect is that there exists a threshold coupling, $|V_{\rm th}|$, 
necessary for $2n$-body condensation to occur {at $T=0$~K}, and a reentrant of the superfluid state
{i}s expected as the 
temperature $T$ is decreased in the case of $|V|>|V_{\rm th}|$.  On the other hand, if $V<0$, the Cooper 
pair condensation is always possible as $T$ is sufficiently reduced, no matter how the $T_{\rm c}$ is 
low, because $\chi_{2}(T)=K_{1}(T)$ diverges logarithmically in the limit $T\to 0$.  
This is consistent with the result for the stability of the quartet condensation against the 
Cooper pair condensation at the level of the ``Cooper problem" discussed in Ref.\ \citen{Kamei}
{, in which the quartet state has a lower energy than two Cooper pairs only in the 
intermediate- or strong-coupling region.}  

\begin{figure}[h]
\begin{center}
\rotatebox{0}{\includegraphics[width=0.7\linewidth]{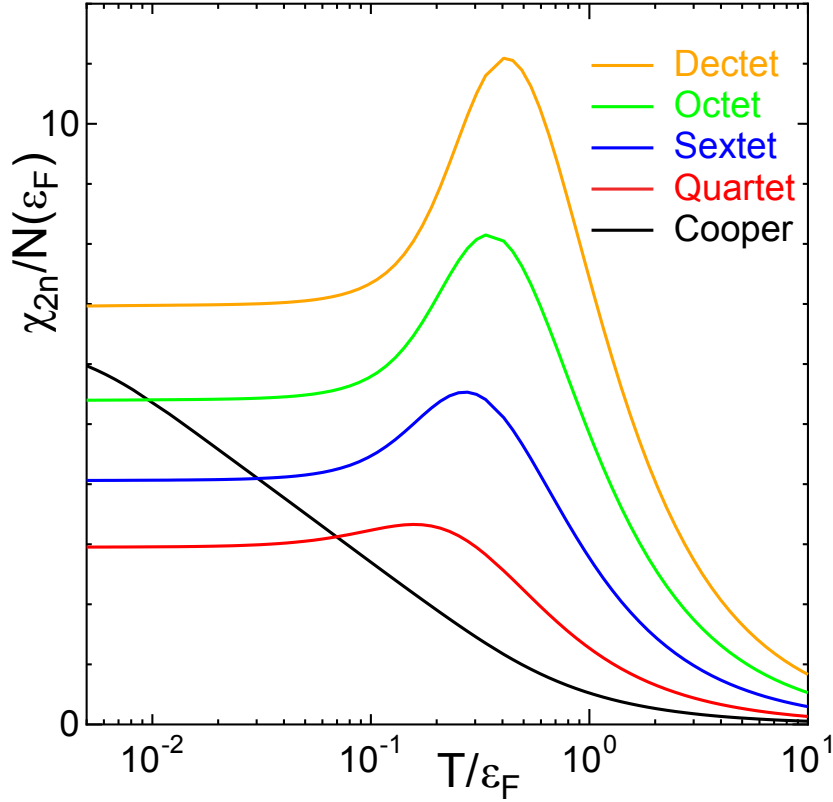}}
\end{center}
\caption{(Color) ``$2n$-body condensation susceptibility" $\chi_{2n}(T)/N(\varepsilon_{\rm F})$ as a 
function of temperature $T/\varepsilon_{\rm F}$ in a logarithmic scale. }
\label{Fig:6}
\end{figure}

\begin{figure}[h]
\begin{center}
\rotatebox{0}{\includegraphics[width=0.7\linewidth]{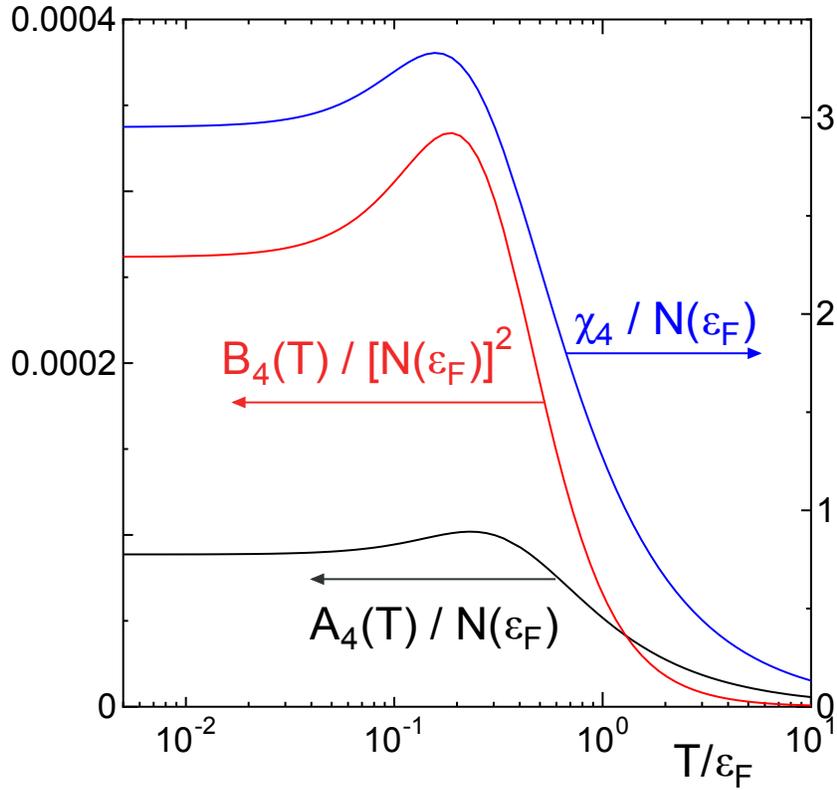}}
\end{center}
\caption{
(Color online) 
{$A_{4}(T)/N(\varepsilon_{\rm F})$, $B_{4}(T)/[N(\varepsilon_{\rm F})]^{2}$, and 
$\chi_{4}(T)/N(\varepsilon_{\rm F})$ as a 
function of temperature $T/\varepsilon_{\rm F}$ in a logarithmic scale. }
}
\label{Fig:6a}
\end{figure}

{
Here, we discuss why $\chi_{2n}(T)$'s ($n\ge 2$) exhibit peaks at 
$T\simeq (0.2\sim 0.4)\varepsilon_{\rm F}$, as shown in Fig.\ \ref{Fig:6}.  For an explicit discussion, 
we discuss the case of the quartet ($n=2$) condensation.  First, we note that 
$A_{4}(T)$, given by Eq.\ (\ref{Q:8}), is the ``bare" susceptibility of the quartet condensation, 
as shown in 
Fig.\ \ref{Fig:1}.  The $T$ dependence of $A_{4}(T)$ is shown in Fig.\ \ref{Fig:6a}, in which one can see 
that $A_{4}(T)$ exhibits a peak at $T\simeq 0.2\varepsilon_{\rm F}$.  Therefore, 
the quartet susceptibility has a tendency of exhibiting a peak structure at around 
$T=0.2\varepsilon_{\rm F}$.  In the high-$T$ region, $T\gg \varepsilon_{\rm F}$, 
$A_{4}(T)\propto T^{-1}$, so that $A_{4}(T)$ increases as $T$ decreases 
$T=\varepsilon_{\rm F}$ because a restriction on momentum integrations due to the momentum 
conservation law in Eq.\ (\ref{Q:8}) is less severe in the classical region 
($T\gsim \varepsilon_{\rm F}$) than in the Fermi degenerate region ($T\ll \varepsilon_{\rm F}$).  
On the other hand, in the low-$T$ region, i.e., $T\ll \varepsilon_{\rm F}$, 
$A_{4}(T)$ should decrease (to a certain finite value) as $T$ decreases because the restriction 
due to the momentum 
conservation law becomes crucial owing to the effect of Fermi degeneracy, which suppresses 
the available momentum space.  As a result, the peak structure in $A_{4}(T)$ is expected to appear.  
This is in marked contrast to the case of the Cooper pair condensation, for which 
$\chi_{2}(T)$ is given by $A_{2}(T)$, given by Eq.\ (\ref{Q:25}).  Since $A_{2}(T)$ is free from such 
an extra restriction due to the momentum conservation law, $A_{2}(T)$ increases monotonically 
(logarithmically) as $T$ decreases .  
Similarly, $B_{4}(T)$, given by Eq.\ (\ref{Q:13}), appearing 
in the numerator of $\chi_{4}(T)$, given by Eq.\ (\ref{R:4}), also exhibits 
a more pronounced peak structure than $A_{4}(T)$, as shown in Fig.\ \ref{Fig:6a}.  
This is because $B_{4}(T)\propto T^{-2}$ at $T\gg \varepsilon_{\rm F}$ so that 
$B_{4}(T)$ increases more sharply than $A_{4}(T)$ as $T$ decreases , making the peak height much 
higher.  As a result, a peak structure in $\chi_{4}(T)=B_{4}(T)/A_{4}(T)$ appears 
at around $T=0.2\varepsilon_{\rm F}$.  
}

\begin{figure}[h]
\begin{center}
\rotatebox{0}{\includegraphics[width=0.7\linewidth]{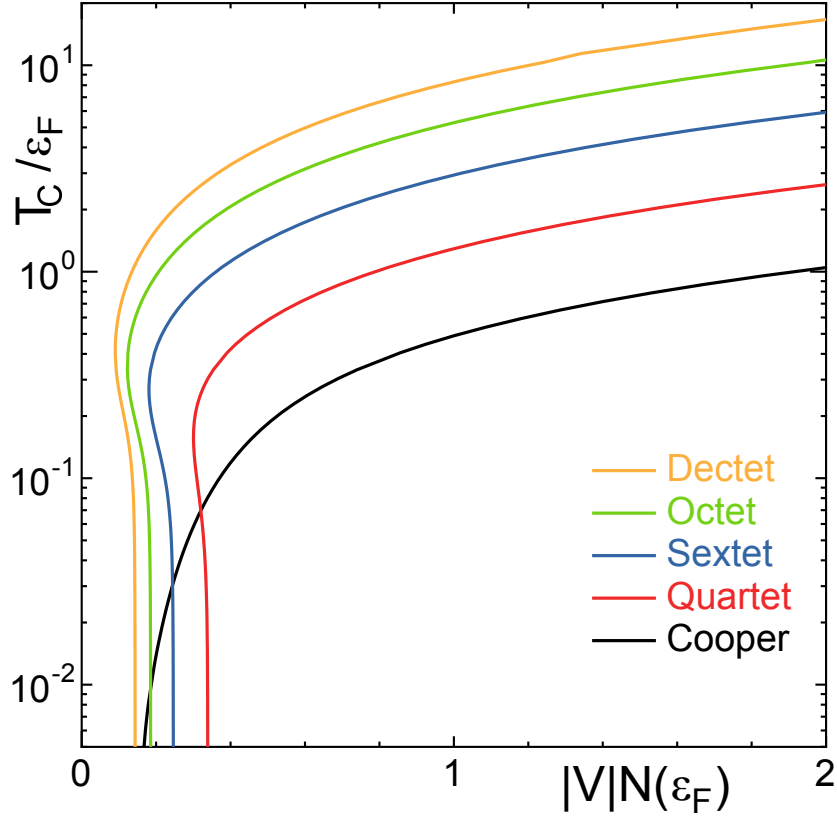}}
\end{center}
\caption{(Color) Phase diagram in the $T_{\rm c}/\varepsilon_{\rm F}-|V|N(\varepsilon_{\rm F})$ 
plane for the $2n$-body condensation ($n\ge 2$) 
and the Cooper pair condensation.}
\label{Fig:7}
\end{figure}

The transition temperature $T_{\rm c}/\varepsilon_{\rm F}$ determined by Eq.\ (\ref{R:3}) 
is shown in Fig.\ \ref{Fig:7} as a function of the strength of attractive interaction 
$|V|N(\varepsilon_{\rm F})$ for a series of $2n$-body condensations.  
The $T_{\rm c}$ of $2n$-body condensation ($n\ge 2$) is higher than that of the Cooper pair 
condensation in the intermediate-coupling region, $|V|N(\varepsilon_{\rm F})\lsim 1$ and 
strong-coupling region, $|V|N(\varepsilon_{\rm F})>1$.  Namely, the 
$2n$-body ($n\ge2$) condensed state is stabilized against the Cooper pair condensed state 
in such regions.  
{For the attractive interaction $V\sim V_{\rm th}$, $T_{\rm c}$ exhibits 
a reentrant behavior.  However, such a region of $|V|$ is restricted in a very narrow region 
above $|V_{\rm th}|$.  The threshold strengths of $V$ are  
$|V_{\rm th}|N(\varepsilon_{\rm F})\simeq 0.1,\,0.15,\,0.2,$ and $0.3$ for the dectet, octet, 
sextet, and quartet condensations, respectively.  Indeed,  in a wide region $|V|>|V_{\rm th}|$, 
$2n$-body ($n\ge 2$) condensation{s dominate} the Cooper pair condensation. 
}

On the other hand, in the strong-coupling region $|V|N(\varepsilon_{\rm F})>1$, 
we have to take into account the effect of the center-of-mass 
motion of such molecules beyond the mean field approximation adopted in previous sections, 
in which the center of mass is assumed to be at rest.  Then, $T_{\rm c}$ is determined by 
the Bose-Einstein condensation temperature $T_{\rm BEC}$, which is higher in the case of 
diatomic molecules than in the case of $2n$-atomic molecules.  This is because the mass of a $2n$-atomic 
molecules is $n$ times larger than that of a diatomic molecule, and the number density 
$N/2n$ of $2n$-atomic molecules is $1/n$ times smaller than that of diatomic molecules, 
resulting in the $T_{\rm c}$ of a $2n$-atomic molecule gas being $1/n^{5/3}$ times smaller than 
that of a diatomic  molecule gas, since $T_{\rm BEC}$ is given as 
$k_{\rm B}T_{\rm BEC}\sim (\hbar^{2}/m_{\rm b})\times(N_{\rm b}/V)^{2/3}$, $m_{\rm b}$ and 
$N_{\rm b}/V$ being the mass and number density of a composite boson.   In this 
strong-coupling region, we need to extend the theory so as to take into account the center-of-mass 
motion, as in the theory of Nozi\`eres and Schmitt-Rink for the BCS-BEC crossover of the transition 
temperature~\cite{NSR}.  However, this is beyond the scope of the present study, and is left for future 
{studies}.  

\section{Two-Dimensional Square Lattice}
In this section, we discuss the problem in the two-dimensional tight binding model 
on the square lattice with nearest-neighbor transfer.  The energy dispersion of this model 
is well known: 
\begin{equation}
\varepsilon_k=-2t(\cos\,k_x a+\cos\,k_y a),
\label{S:0}
\end{equation}
where $t$ is the transfer integral among nearest-neighbor sites and $a$ is the lattice 
constant.  In the lattice model, the attractive interaction at the on-site is denoted as $-U$, 
which should be distinguished from the Fourier component $V$ of the interaction in the continuum model 
in three-dimensional free space discussed in previous sections and Appendix A.  
Corresponding to Eq.\ (\ref{R:1}), the Matsubara Green function 
$G({\bf r}_{i},\tau)$ at the lattice point ${\bf r}_{i}$ and the imaginary time $\tau$ is given by
\begin{equation}
G({\bf r}_{i},\tau)=\displaystyle \frac{1}{N_L} \displaystyle \sum 
\limits_{\bf k} T \displaystyle \sum \limits _{\epsilon_n}
\frac{1}{{\rm i}\epsilon_n -\xi_k}e^{{\rm i}({\bf k}\cdot{\bf r}_{i}-\epsilon_n\tau)},
\end{equation}
where $N_L$ is the number of lattice points and $\xi_{k}\equiv \varepsilon_{k}-\mu$.  
Note that $G({\bf r}_{i},\tau)$ is a real quantity because it is given by an expression 
similar to Eq.\ (\ref{B:4}), which is real since the term including $\cos({\bf p}\cdot{\bf r}_{i})$ 
vanishes.  
In order to apply the technique of fast Fourier transformation (FFT) to the calculation 
{of} the coefficients $A_{2n}(T)$ and $B_{2n}(T)$ given in Sect. 2, 
let us introduce the following quantity: 
\begin{equation}
X_{m}({\bf r}_{i},\tau)\equiv[G({\bf r}_{i},\tau)]^{m}.
\label{S:1}
\end{equation} 
Note that $X_{m}({\bf r}_{i},\tau)$ is a real quantity and expressed by the Fourier series as 
(in the case where $m$ is an even natural number)
\begin{equation}
X_{m}({\bf r}_{i},\tau)={T\over N_L}\sum_{{\bf k}_j}\sum_{\omega_{n}}
X_{m}({\bf k}_{j},{\rm i}\omega_{n})
e^{{\rm i}({\bf k}_{j}\cdot{\bf r}_{i}-\omega_{n}\tau)}, 
\label{S:2}
\end{equation}  
where the Fourier component $X_{m}({\bf k}_{j},{\rm i}\omega_{n})$ is 
defined as 
\begin{equation}
X_{m}({\bf k}_{j},{\rm i}\omega_{n})\equiv \sum_{{\bf r}_{i}}\int_{0}^{\beta}
d\tau\, X_{m}({\bf r}_{i},\tau)
e^{-{\rm i}({\bf k}_{j}\cdot{\bf r}_{i}-\omega_{n}\tau)}, 
\label{S:3}
\end{equation}  
where $\omega_{n}\equiv 2\pi n T$ is the bosonic Matsubara frequency because 
$X_{m}({\bf r}_{i},\tau+\beta)=X_{m}({\bf r}_{i},\tau)$.  The coefficients $A_{2n}(T)$, given by 
Eqs.\ (\ref{Q:11}) and (\ref{Q:22}), and $B_{2n}(T)$,given by Eqs.\ (\ref{Q:14}) 
and (\ref{Q:24}), are expressed in terms of $X_{m}({\bf r}_{i},\tau)$, given by Eq.\ (\ref{S:1}), 
as follows: 
\begin{equation}
A_{2n}(T)=\sum_{{\bf r}_{i}}\int_{0}^{\beta}
d\tau\, X_{2n}({\bf r}_{i},\tau), 
\label{S:4}
\end{equation} 
and 
\begin{equation}
B_{2n}(T)=\,_{2n}C_{2}\,\sum_{{\bf r}^{(1)}_{i}}\sum_{{\bf r}^{(2)}_{i}}
\int_{0}^{\beta}d\tau_{1}\int_{0}^{\beta}d\tau_{2}
\, X_{2}({\bf r}^{(1)}_{i},\tau_{1})X_{2}({\bf r}^{(2)}_{i},\tau_{2})
X_{2n-2}({\bf r}^{(1)}_{i}+{\bf r}^{(2)}_{i},\tau_{1}+\tau_{2}). 
\label{S:5}
\end{equation} 
Substituting Eq.\ (\ref{S:2}) into Eqs.\ (\ref{S:4}) and (\ref{S:5}), and taking 
summations with respect to ${\bf r}_{i}$, ${\bf r}^{(1)}_{i}$, and ${\bf r}^{(2)}_{i}$ 
and performing integration with respect to $\tau$, $\tau_{1}$, and $\tau_{2}$, 
these quantities are expressed in terms of the Fourier component in Eq.\ (\ref{S:3}) as 
\begin{equation}
A_{2n}(T)={T\over N_L}\sum_{{\bf k}_j}\sum_{\omega_{n}}
X_{2n}({\bf k}_{j},{\rm i}\omega_{n})
\label{S:6}
\end{equation} 
and 
\begin{equation}
B_{2n}(T)=\,_{2n}C_{2}\,{T\over N_L}\sum_{{\bf k}_j}\sum_{\omega_{n}}
\left[X_{2}(-{\bf k}_{j},-{\rm i}\omega_{n})\right]^{2}
X_{2n-2}({\bf k}_{j},{\rm i}\omega_{n}).
\label{S:7}
\end{equation} 
A number of ${\bf k}$-points in the two-dimensional Brillouin zone is taken as 
{$2^{5}\times 2^{5}$}, and that of the {bosonic (fermionic)} Matsubara frequency 
{$\omega_{n}=2n\pi T$ ($\epsilon_{n}=(2n+1)\pi T$)} is restricted 
within the region $-2^{10}\le n \le 2^{10}$.  One may suspect that this number of meshes 
{$2^{5}\times 2^{5}$} is not sufficiently large to maintain the accuracy of the results.  
However, we have verified that this number gives sufficient 
accuracy by performing calculations for a series of numbers of meshes by relaxing the cut in 
the Matsubara frequency, which is much more important for maintaining the accuracy of calculations.  
{
Nevertheless, this mesh size gives a restriction on temperature above which the accuracy of 
calculations of $A_{2n}(T)$ and $B_{2n}(T)$ is guaranteed.  The lower limit of the temperature 
${\tilde T}_{\rm LL}$ is estimated as follows: 
${\tilde T}_{\rm LL}=8t/(2^{4}\times 2^{4})$, where $8t$ is the bandwidth of dispersion of 
Eq.\ (\ref{S:0}) and $2^{4}\times 2^{4}$ is the number of meshes in the first quadrant 
in the Brillouin zone.  This restriction for temperature, $T>{\tilde T}_{\rm LL}$, is expected to 
give a more severe effect in the case with a low filling of particles compared with half-filling. }

Then, we only have to perform summations with respect to 
${\bf r}_{i}$ and $\tau$ or ${\bf k}$ and $\omega_{n}$ (or $\epsilon_{n}$) 
several times, instead of directly performing multiple integrations with respect to 
${\bf r}_{i}$ and $\tau_{i}$.  The latter calculation needs a much longer time than 
the present FFT technique, and it is technically impossible to use it for 
integrations and summations for   
$C_{i}(T)$ ($i=1,2$) and $D_{i}(T)$ ($i=1\,{\sim 9}$), which are the coefficients of the quartic terms 
in $\Delta$ and $\Delta^{*}$, as discussed in Appendix C.

\begin{figure}[h]
\begin{center}
\rotatebox{0}{\includegraphics[width=0.7\linewidth]{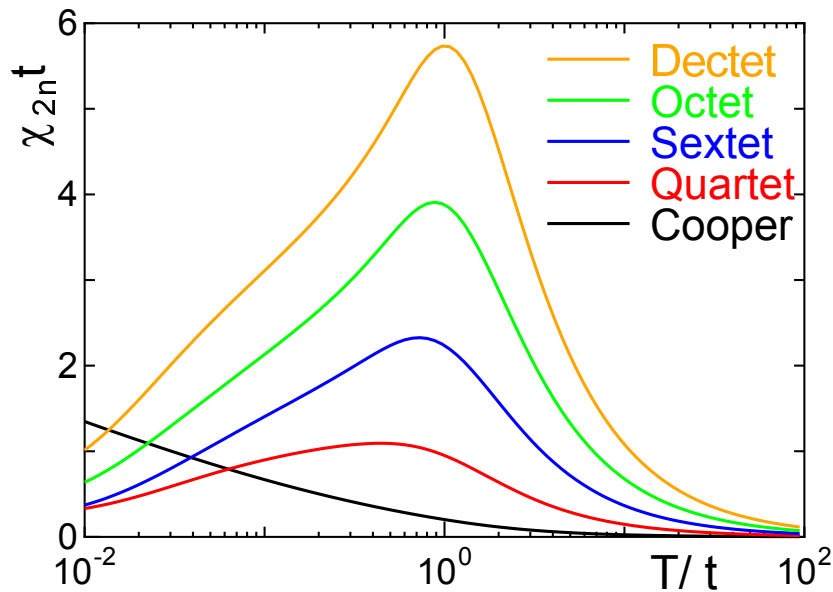}}
\end{center}
\caption{(Color) 
Temperature dependence of ``$2n$-body condensation susceptibility" $\chi_{2n}(T)$ with $n\ge 2$,  
and the Cooper pair susceptibility {at half-filling}.  The units 
of energy and temperature are chosen as $t$, the transfer integral among the nearest-neighbor sites. }
\label{Fig:8}
\end{figure}

The transition temperature $T_{\rm c}$ of ``$2n$-body condensation" is given by  
 Eq.\ (\ref{R:3}) with 
the ``$2n$-body condensation susceptibility" $\chi_{2n}(T)$, given by Eq.\ (\ref{R:4}).  
Figure\ \ref{Fig:8} shows the 
temperature dependence of $\chi_{2n}(T)$ in the cases from the quartet ($n=2$) condensation 
to the dectet ($n=5$) condensation together with the case of the Cooper pair condensation ($n=1$).  
Owing to a restriction on the size of the number of Matsubara frequencies, there exists a lower 
limit of temperature, $T_{\rm LL}$, below which the FFT calculation becomes {inaccurate}. 
Therefore, we show $\chi_{2n}(T)$ {for $T>T_{\rm LL}$} in Fig.\ \ref{Fig:8}.  
The filling of fermionic atoms is fixed {at the} half-filling ($n_{\rm A}=1$). 
{
Here, the filling $n_{\rm A}$ is defined by the ratio of twice the number of occupied states in the 
${\bf k}$-space (in the hypothetical normal ground state) and the total number of ${\bf k}$ points 
in the Brillouin zone.
}

Note that $\chi_{2n}(T)$'s for $n\ge 2$ have peaks at around $T\simeq t$, and 
are larger than that for $n=1$ (Cooper pair susceptibility) 
in the high-temperature region $T\gsim 10^{-1}t$, while the tendency is reversed in the low-temperature 
region, i.e., the Cooper pair susceptibility $\chi_{2}$ dominates $\chi_{2n}(T)$ 
for $n\ge 2$ at $T\lsim 10^{-2}t$.  This is consistent with the result in the case of 
three-dimensional free space shown in Fig.\ \ref{Fig:6}.  Also note that the combination factor 
$_{2n}C_{2}$ is crucial for $\chi_{2n}(T)$ with $n\ge 2$, which exceeds $\chi_{2}(T)$ 
in the high-temperature region.  Indeed, without the factor $_{2n}C_{2}$, 
$\chi_{2}(T)$ is larger than $\chi_{2n}(T)$ with $n\ge 2$ in the entire temperature 
region, although we do not explicitly show the result here.  

{
$T_{\rm LL}$ is estimated as follows.  The maximum magnitude of the 
Matsubara frequencies is $2\pi\times 2^{10}\,T$.  $T_{\rm LL}$ is defined by the 
condition $2\pi\times 2^{10}\,T_{\rm LL}=40t$, where $40t$ is 10 times half the 
bandwidth $4t$, i.e., $T_{\rm LL}=40t/[{2\pi\times 2^{10}}] \simeq 6.2\times 10^{-3}t$.  
}

\begin{figure}[h]
\begin{center}
\rotatebox{0}{\includegraphics[width=0.7\linewidth]{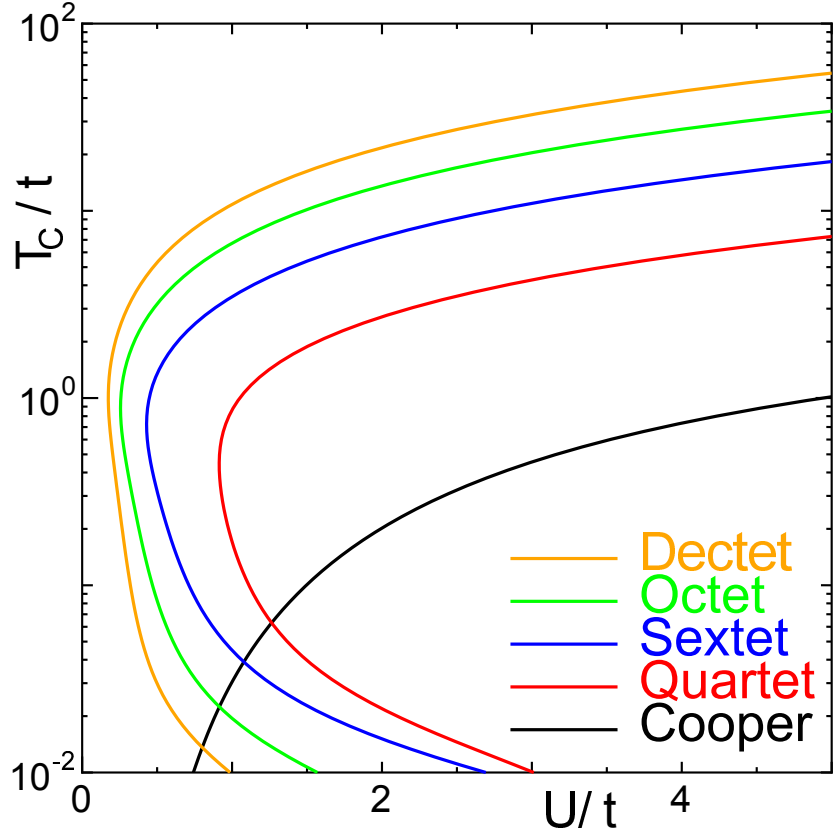}}
\end{center}
\caption{(Color) $T_{\rm c}/t$ vs $U/t$ at half-filling ($n_{\rm A}=1$) }
\label{Fig:9}
\end{figure}

Figure \ref{Fig:9} shows the relationships between the strength $U$ of the attractive interaction and 
the transition temperature $T_{\rm c}$, which is also obtained by solving Eq.\ (\ref{R:3}) 
in the case of half-filling.  
{Here, we show only $T_{\rm c}$ such that $T_{\rm c}>T_{\rm LL}$, as in Fig.\ \ref{Fig:8}.}  
In order for the ``$2n$-body condensation"  with $n\ge 2$ 
to appear, the attractive interaction needs to exceed a threshold, while the Cooper pair 
condensation is always possible, if the temperature is reduced sufficiently, owing to a logarithmic 
divergence of $\chi_{2}(T)\propto -\log\ T$ in the limit $T\to 0$.  This behavior is also 
consistent with the result in the case of three-dimensional free space shown in 
Fig.\ \ref{Fig:7}.  

\section{Properties of Quartet Condensation on Square Lattice}
In this section, some aspects of the quartet condensation on the square lattice are 
discussed{.}  All the calculations in this section are performed by taking into 
account the $T$ dependence of the chemical potential $\mu$ in a noninteracting system.   

\subsection{Dependence on filling of fermionic atom}
\begin{figure}[h]
\begin{center}
\rotatebox{0}{\includegraphics[width=0.7\linewidth]{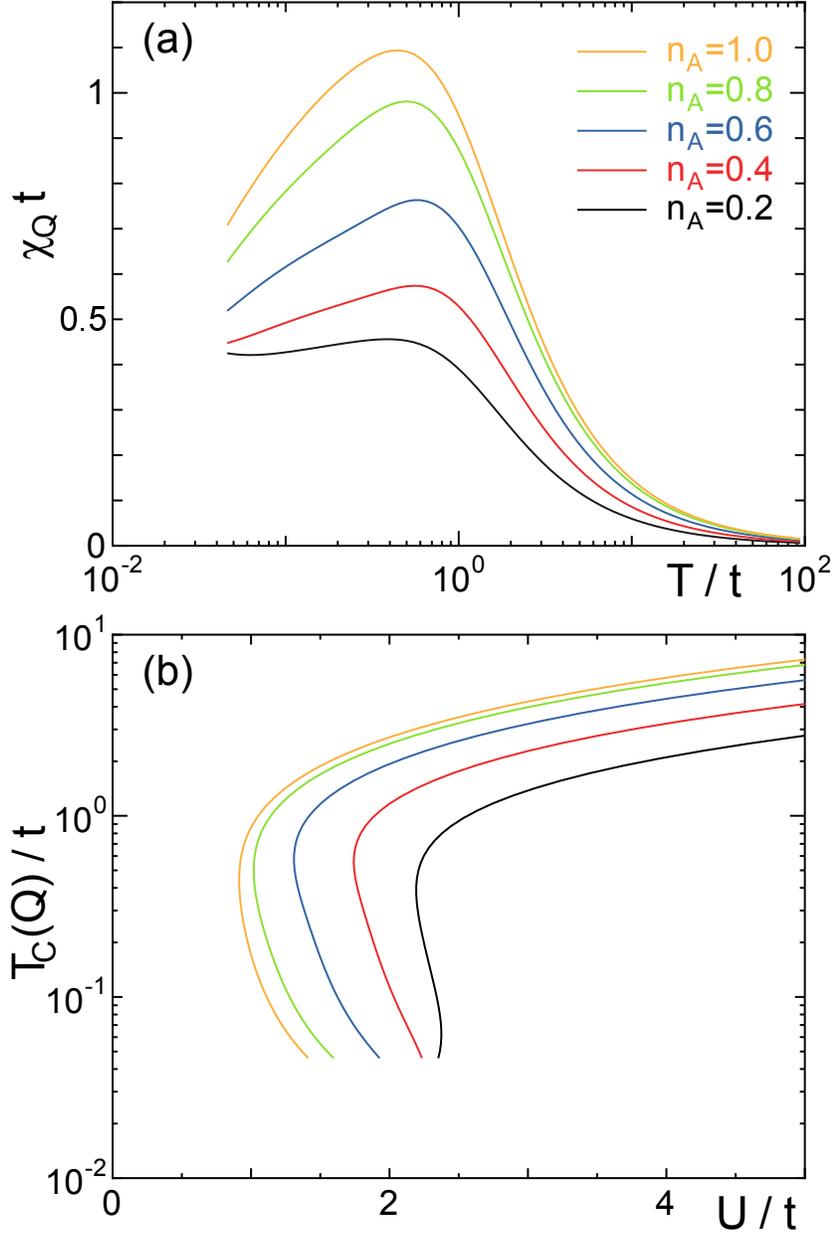}}
\end{center}
\caption{(Color) (a) $T$ dependence of {$\chi_{\rm Q}(T)$}, and 
(b) {$T_{\rm c}({\rm Q})/t$ vs $U/t$}, 
for a series of fillings $n_{\rm A}=0.2,\,0.4,\,0.6,\,0.8,\,1.0$.  
{Data in both figures are restricted to those at $T>{\tilde T}_{\rm LL}$.}}
\label{Fig:10}
\end{figure}

Figure\ \ref{Fig:10}(a) shows the temperature dependence of {$\chi_{\rm Q}(T)\,[=\chi_{4}(T)]$}, 
and Fig.\ \ref{Fig:10}(b) shows the relationship between the transition temperature $T_{\rm c}$ and 
the strength of attractive interaction, $U/t$, for the quartet condensation 
for a series of fillings $n_{\rm A}$ of fermionic atoms.  This result implies that $T_{\rm c}$ increases 
as the filling increases, 
{which is consistent with the results in Refs.\ \citen{Ropke} and \citen{Sogo1}}{.}

\begin{figure}[h]
\begin{center}
\rotatebox{0}{\includegraphics[width=0.7\linewidth]{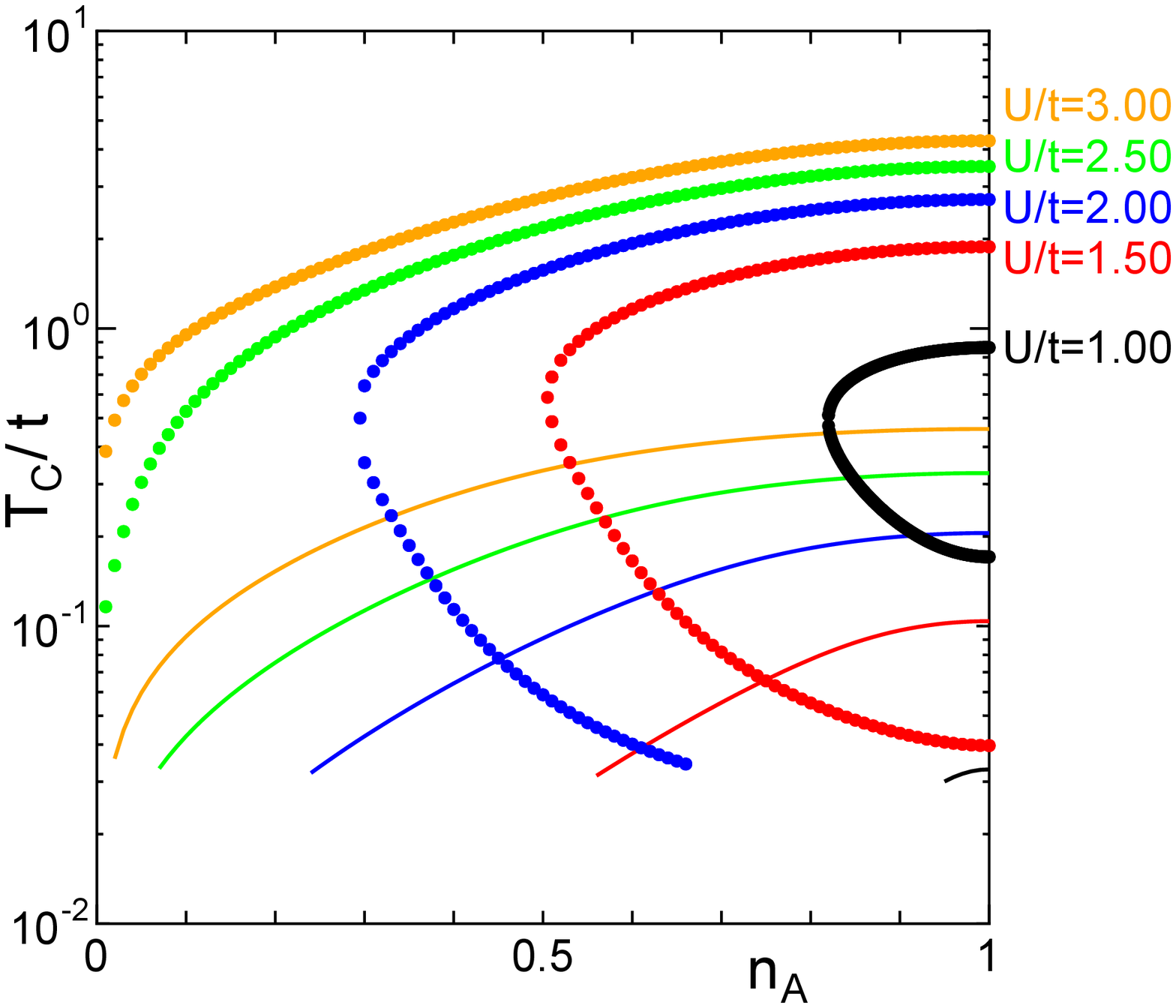}}
\end{center}
\caption{(Color) 
$n_{\rm A}$ vs $T_{\rm c}$ for quartet (dots) and Cooper pair (lines) condensations 
for a series of strengths of attractive interaction $U/t$. 
{Data are restricted to those at $T>{\tilde T}_{\rm LL}$.}}
\label{Fig:11}
\end{figure}

Figure\ \ref{Fig:11} shows the relationship between $n_{\rm A}$ and $T_{\rm c}/t$ of 
the quartet condensation (shown by dots) together with that of the Cooper pair condensation 
(shown by lines) for a series of strengths $U$ of the attractive interaction.   
One can see that the {region with the condensation} extends to the region of low density ($n_{\rm A}$) as 
$U$ increases.  
{For $U/t\gsim 2.5$, the $T_{\rm c}$ of 
the quartet condensation is higher than that of the Cooper pair condensation 
for any filling $0<n_{\rm A}\le 1$.  
This is consistent with the result of the ``Cooper problem" in the quartet case, 
in which the quartet state is stabilized in the intermediate- or strong-coupling region 
and in the low-density region~\cite{Kamei}, and also {consistent} 
with those for the $T_{\rm c}$ of the $\alpha$-condensation in the nuclear matter 
discussed in Refs.\ \citen{Ropke} and \citen{Sogo1}.  
On the other hand, in the case of weak and intermediate couplings $U/t\lsim 2.25$, 
the condensed state appears only in the region $n_{\rm A}>n_{\rm A}^{\rm th}$, 
where $n_{\rm A}^{\rm th}$ denotes a threshold filling, and $T_{\rm c}$ exhibits 
a reentrant behavior near the threshold $n_{\rm A}\gsim n_{\rm A}^{\rm th}$.  
This is somewhat different from the results shown in Refs.\ \citen{Ropke} and \citen{Sogo1}, 
where the $T_{\rm c}$ 
of the Cooper pair condensed state is higher than that of the quartet state in the 
high-density region, and also from the result for the ``Cooper problem" discussed in 
Ref.\ \citen{Kamei}. 
}

\subsection{Quartet ordered state in {GL} region}
By extending the expression (\ref{Q:15}) in Sect. 2, the {GL} free energy of the 
quartet condensation is given in its usual form as 
\begin{equation}
\tilde{\Omega}(\Delta)\simeq
\Omega_{0}+a(T)|\Delta|^{2}
+{1\over 2}b{(T)}|\Delta|^{4}+\cdots\,,
\label{S:12}
\end{equation} 
where the coefficients $a$ and $b$ are defined as 
\begin{eqnarray}
& & a(T)\equiv \left[A_{4}(T){+V}\,B_{4}(T)\right]=A_{4}(T){\left[1+V\chi_{4}(T)\right]}
\nonumber
\\
& &\qquad
\simeq {V}A_{4}(T_{\rm c}){\left[{d\chi_{4}(T)\over dT}\right]_{T=T_{\rm c}}}
\times (T-T_{\rm c}),
\label{S:13}
\end{eqnarray}
and 
\begin{equation}
b{(T)}= b_{1}(T)+Vb_{2}(T),
\end{equation}
with $b_{1}(T)$ and $b_{2}(T)$ defined as 
\begin{eqnarray}  
& &b_{1}(T)\equiv 2\sum_{i=1}^{3}\left[C_{i}({T})+D_{i}({T})\right],
\label{S:14a}
\\
& &b_{2}(T)\equiv 2\sum_{i=4}^{9}\frac{D_{i}(T)}{V},
\label{S:14b}
\end{eqnarray}
where $C_{i}(T)\,(i=1\sim 3)$ and $D_{i}(T)\,(i=1\sim 9)$ are explicitly given in Appendix C.  
Note that the interaction $V$ is equal to $-U$ in Sect. 6.1. 
It turns out that $b(T)$ is positive by explicit calculations below.  
Therefore, the standard treatment for the second-order phase transition is possible.  

\begin{figure}[h]
\begin{center}
\rotatebox{0}{\includegraphics[width=0.5\linewidth]{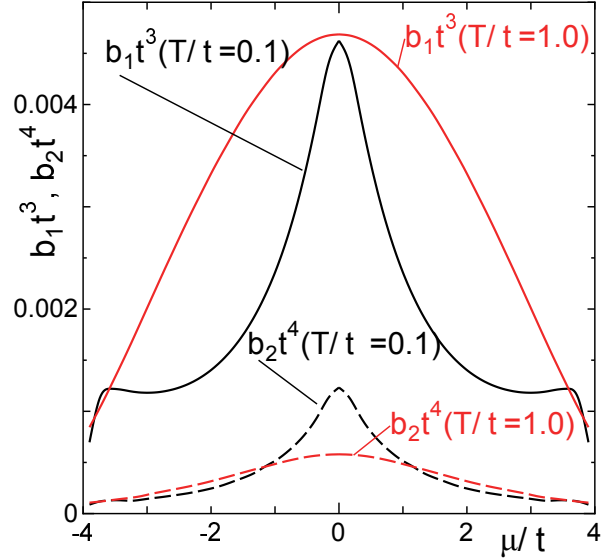}}
\end{center}
\caption{(Color online) 
Filling (chemical potential $\mu$) dependences of the coefficients $b_{1}(T)\,{t^{3}}$ 
and {$b_{2}(T)\,t^{4}$} at $T=0.1\,t$ and $T=t$.
}
\label{Fig:12a}
\end{figure}

Indeed, $C_{i}(T)$'s and $D_{i}(T)$'s are calculated in Appendix C as follows: 
$C_{1}(T)$ and $C_{2}(T)=C_{3}(T)$ are given by Eqs.\ (\ref{Q:18B}) and (\ref{S:11}), respectively;  
$D_{1}(T)=-2C_{1}(T)$ and $D_{2}(T)=D_{3}(T)=-2C_{2}(T)$; other coefficients 
$D_{4}(T)$, $D_{5}(T)$, $D_{6}(T)$, 
$D_{7}(T)$, $D_{8}(T)$, and $D_{9}(T)$ are given by Eqs.\ (\ref{Q:11A}), (\ref{C:14A}), 
(\ref{C:19}), (\ref{C:23}), (\ref{C:27}), and (\ref{C:31}), respectively. 
The results of the filling (chemical potential $\mu$) dependences of 
$b_{1}(T){\,t^{3}}$ and $b_{2}(T){\,t^{4}}$ at {$T=0.1\,t$ and $T=t$} 
are shown in Fig.\ \ref{Fig:12a}.  
Meshes of summations in these formulas are taken as {$2^{6}\times 2^{6}$} for 
summations in wave numbers {over the whole Brillouin zone of the square lattice}, 
and as {$2^{6}$} for those in the Matsubara frequencies 
{$-2\pi 2^{4}T\le \omega_{m}\le 2\pi 2^{4}T$ and 
$-\pi{(2\times 2^{4}-1)T\le \epsilon_{n}\le \pi(2\times 2^{4}-1)}T$. 
A lower limit of temperature, $T_{\rm LL}^{*}$, above which the accuracy of calculations is guaranteed, 
is defined by the condition $2\pi 2^{4}T_{\rm LL}^{*}=40t$ as in Sect. 5, i.e., 
$T_{\rm LL}^{*}=40t/2^{5}\pi\simeq 4.0\times 10^{-1}t$.  We have verified in the case of 
$\mu/t=0$ that the accuracy of the temperature dependences of $b_{1}(T)\,t^{3}$ and $b_{2}(T)\,t^{4}$ 
is maintained up to 90\% of those obtained for meshes $2^{8}\times 2^{8}$ for 
summations in wave numbers and $2^{8}$ for those in the Matsubara frequencies, 
which {corresponds to} the lower limit of temperature of $T_{\rm LL}^{*}=5.0\times 10^{-2}t$.}  

\begin{figure}[h]
\begin{center}
\rotatebox{0}{\includegraphics[width=0.5\linewidth]{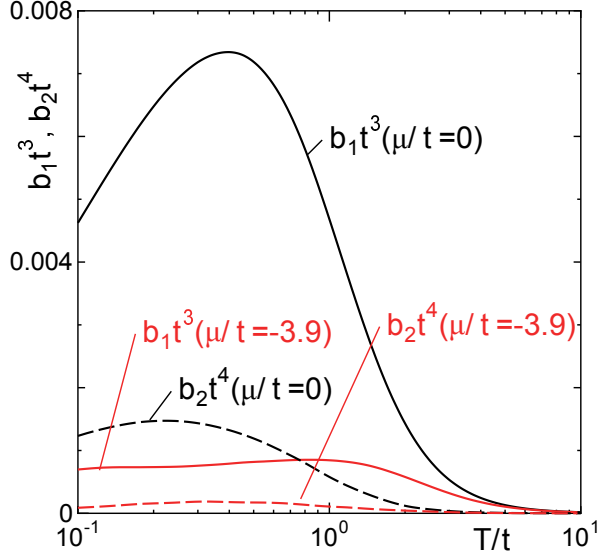}}
\end{center}
\caption{(Color online) 
Temperature ($T$) dependences of $b_{1}(T){\,t^{3}}$ and {$b_{2}(T)\,t^{4}$} 
for fillings of fermions, {$\mu/t=0$ and $\mu/t=3.9$}.
}
\label{Fig:13a}
\end{figure}

One can see {in Fig.\ \ref{Fig:12a}} that {$b(T)=b_{1}(T)+Vb_{2}(T)$ is positive,  
at least in the region of attractive interaction giving $T_{\rm c}\le t$ (see Fig.\ \ref{Fig:9})}.  
Therefore, the phase transition is of the second kind, as in the case of the Cooper pair condensation. 
The results of the temperature ($T$) dependences of {$b_{1}(T)t^{3}$ and $b_{2}(T)t^{4}$} are shown in 
Fig.\ \ref{Fig:13a} for { the filling corresponding to $\mu/t=0$ and $\mu/t=3.9$}. This also shows that 
{$b(T)=b_{1}(T)+Vb_{2}(T)$ is positive for relevant parameter sets giving a reasonable $T_{\rm c}$, 
as shown in Fig.\ \ref{Fig:9}, guaranteeing the second-order phase transition.}

Of course, there is no technical difficulty in calculating $b(T_{\rm c})$ with $T_{\rm c}$ 
determined by the condition $a(T_{\rm c})=0$, i.e., $A_{4}(T_{\rm c})={|V|}B_{4}(T_{\rm c})$, 
in Eq.\ (\ref{S:13}).  The coefficients $b_{1}(T)$ and $b_{2}(T)$ of the quartic term {in} $\Delta$ and 
$\Delta^{*}$ can be calculated with the required accuracy.  Thermodynamic analysis based on the 
{GL} thermodynamic potential is left for future studies.

\section{Possibility of Sextet Condensation in $^{173}$Yb Atomic Gas}
It has been reported that $^{173}$Yb atomic gas is cooled down below the Fermi degeneracy 
temperature $T_{\rm F}$ by means of evaporative cooling in an optical trap~\cite{Fukuhara}.  
The neutral atom of $^{173}$Yb 
has sextuplet degeneracy owing to the degrees of freedom of nuclear spin $I=5/2$ {with 
electron spins being quenched in the singlet state $S=0$}.  
Then, the sextet condensation is 
possible if a sufficiently attractive interaction works between two atoms in the dilute gas state.  
It has also been reported that the $s$-wave scattering length $a_{s}$ in the low energy limit 
of scattering atoms is positive and $a_{s}\simeq 10.6\,{\rm nm}$, which is fairly long compared 
with the range of a two-atomic interaction~\cite{Kitagawa}.  This implies that there exists a shallow 
two-body bound state with the binding energy 
\begin{equation}
E_{0}=-{\hbar^{2}\over ma_{s}^{2}}. 
\label{Y:1}
\end{equation}
Then, according to the Nagaoka-Usui theorem~\cite{Nagaoka}, the ground {state} of a six-particle 
system is fully symmetric in space coordinates and anti-symmetric in spin coordinates.  This 
state is not {an aggregation} of two-atomic bound states, but is a coherent object formed by six 
particles.  Of course, the situation is different in macroscopic systems.~\cite{Bruch}  
Nevertheless, there may be a chance that the sextet condensation is much more favorable than 
the Cooper pair condensation in some regions of temperature and atomic number density, 
as discussed in Sect. 4.  

The binding energy, given by Eq.\ (\ref{Y:1}), with $a_{s}=10.6\,{\rm nm}$, is estimated as 
$|E_{0}|/k_{\rm B}\simeq 25\,\mu$K.  This is higher than the Fermi temperature 
$T_{\rm F}\simeq 5\,\mu{\rm K}$ of Yb gas {attained from that with} 
a temperature $T\simeq 100\,\mu{\rm K}$ and an 
atomic number density $N/V\simeq 7.3\times10^{15}/{\rm cm}^{3}$ at the initial stage of cooling.  
The Yb gas is finally {cooled} to 
$T\simeq 75\,{\rm nK}$ and $N/V\simeq 6.0\times10^{13}/{\rm cm}^{3}$ by evaporation.  
Thus, the  cooling is accompanied by the 
dilution of the atomic number density, which decreases $T_{\rm F}$.  In the final stage  
of cooling, $T\simeq 0.37\,T_{\rm F}$, with $T_{\rm F}\simeq 203\,{\rm nK}$.  
Therefore, in the intermediate stage of cooling, there is a chance that both $T_{\rm F}$ and 
$T$ of the system are comparable to or smaller than $|E_{0}|/k_{\rm B}$.

Here, let us estimate the strength of the attractive interaction potential, $V_{q}$, discussed in Sect. 4.  
We assume $V_{q}$ as follows: 
\begin{equation}
V_{q}=\begin{cases}
V,&0<q<k_{\rm c};\\
                  0,&k_{\rm c}<q.\\
                  \end{cases}
\label{Y:2}
\end{equation}
Since $V_{q}$ is a matrix element of scattering{,} (${\bf p}$, ${\bf p}^{\prime}$) $\to$ 
(${\bf p}+{\bf q}$, ${\bf p}^{\prime}-{\bf q}$), and the scattering with 
$|{\bf p}|$, $|{\bf p}^{\prime}|$, $|{\bf p}+{\bf q}|$, $|{\bf p}^{\prime}-{\bf q}|$ 
$\sim$ $k_{\rm F}$ is important, it may be reasonable to take $k_{\rm c}\sim k_{\rm F}$.  
Then, the strength of the attractive interaction $V^{*}$ ($<0$) in real space is related to 
$V$ as 
\begin{equation}
V\simeq V^{*}\left({\pi\over k_{\rm F}}\right)^{3}.
\label{Y:3}
\end{equation} 
The strength of $|V^{*}|$ should be larger than $|E_{0}|$, the binding energy of the two-body 
bound state, i.e., $|V^{*}|>|E_{0}|$. Then, by using Eq.\ (\ref{Y:3}) and 
$N(\varepsilon_{\rm F})=mk_{\rm F}/2\pi^{2}\hbar^{2}$, 
\begin{equation}
|V|N(\varepsilon_{\rm F})>{\pi\over 4}{|E_{0}|/k_{\rm B}\over T_{\rm F}}.
\label{Y:4}
\end{equation} 
Therefore, it is really possible for the strong coupling region, $|V|N(\varepsilon_{\rm F})>1$, 
to be reached in the course of cooling and in the region of $T$ where the sextet condensation 
is realized, as shown in Fig.\ \ref{Fig:7}.  

As discussed partly in Sect.4, the physical picture in the strong-coupling region is not simple.  
The binding energy of the 6-body bound state is larger than that of three 2-body bound states, so that 
6-atomic molecules are formed as $T$ decreases.  On the other hand, the $T_{\rm BEC}$ of 
6-atomic molecules is lower than that of diatomic molecules.  Therefore, when the temperature is 
decreased from the normal state, the transition to the Bose-Einstein condensation of diatomic  
{molecules} would occur first if the diatomic molecules were formed at that temperature.  
However, 6-atomic molecules are formed first when the temperature is decreased from the 
high-temperature side.  
Then, the formation of diatomic  molecules is prohibited energetically, so that the Bose-Einstein 
condensation of diatomic  molecules does not occur.    

\section{Summary}
{
We have developed a mean-field theory for $2n$-body ($n\ge 2$) condensation of the 
Ginzburg-Landau ({GL}) type, on the basis of the idea of variational principles on which the 
{GL} theory is based.  We have found that the transition temperature $T_{\rm c}$ is expressed 
in concise form, which is numerically tractable for any number of $n\ge2$. Namely, the $T_{\rm c}$'s 
for the quartet, sextet, octet, and dectet condensation{s} have been calculated for fermions with 
internal degrees of freedom, 4, 6, 8, and 10, respectively, not only in three-dimensional free space 
but also in a two-dimensional square lattice.  We have also calculated the $T_{\rm c}$ 
for the Cooper pair condensation with the same formalism of numerical calculations.  The results are 
summarized as follows: 

1) There exists a threshold $|V_{\rm th}|$ of the strength of an 
attractive interaction $V$ for the $2n$-body ($n\ge 2$) condensation to be realized, and 
the $T_{\rm c}$'s exhibit the reentrant behavior for $|V|$ near the threshold $|V_{\rm th}|$.  In the 
case of three-dimensional free space, the threshold values extend as 
$|V_{\rm th}|N(\varepsilon_{\rm F})=0.1\sim0.3$ from the dectet condensation to the quartet condensation. 
In the region of $|V_{\rm th}|N(\varepsilon_{\rm F})$ in which $2n$-body condensation has a 
finite $T_{\rm c}$, $T_{\rm c}$'s are higher than that of the Cooper pair condensation. 
However, in the weak-coupling region $|V_{\rm th}|N(\varepsilon_{\rm F})\lsim 0.1$, 
the quartet condensation is not possible, while the Cooper pair condensation is always 
possible if $V$ is attractive no matter how small $|V|N(\varepsilon_{\rm F})$ is.  

2) A similar trend is obtained in the case of a two-dimensional square lattice.  
A new aspect is the filling dependence of $T_{\rm c}$ for the quartet condensation.  
$T_{\rm c}$ increases as the filling $n_{\rm A}$ increases.  
In the strong-coupling region $U/t\gsim 2.5$, $T_{\rm c}${'s} are higher 
than that of the Cooper 
pair condensation for any filling $0<n_{\rm A}\le 1$.  The transition to the quartet 
condensed state is shown to {be} of the second order by an explicit calculation of the 
quartic terms in the {GL} thermodynamic potential.

3) The sextet condensation is possible in a cold atom system of 
$^{173}$Yb, which has a shallow two-body $s$-wave bound state implying that the 
attractive interaction satisfies the condition for the sextet condensation to 
occur dominating the Cooper pair condensation.     

Our {GL}-type formalism also makes it possible to search for the thermodynamic properties of 
$2n$-body ($n\ge 2$) condensation near $T_{\rm c}$, as discussed 
in Sect. 6.2 {for the quartet ($n=2$) condensation}.  
However, detailed discussions {for $2n$-body ($n\ge 3$) condensation} are left for future studies.  
{
The present results are valid near the transition temperature $T_{\rm c}$ because they are derived 
on the basis of the GL-type formalism.  Therefore, it is not self-evident whether the 
2$n$-body ($n\ge 2$) condensed state remains as the most stable ground state even though 
$T_{\rm c}$'s are higher than that of the Cooper pairing state.  
}
Another important issue is how to treat the effect of the center-of-mass motion of 
$2n$-body ($n\ge 2$) molecules in the strong-coupling regime, in order to discuss 
the crossover to the Bose-Einstein condensation of such molecules, as discussed by 
Nozi\`eres and Schmitt-Rink in clarifying the problem of the BCS and BEC 
crossover {phenomenon}~\cite{NSR}.   
}

\section*{Acknowledgments}
{We are} grateful to T. Sogo for stimulating discussions on the 
quartet condensation and for informative conversations on the recent development of 
his theory.  We also acknowledge S. Watanabe for his question that prompted us to correct 
an error in factors included in an earlier version of Appendix A.  
This work is supported by a Grant-in-Aid for Scientific 
Research on Innovative Areas ``Topological Quantum Phenomena" 
(No. 22103003) from the Ministry of Education, Culture, Sports, Science and
Technology of Japan, and by a Grant-in-Aid for Scientific 
Research (No. 25400369) from the Japan Society for the Promotion of Science. 

\newpage
\appendix
\section{Ginzburg-Landau Formalism Revisited}
In this Appendix, we reformulate the Ginzburg-Landau ({GL}) theory~\cite{GL} for a uniform 
($s$-wave spin singlet) pair condensed state by using the Feynman diagram 
representation. 

Let us start with the Feynman inequality for the thermodynamic potential $\Omega$~\cite{Feynman}:   
\begin{equation} 
\Omega \le \Omega_{\rm mf}+\langle H-H_{\rm mf} \rangle_{\rm mf},
\label{A:1}
\end{equation}
where $H$ is the Hamiltonian of the system in consideration, $H_{\rm mf}$ is a mean-field 
Hamiltonian, and ${\Omega}_{\rm mf}$ is the thermodynamic potential for the system 
described by $H_{\rm mf}$.  Let us define the right-hand side of Eq.\ (\ref{A:1}) as 
$\tilde{\Omega}$, which is finally identified {with} the {GL} thermodynamic potential.  Namely, 
\begin{equation}
\tilde{\Omega}\equiv \Omega_{\rm mf}+\langle H-H_{\rm mf} \rangle_{\rm mf}.
\label{A:2}
\end{equation}

The Hamiltonian of the fermion system with a pairing interaction $V_{{\bf k},{\bf k}^{\prime}}$ 
is expressed as 
\begin{equation}
H=\sum_{{\bf k},\sigma} \xi_k a^{\dagger}_{{\bf k}\sigma} a_{{\bf k} \sigma}
+\sum_{\bf q}\sum_{{\bf k},{\bf k}^{\prime}}V_{{\bf k},{\bf k}^{\prime}}
a_{{\bf k}+{\bf q}/2,\uparrow}^{\dagger}a_{-{\bf k}+{\bf q}/2, \downarrow}^{\dagger}
a_{-{\bf k}^{\prime}+{\bf q}/2, \downarrow}a_{{\bf k}^{\prime}+{\bf q}/2, \uparrow},
\label{A:3}
\end{equation} 
where $\xi_{k}$ is the dispersion of quasiparticles measured from the chemical potential, and 
$a^{\dagger}_{{\bf k}\sigma}$ ($a_{{\bf k} \sigma} $) is the creation (annihilation) operator 
of quasiparticles with a wave vector ${\bf k}$ and a spin $\sigma$ (=$\uparrow$, $\downarrow$).  
Hereafter, $V_{{\bf k},{\bf k}^{\prime}}$ is assumed to be constant $V\,(<0)$.  
The mean-field Hamiltonian with the mean-field gaps $\Delta_{\bf k}$ and 
$\Delta^{{*}}_{{\bf k}^{\prime}}$ is 
given by 
\begin{equation}
H_{\rm mf}=\sum_{{\bf k},\sigma} \xi_k a^{\dagger}_{{\bf k}\sigma} a_{{\bf k} \sigma}
-\sum_{\bf k}\left(\Delta_{\bf k}^{*}a_{-{\bf k} \downarrow}a_{{\bf k} \uparrow}+
\Delta_{{\bf k}}a_{{\bf k} \uparrow}^{\dagger}a_{-{\bf k} \downarrow}^{\dagger}\right).
\label{A:4}
\end{equation} 
Therefore, the operator corresponding to the second term in Eq.\ (\ref{A:2}) is given by 
\begin{equation}
H-H_{\rm mf}=\sum_{{\bf k},{\bf k}^{\prime}}V_{{\bf k},{\bf k}^{\prime}}
a_{{\bf k} \uparrow}^{\dagger}a_{-{\bf k} \downarrow}^{\dagger}
a_{-{\bf k}^{\prime} \downarrow}a_{{\bf k}^{\prime} \uparrow}
+\sum_{\bf k}\left(\Delta_{\bf k}^{*}a_{-{\bf k} \downarrow}a_{{\bf k} \uparrow}+
\Delta_{{\bf k}}a_{{\bf k} \uparrow}^{\dagger}a_{-{\bf k} \downarrow}^{\dagger}\right).
\label{A:5}
\end{equation} 

\begin{figure}[h]
\begin{center}
\rotatebox{0}{\includegraphics[width=0.65\linewidth]{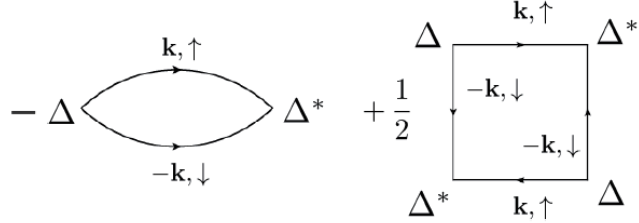}}
\caption{
Feynman diagram for $\Omega_{\rm mf}$ up to the quartic terms in $\Delta$ and 
$\Delta^{*}$.  
}
\label{Fig:A1}
\end{center}
\end{figure}

First, we calculate $\Omega_{\rm mf}$ by perturbation expansion with respect to the $s$-wave gap 
$\Delta$ (without the ${\bf k}$ dependence) in the Hamiltonian (\ref{A:4}) up to the quartic term 
in $\Delta$ {and $\Delta^{*}$}.  The result is given by 
\begin{equation}
\Omega_{\rm mf}\simeq \Omega_{0}-K_{1}(T)|\Delta|^{2}{+}{1\over 2}K_{2}(T)|\Delta|^{4}+\cdots. 
\label{A:6}
\end{equation} 
Here, $\Omega_{0}$ is the thermodynamic potential in the normal state, the coefficients 
$K_{i}(T)$ ($i=1,\,2$) are given by the Feynman diagrams shown in Fig.\ \ref{Fig:A1}, and their 
analytical expressions are given as  
\begin{eqnarray}
& &K_{1}(T)=T\sum_{\epsilon_{n}}\sum_{\bf k}G({\bf k},{\rm i}\epsilon_{n})
G(-{\bf k},-{\rm i}\epsilon_{n})
\nonumber
\\
& &\qquad \quad
\simeq{N_{\rm F}} \displaystyle \int^{\varepsilon_c}_{-\varepsilon_c}\displaystyle 
\frac{d\xi}{2\xi} 
\tanh \left(\displaystyle \frac{\beta \xi}{2}\right)
\nonumber
\\
& &\qquad \quad
\simeq{N_{\rm F}} \log\left(\frac{2\varepsilon_{\rm c}\gamma}{\pi T}\right)
,
\label{A:7}
\end{eqnarray}
and 
\begin{eqnarray}
& &K_{2}(T)=T\sum_{\epsilon_{n}}\sum_{\bf k}[G({\bf k},{\rm i}\epsilon_{n})
G(-{\bf k},-{\rm i}\epsilon_{n})]^{2}
\nonumber
\\
& &\qquad \quad
\simeq 
T\sum_{\epsilon_{n}}{N_{\rm F}} 
\displaystyle \int^{\varepsilon_c}_{-\varepsilon_c}\displaystyle d\xi 
\displaystyle \frac{1}{(\xi^{2}+\epsilon_{n}^{2})^{2}}
\nonumber
\\
& &\qquad \quad
\simeq \frac{{N_{\rm F}}}{(\pi T)^{2}}\frac{7\zeta(3)}{8},
\label{A:8}
\end{eqnarray}
where $G$ is the Matsubara Green function of quasiparticles in the normal state, 
${N_{\rm F}}$ is the 
density of states of quasiparticles at the Fermi level {per spin}, 
and $\varepsilon_{\rm c}$ is the energy cutoff 
of the pairing interaction.  $\gamma$ is the Euler number and $\zeta(z)$ is the Riemann $\zeta$-function.

\begin{figure}[h]
\begin{center}
\rotatebox{0}{\includegraphics[width=0.9\linewidth]{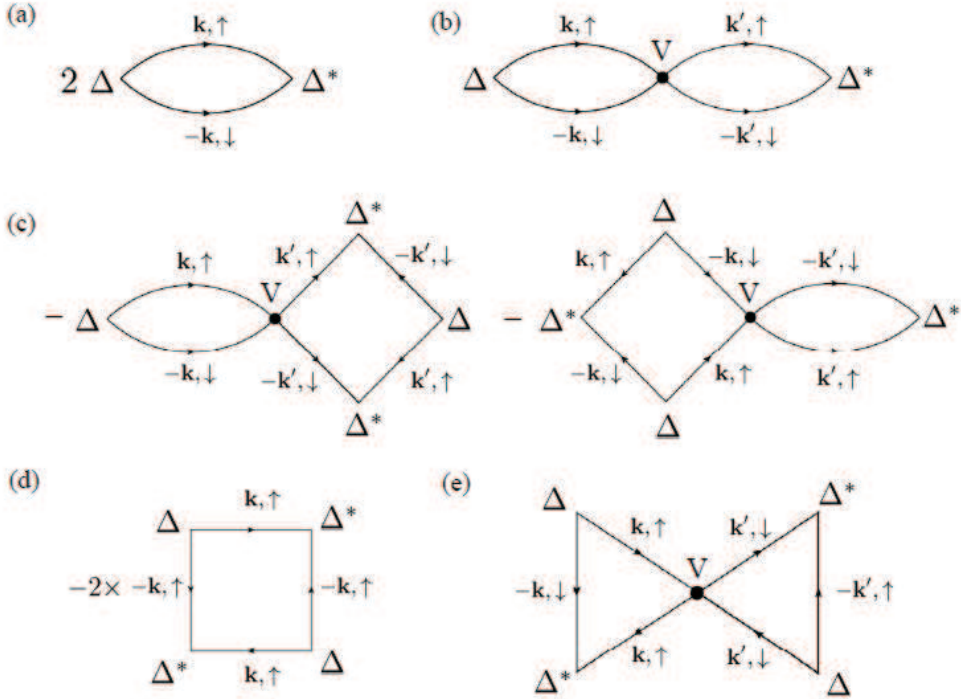}}
\caption{
Feynman diagrams for $\langle H-H_{\rm mf} \rangle_{\rm mf}$ up to the quartic terms 
with respect to $\Delta$ and $\Delta^{*}$.  
}
\label{Fig:A2}
\end{center}
\end{figure}

Next, we calculate the grand canonical average of Eq.\ (\ref{A:5}) with the mean-field 
Hamiltonian (\ref{A:4}) up to the quartic terms in the gaps $\Delta$ and $\Delta^{*}$.  
These terms are given by the Feynman diagrams shown in Figs.\ A2(a)$\sim$A2(e), and their 
analytical expressions are given as
\begin{equation}
\langle H-H_{\rm mf} \rangle_{\rm mf}\simeq
2K_{1}(T)|\Delta|^{2}+VK_{1}(T)^{2}|\Delta|^{2}{-2}VK_{1}(T)K_{2}(T)|\Delta|^{4}
{-2}K_{2}(T)|\Delta|^{4}+\cdots,
\label{A:9}
\end{equation}
where the first term corresponds to Fig.\ \ref{Fig:A2}(a), the second term to Fig.\ \ref{Fig:A2}(b), 
the third term to 
Fig.\ \ref{Fig:A2}(c), and the fourth term to Fig.\ \ref{Fig:A2}(d), while the term in 
Fig.\ \ref{Fig:A2}(e) vanishes in the case where the particle-hole symmetry is maintained, as 
usually assumed in the weak-coupling treatment of the Cooper pair condensation. {Indeed}, 
an explicit expression for the triangle of the last term in Fig.\ \ref{Fig:A2}(e) is 
given as 
\begin{equation}
T\sum_{\epsilon_{n}}\sum_{\bf k}\frac{1}{{\rm i}\epsilon_{n}-\xi_{\bf k}}
\frac{1}{(-{\rm i}\epsilon_{n}-\xi_{-\bf k})^{2}},
\label{A:9a}
\end{equation}
which is easily shown to be zero owing to the even-oddness of the integrand with respect to the 
inversion of $\epsilon_{n}\to -\epsilon_{n}$ and $\xi \to -\xi$. 

Therefore, by adding Eq.\ (\ref{A:6}), 
the {GL} thermodynamic 
potential $\tilde{\Omega}(\Delta)$ is expressed as  
\begin{equation}
\tilde{\Omega}(\Delta)\simeq
\Omega_{0}+\left[1+VK_{1}(T)\right]\left[K_{1}(T)|\Delta|^{2}{-2}K_{2}(T)|\Delta|^{4}\right]
+{1\over 2}K_{2}(T)|\Delta|^{4}+\cdots.
\label{A:10}
\end{equation}
This is nothing but the {GL} thermodynamic potential.  Indeed, $\tilde{\Omega}(\Delta)$ 
is exactly the same as $\Omega_{\rm GL}(\Delta)$ given by Leggett in Sect. 5.E of Ref.\ \citen{Leggett}.
The transition temperature 
$T_{\rm c}$ is given by the condition that the coefficient of $|\Delta|^{2}$ term is zero: 
\begin{equation}
1=|V|K_{1}(T_{\rm c}).
\label{A:11}
\end{equation}
By using Eq.\ (\ref{A:7}), an explicit form of Eq.\ (\ref{A:11}) is reduced to 
the BCS formula 
\begin{equation}
1=|V|{N_{\rm F}} \log\left(\frac{2\varepsilon_{\rm c}\gamma}{\pi T_{\rm c}}\right)
.
\label{A:12}
\end{equation}
Note that the quartic term of $\tilde{\Omega}(\Delta)$ is given essentially by the third term 
of Eq.\ (\ref{A:10}) because the quartic term in the second term is not effective near the 
transition temperature $T_{\rm c}$ where the factor $[1+VK_{1}(T)]$ vanishes.  

The equation determining the gap $\Delta$ in the equilibrium at $T\lsim T_{\rm c}$ is given by 
the condition $\partial {\tilde \Omega}(\Delta)/\partial \Delta=0$, the explicit form of which is 
expressed as 
\begin{equation}
\left[1+VK_{1}(T)\right]K_{1}(T_{\rm c})\Delta^{*}+K_{2}(T_{\rm c})|\Delta|^{2}\Delta^{*}=0.
\label{A:13}
\end{equation}
By using the explicit forms of $K_{1}(T)$, i.e., Eq.\ (\ref{A:7}), and that of 
$K_{2}(T)$, i.e., Eq.\ (\ref{A:8}), this equation is reduced to 
\begin{equation}
{N_{\rm F}}\left[\frac{T-T_{\rm c}}{T_{\rm c}}+\frac{7\zeta(3)}{8(\pi T_{\rm c})^{2}}
|\Delta|^{2}\right]\Delta^{*}=0.
\label{A:14}
\end{equation}
This is exactly the same form as that given by Gor'kov on the basis of the field theoretical 
method.~\cite{Gorkov, AGD}

\newpage
\section{Calculation of ${\tilde G}({\bf r},\tau)$} 
In this Appendix, we calculate the quantity ${\tilde G}({\bf r},\tau)$ in the square brackets 
of Eq.\ (\ref{Q:11}){:}   
\begin{equation}
{\tilde G}({\bf r},\tau)=\displaystyle \int \displaystyle \frac{d {\bf p}}{(2\pi)^3} \, 
|f({\bf p})|^{2}T\displaystyle \sum \limits _{\epsilon_n} e^{-{\rm i}\epsilon_n \tau}
\displaystyle \frac{1}{{\rm i}\epsilon_n -\xi_p}e^{{\rm i}{\bf p} \cdot {\bf r}}.
\label{B:1}
\end{equation}
Considering the periodicity of the Matsubara Green function, we restrict the variable region 
of $\tau$ {within} $0\le\tau\le\beta$.  Then, the summation with respect to $\epsilon_{n}$ 
is performed {in a standard manner} as 
\begin{equation}
T \displaystyle \sum \limits _{\epsilon_n} e^{-{\rm i}\epsilon_n \tau}
\displaystyle \frac{1}{{\rm i}\epsilon_n -\xi_p}=\oint \displaystyle
 \frac{dz}{2\pi {\rm i}} \, \displaystyle  \frac{e^{-z\tau}}{e^{-\beta
z}+1} \quad \displaystyle \frac{1}{z-\xi_p}=- \displaystyle
\frac{e^{-\tau \xi _p}}{e^{-\beta \xi _p}+1}. 
\label{B:2}
\end{equation}
Therefore, 
\begin{equation}
{\tilde G}({\bf r},\tau)=\displaystyle \int \displaystyle \frac{d {\bf p}}{(2\pi)^3}
\, |f({\bf p})|^{2}e^{{\rm i}{\bf p} \cdot {\bf r}}\,(-1)\,
\displaystyle \frac{e^{(\beta-\tau)\xi_p}}{e^{\beta \xi_p}+1}.
\label{B:3}
\end{equation}
After integrating with respect to the angular variables of ${\bf p}$, we obtain
\begin{equation}
{\tilde G}({\bf r},\tau)=-\displaystyle \frac{1}{2\pi ^2}\,
\displaystyle \frac{1}{r} \displaystyle \int ^\infty _0 dp \quad |f({\bf p})|^{2}\,p
\,\sin(pr)\,\displaystyle \frac{e^{(\beta -\tau) \xi_p}}{e^{\beta \xi_p}+1}{.}
\label{B:4}
\end{equation}
Changing the integration variable from $p$ to 
$x=\xi_{p}/\varepsilon_{\rm F}=(p^{2}/2m\varepsilon_{\rm F})-1$, and using the form of 
$f({\bf p})$ given by Eq.\ (\ref{fp}), we obtain
\begin{equation}
{\tilde G}({\bf r},\tau)=
-\displaystyle \frac{m}{2\pi ^2}\,
\frac{\varepsilon_F}{r} \displaystyle \int ^{x_{\rm c}}_{-1} dx
\sin[\sqrt{x+1}(k_F r)]\, 
\frac{e^{(\beta -\tau) \varepsilon_{\rm F} x}}{e^{\beta \varepsilon_{\rm F} x}+1}\,,
\label{B:5}
\end{equation}
where $x_{\rm c}\equiv(\varepsilon_{\rm c}/\varepsilon_{\rm F})-1$, and 
we have used an approximation $\mu\simeq\varepsilon_{\rm F}$.

\newpage
\section{Calculations of quartic terms in $\Delta$ and $\Delta^{*}$} 
In this Appendix, we give the explicit {expressions} of the quartic terms in $\Delta$ 
and $\Delta^{*}$ of the {GL} expansion for thr quartet condensation in a two-dimensional 
square lattice where the function 
$f({\bf p}_{i})$ is set to unity, i.e., $f({\bf p}_{i})=1$.  

\subsection*{$\Omega_{\rm mf}$ as function of $\Delta$ and $\Delta^{*}$}
Of the quartic terms in $\Delta$ and $\Delta^{*}$, those for $\Omega_{\rm mf}$ 
are given by the Feynman diagrams shown in Fig.\ \ref{Fig:3A}.  

The analytical expression $C_{1}(T)$ for the diagram shown in Fig.\ {\ref{Fig:3A}(a) is 
given as  
\begin{eqnarray}
& &C_{1}(T)=-{\,_{4}{\rm C}_{2}\over 2}\,T^5 
\prod_{i=1}^{8}
\displaystyle \displaystyle \frac{1}{N_{L}}\sum_{{\bf p}_i}
\displaystyle \sum \limits _{\epsilon_{n_i}} 
G({\bf p}_{i},{\rm i}\epsilon_{n_i})
\nonumber \\
 & &\qquad\qquad\qquad
\times \delta\left({\bf p}_1+{\bf p}_2+{\bf p}_3+{\bf p}_4\right)  
            \times  \delta_{\epsilon_{n_1}+\epsilon_{n_2}+\epsilon_{n_3}+\epsilon_{n_4},0}
\nonumber \\
 & &\qquad\qquad\qquad
  \times \delta\left({\bf p}_1+{\bf p}_2+{\bf p}_5+{\bf p}_6\right)  
            \times  \delta_{\epsilon_{n_1}+\epsilon_{n_2}+\epsilon_{n_5}+\epsilon_{n_6},0}
            \nonumber \\
 & &\qquad\qquad\qquad
  \times \delta\left({\bf p}_5+{\bf p}_6+{\bf p}_7+{\bf p}_8\right)  
            \times  \delta_{\epsilon_{n_5}+\epsilon_{n_6}+\epsilon_{n_7}+\epsilon_{n_8},0}.
\label{Q:17}
\end{eqnarray}
Here, the combination factor $-_4$C$_2/2$ comes from the number of combinations for perturbation expansion 
and the Wick theorem.  Namely, 
\begin{equation}
-\frac{1}{4!}\times {_4{\rm C}_2}\times 2\times {_4{\rm C}_2}\times (+1)=-\frac{_4{\rm C}_2}{2},
\label{Q:17A}
\end{equation}
where the factor $(-1/4!)$ comes from the perturbation expansion of $\Omega_{\rm {m}f}$ 
to the 4th order in $\Delta$ and $\Delta^{*}$, 
the factor $ _4{\rm C}_{2}$ is the  number of ways of choosing two $\Delta$'s from four products of the 
perturbation terms in $H_{\rm mf}$, given by Eq.\ (\ref{Q:3}), 
the factor 2 is the  number of ways of choosing two $\Delta^{*}$'s from two products of 
the perturbation terms in $H_{\rm mf}$, also given by Eq.\ (\ref{Q:3}), another factor $_{4}{\rm C}_{2}$ is the number of 
combinations how to choose 2 spin states of Green functions connecting a certain pair of $\Delta$ and 
$\Delta^{*}$ from $\alpha$, $\beta$, $\gamma$, and $\delta$ and 
the factor (+1) represents that the number of interchanges 
of Fermion operators is even in the Wick expansion. Other assignments of 
the spin variables  $\alpha$, $\beta$, $\gamma$, and $\delta$ to Green functions are  
automatically determined by the conservation law of spins.  

In the case of lattice systems, instead of the relation Eq.\ (\ref{Q:9}), the following relation 
holds: 
\begin{equation}
\delta\left({\bf p}_1+{\bf p}_2+{\bf p}_3+{\bf p}_4\right)
=\displaystyle \frac{1}{N_L}\sum_{{\bf r}_i} 
e^{{\rm i}({\bf p}_1+{\bf p}_2+{\bf p}_3+{\bf p}_4) \cdot {\bf r}_{i}}{,}
\label{Q:9L}
\end{equation}
{where $N_{L}$ is the number of lattice points.}
Then, by using Eqs.\ (\ref{Q:9L}) and (\ref{Q:10}), the coefficient $C_{1}(T)$ 
is reduced to 
\begin{eqnarray}
& &C_{1}(T)= \displaystyle -\frac{_{4}{\rm C}_{2}}{2}
\prod_{i=1}^{3}\displaystyle \int^\beta _0 d\tau_{i}
\displaystyle \sum_{{\bf r}_i}
\left[\displaystyle \frac{1}{N_{L}}\sum_{{\bf q}_1}
T\sum \limits_{\epsilon_{m_1}} G({\bf q}_1,{\rm i}\epsilon_{m_1}) 
e^{{\rm i}({\bf q}_1\cdot{\bf r}_1-\epsilon_{m_1} \tau_1)}\right]^2 
\nonumber 
\\
& &\qquad\qquad\qquad
\times 
\left[\displaystyle\frac{1}{N_L}\sum_{{\bf q}_2}
T\sum \limits_{\epsilon_{m_2}} G({\bf q}_2,{\rm i}\epsilon_{m_2}) 
e^{{\rm i}({\bf q}_2\cdot{\bf r}_2-\epsilon_{m_2} \tau_2)}\right]^2
\nonumber 
\\
& &\qquad\qquad\qquad
\times 
\left[\displaystyle\frac{1}{N_L}\sum_{{\bf q}_3}
T\sum \limits _{\epsilon_{m_3}} 
G({\bf q}_3,{\rm i}\epsilon_{m_3}) e^{{\rm i}({\bf q}_3\cdot ({\bf r}_1+{\bf r}_3)
-\epsilon_{m_3} (\tau_1+\tau_3))}\right]^2
\nonumber 
\\
& &\qquad\qquad\qquad
\times 
\left[\displaystyle\frac{1}{N_L}\sum_{{\bf q}_4}
T\sum \limits _{\epsilon_{m_4}} 
G({\bf q}_4,{\rm i}\epsilon_{m_4}) e^{{\rm i}({\bf q}_4\cdot ({\bf r}_2+{\bf r}_3)
-\epsilon_{m_4} (\tau_2+\tau_3))}\right]^2.
\label{Q:18}
\end{eqnarray}

This expression is managed easily using the FFT algorithm, as discussed in Sect. 5.  
In terms of $X_{m}$ defined in Eq.\ (\ref{S:1}), (\ref{Q:18}) is expressed as 
\begin{eqnarray}
& &C_{1}(T)=-{\,_{4}{\rm C}_{2}\over 2}\sum_{{\bf r}^{(1)}_{i}}
\sum_{{\bf r}^{(2)}_{i}}\sum_{{\bf r}^{(3)}_{i}}
\int_{0}^{\beta}d\tau_{1}\int_{0}^{\beta}d\tau_{2}\int_{0}^{\beta}d\tau_{3}
\, X_{2}({\bf r}^{(1)}_{i},\tau_{1})X_{2}({\bf r}^{(2)}_{i},\tau_{2})
\nonumber 
\\
& &\qquad\qquad
\times X_{2}({\bf r}^{(1)}_{i}+{\bf r}^{(3)}_{i},\tau_{1}+\tau_{3})
X_{2}({\bf r}^{(2)}_{i}+{\bf r}^{(3)}_{i},\tau_{2}+\tau_{3}).
\label{Q:18A}
\end{eqnarray} 
Then, by calculations similar to those leading to {Eq.\ }(\ref{S:7}) from Eq.\ (\ref{S:5}), 
the expression \ (\ref{Q:18A}) is reduced to  
\begin{equation}
C_{1}(T)=-{\,_{4}{\rm C}_{2}\over 2}{T\over N_L}\sum_{{\bf k}}\sum_{\omega_{n}}
\left[X_{2}(-{\bf k},-{\rm i}\omega_{n})X_{2}({\bf k},{\rm i}\omega_{n})\right]^{2}, 
\label{Q:18B}
\end{equation} 
where $\omega_{n}$ is the bosonic Matsubara frequency.  
Note that $X_{m}({\bf r}_{i},\tau)$ with the odd natural number $m$ is expanded 
into the Fourier series with the component $X_{m}({\bf r}_{i},{\rm i}\epsilon_{n})$ with 
a fermionic Matsubara frequency $\epsilon_{n}=(2n+1)\pi T$ because 
$X_{m}({\bf r}_{i},\tau+\beta)=-X_{m}({\bf r}_{i},\tau)$ for the odd natural number $m$.

The analytical expression $C_{2}(T)$ for the Feynman diagram shown in Figs.\ \ref{Fig:3A}(b) is given as 
\begin{eqnarray}
& &C_{2}(T)=\frac{_{4}{\rm C}_{3}}{2}\,T^5 
\prod_{i=1}^{8}
\displaystyle \displaystyle \frac{1}{N_{L}}\sum_{{\bf p}_i}
\displaystyle \sum \limits _{\epsilon_{n_i}} 
G({\bf p}_{i},{\rm i}\epsilon_{n_i})
\nonumber \\
 & &\qquad\qquad\qquad
\times \delta\left({\bf p}_1+{\bf p}_2+{\bf p}_3+{\bf p}_4\right)  
            \times  \delta_{\epsilon_{n_1}+\epsilon_{n_2}+\epsilon_{n_3}+\epsilon_{n_4},0}
\nonumber \\
 & &\qquad\qquad\qquad
  \times \delta\left({\bf p}_2+{\bf p}_3+{\bf p}_4+{\bf p}_5\right)  
            \times  \delta_{\epsilon_{n_2}+\epsilon_{n_3}+\epsilon_{n_4}+\epsilon_{n_5},0}
            \nonumber \\
 & &\qquad\qquad\qquad
  \times \delta\left({\bf p}_5+{\bf p}_6+{\bf p}_7+{\bf p}_8\right)  
            \times  \delta_{\epsilon_{n_5}+\epsilon_{n_6}+\epsilon_{n_7}+\epsilon_{n_8},0}.
\label{Q:17bc}
\end{eqnarray}
Here, the combination factor ${_4{\rm C}_3}/2$ comes from the number of combinations for perturbation 
expansion and the Wick theorem.  Namely, 
\begin{equation}
-\frac{1}{4!}\times {_4{\rm C}_2}\times 2\times {_4{\rm C}_{3}}\times (-1)=
\frac{_{4}{\rm C}_{3}}{2},
\label{Q:17B}
\end{equation}
where the factor $(-1/4!)$ comes from the perturbation expansion of $\Omega_{\rm {m}f}$ 
to the 4th order, 
the factor $_4{\rm C}_{2}$ is the  number of ways of choosing two $\Delta$'s from four products of the 
perturbation terms in $H_{\rm mf}$, given by Eq.\ (\ref{Q:3}), 
the factor 2 is the  number of ways of choosing two $\Delta^{*}$'s from two products of 
the perturbation terms in $H_{\rm mf}$, also given by Eq.\ (\ref{Q:3}), the factor $_{4}{\rm C}_{3}$ is the number of 
combinations how to choose 3 spin states of Green functions connecting a certain pair of $\Delta$ and 
$\Delta^{*}$ from $\alpha$, $\beta$, $\gamma$, and $\delta$, and 
the factor (-1) represents that the number of interchanges 
of Fermion operators is odd in the Wick expansion. Other assignments of 
the spin variables  $\alpha$, $\beta$, $\gamma$, and $\delta$ to Green functions are  
automatically determined by the conservation law of spins.  

By using Eqs.\ (\ref{Q:9L}) and (\ref{Q:10}) and similar ones, the coefficient $C_{2}(T)$ 
is reduced to 
\begin{eqnarray}
& &C_{2}(T)= \displaystyle \frac{_{4}{\rm C}_{3}}{2}
\prod_{i=1}^{3}\displaystyle \int^\beta _0 d\tau_{i}
\displaystyle \sum_{{\bf r}_i}
\left[\displaystyle \frac{1}{N_{L}}\sum_{{\bf q}_1}
T\sum \limits_{\epsilon_{m_1}} G({\bf q}_1,{\rm i}\epsilon_{m_1}) 
e^{{\rm i}({\bf q}_1\cdot{\bf r}_1-\epsilon_{m_1} \tau_1)}\right]
\nonumber 
\\
& &\qquad\qquad\qquad
\times 
\left[\displaystyle\frac{1}{N_L}\sum_{{\bf q}_2}
T\sum \limits_{\epsilon_{m_2}} G({\bf q}_2,{\rm i}\epsilon_{m_2}) 
e^{{\rm i}({\bf q}_2\cdot{\bf r}_2-\epsilon_{m_2} \tau_2)}\right]^3
\nonumber 
\\
& &\qquad\qquad\qquad
\times 
\left[\displaystyle\frac{1}{N_L}\sum_{{\bf q}_3}
T\sum \limits _{\epsilon_{m_3}} 
G({\bf q}_3,{\rm i}\epsilon_{m_3}) e^{{\rm i}({\bf q}_3\cdot ({\bf r}_1+{\bf r}_3)
-\epsilon_{m_3} (\tau_1+\tau_3))}\right]^3
\nonumber 
\\
& &\qquad\qquad\qquad
\times 
\left[\displaystyle\frac{1}{N_L}\sum_{{\bf q}_4}
T\sum \limits _{\epsilon_{m_4}} 
G({\bf q}_4,{\rm i}\epsilon_{m_4}) e^{{\rm i}({\bf q}_4\cdot ({\bf r}_2+{\bf r}_3)
-\epsilon_{m_4} (\tau_2+\tau_3))}\right].
\label{Q:18bc}
\end{eqnarray}
In terms of $X_{m}$ defined in Eq.\ (\ref{S:1}), {the right-hand side of Eq.}\ (\ref{Q:18bc}) 
is expressed as 
\begin{eqnarray}
& &C_{2}(T)=\frac{_{4}{\rm C}_{3}}{2}
\sum_{{\bf r}^{(1)}_{i}}\sum_{{\bf r}^{(2)}_{i}}\sum_{{\bf r}^{(3)}_{i}}
\int_{0}^{\beta}d\tau_{1}\int_{0}^{\beta}d\tau_{2}\int_{0}^{\beta}d\tau_{3}
\, X_{1}({\bf r}^{(1)}_{i},\tau_{1})X_{3}({\bf r}^{(2)}_{i},\tau_{2})
\nonumber 
\\
& &\qquad\qquad
\times X_{3}({\bf r}^{(1)}_{i}+{\bf r}^{(3)}_{i},\tau_{1}+\tau_{3})
X_{1}({\bf r}^{(2)}_{i}+{\bf r}^{(3)}_{i},\tau_{2}+\tau_{3}). 
\label{S:9}
\end{eqnarray} 
Then, by calculations similar to those leading to {Eq.}\ (\ref{S:7}) from Eq.\ (\ref{S:5}), 
the expression (\ref{S:9}) is reduced to  
\begin{equation}
C_{2}(T)=\frac{_{4}{\rm C}_{3}}{2}{T\over N_L}\sum_{{\bf k}}\sum_{\epsilon_{n}}
\left[X_{1}({\bf k},{\rm i}\epsilon_{n})\right]^{2}
\left[X_{3}(-{\bf k},-{\rm i}\epsilon_{n})\right]^{2},  
\label{S:11}
\end{equation} 
where $\epsilon_{n}$ is the fermionic Matsubara frequency as mentioned just below Eq.\ (\ref{Q:18B}). 

The analytical expression $C_{3}(T)$ for the Feynman diagram shown in {Fig.}\ \ref{Fig:3A}(c) is 
identical to that shown in {Fig.}\ \ref{Fig:3A}(b).  Namely, 
\begin{equation}
C_{3}(T)=C_{2}(T).
\label{C:1}
\end{equation}

\begin{figure}[h]
\begin{center}
\rotatebox{0}{\includegraphics[width=0.85\linewidth]{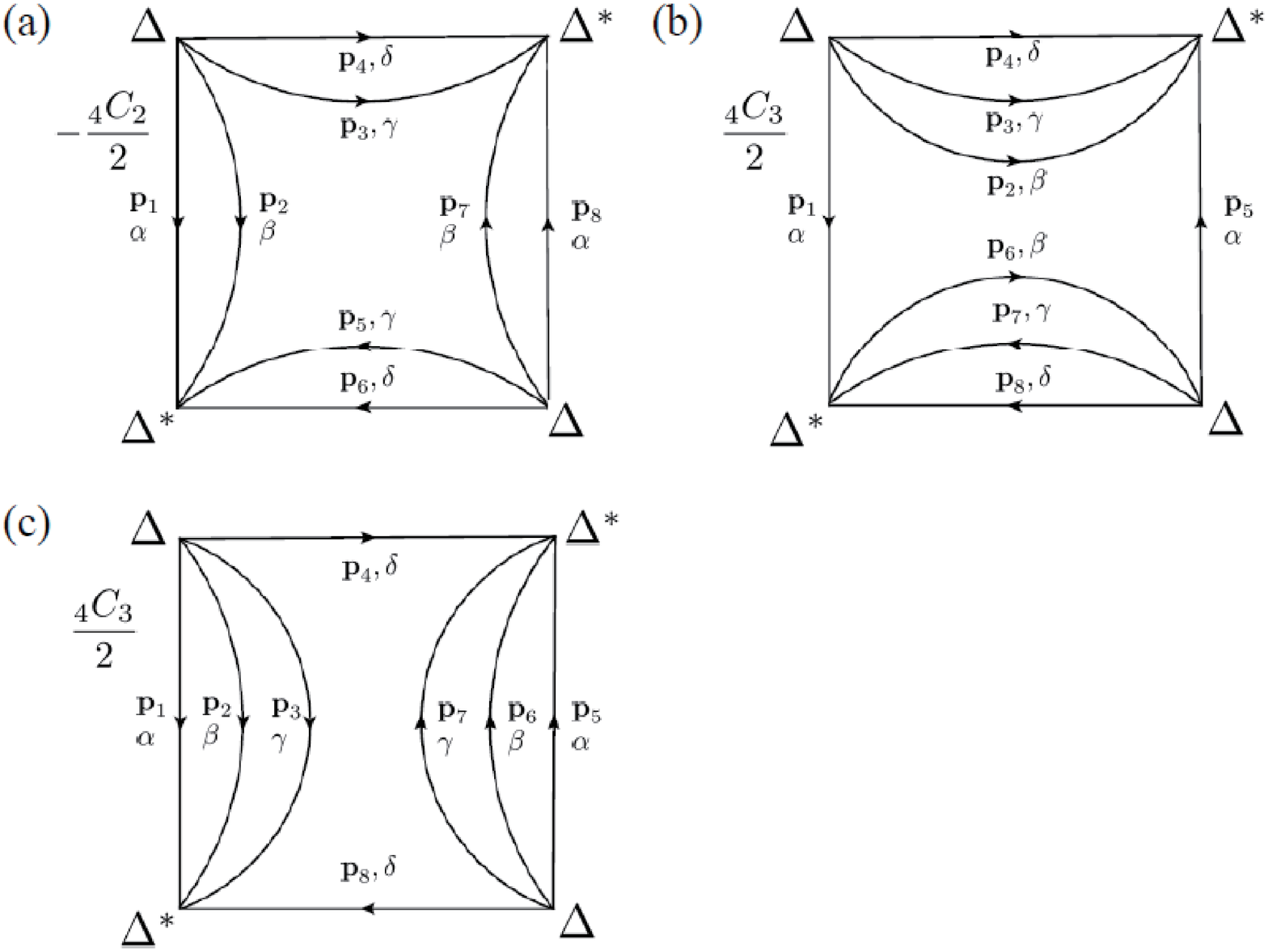}}
\caption{
Feynman diagrams for $\Omega_{\rm mf}$ of the quartic terms with respect to $\Delta$ and $\Delta^{*}$.  
}
\label{Fig:3A}
\end{center}
\end{figure}

\subsection*{Terms without $V$ in $\langle H-H_{\rm mf}\rangle_{\rm mf}$}

Of the quartic terms in $\Delta$ and $\Delta^{*}$ arising from 
$\langle H-H_{\rm mf} \rangle_{\rm mf}$, those without the interaction $V$ are 
given by the Feynman diagrams shown in Fig.\ \ref{Fig:3B}.  

The analytical expression $D_{1}(T)$ for the diagram shown in Fig.\ {\ref{Fig:3B}(a) is 
given as  
\begin{eqnarray}
& &D_{1}(T)=2\times{_4{\rm C}_2}\,T^5 
\prod_{i=1}^{8}
\displaystyle \displaystyle \frac{1}{N_{L}}\sum_{{\bf p}_i}
\displaystyle \sum \limits _{\epsilon_{n_i}} 
G({\bf p}_{i},{\rm i}\epsilon_{n_i})
\nonumber \\
 & &\qquad\qquad\qquad
\times \delta\left({\bf p}_1+{\bf p}_2+{\bf p}_3+{\bf p}_4\right)  
            \times  \delta_{\epsilon_{n_1}+\epsilon_{n_2}+\epsilon_{n_3}+\epsilon_{n_4},0}
\nonumber \\
 & &\qquad\qquad\qquad
  \times \delta\left({\bf p}_1+{\bf p}_2+{\bf p}_5+{\bf p}_6\right)  
            \times  \delta_{\epsilon_{n_1}+\epsilon_{n_2}+\epsilon_{n_5}+\epsilon_{n_6},0}
            \nonumber \\
 & &\qquad\qquad\qquad
  \times \delta\left({\bf p}_5+{\bf p}_6+{\bf p}_7+{\bf p}_8\right)  
            \times  \delta_{\epsilon_{n_5}+\epsilon_{n_6}+\epsilon_{n_7}+\epsilon_{n_8},0}.
\label{C:2}
\end{eqnarray}
This is the same as Eq.\ (\ref{Q:17}) except for a difference in a {pre}factor.  
Here, the factor $2\times{_4{\rm C}_2}$ comes from the number of combinations for perturbation expansion 
and the Wick theorem.  Namely, 
\begin{equation}
\frac{1}{3!}\times 2\times 3\times 2\times {_4{\rm C}_2}\times (+1)=2\times {_4{\rm C}_2},
\label{C:3}
\end{equation}
where the factor $(1/3!)$ comes from the perturbation expansion of $(H-H_{\rm mf})$ to the 3rd order 
in $\Delta$ and $\Delta^{*}$, 
the factor 2 is the  number of ways of choosing $\Delta$ or $\Delta^{*}$ from $(H-H_{\rm mf})$, given by 
Eq.\ (\ref{Q:5}), 
the factor 3 is the  number of ways of choosing two $\Delta^{*}$'s or $\Delta$'s from three products of 
the perturbation terms in $H_{\rm mf}$, given by Eq.\ (\ref{Q:3}), 
another factor 2 is the  number of ways of choosing two $\Delta^{*}$'s from two products of 
the perturbation terms in $H_{\rm mf}$, also given by Eq.\ (\ref{Q:3}), 
the factor $_{4}C_{2}$ is the number of 
combinations how to choose 2 spin states of Green functions connecting a certain pair of $\Delta$ and 
$\Delta^{*}$ from $\alpha$, $\beta$, $\gamma$, and $\delta$, and 
the factor (+1) represents that the number of interchanges 
of Fermion operators is even in the Wick expansion. Other assignments of 
the spin variables  $\alpha$, $\beta$, $\gamma$, and $\delta$ to Green functions are  
automatically determined by the conservation law of spins.  
Therefore, $D_{1}(T)$ is given in terms of $C_{1}(T)$ as
\begin{equation}
D_{1}(T)=-4C_{1}(T).
\label{C:4}
\end{equation}

The analytical expression $D_{2}(T)$ for the Feynman diagram shown in Fig.\ \ref{Fig:3A}(b) is given as 
\begin{eqnarray}
& &D_{2}(T)=-2\times{_4{\rm C}_3}\,T^5 
\prod_{i=1}^{8}
\displaystyle \displaystyle \frac{1}{N_{L}}\sum_{{\bf p}_i}
\displaystyle \sum \limits _{\epsilon_{n_i}} 
G({\bf p}_{i},{\rm i}\epsilon_{n_i})
\nonumber \\
 & &\qquad\qquad\qquad
\times \delta\left({\bf p}_1+{\bf p}_2+{\bf p}_3+{\bf p}_4\right)  
            \times  \delta_{\epsilon_{n_1}+\epsilon_{n_2}+\epsilon_{n_3}+\epsilon_{n_4},0}
\nonumber \\
 & &\qquad\qquad\qquad
  \times \delta\left({\bf p}_2+{\bf p}_3+{\bf p}_4+{\bf p}_5\right)  
            \times  \delta_{\epsilon_{n_2}+\epsilon_{n_3}+\epsilon_{n_4}+\epsilon_{n_5},0}
            \nonumber \\
 & &\qquad\qquad\qquad
  \times \delta\left({\bf p}_5+{\bf p}_6+{\bf p}_7+{\bf p}_8\right)  
            \times  \delta_{\epsilon_{n_5}+\epsilon_{n_6}+\epsilon_{n_7}+\epsilon_{n_8},0}.
\label{C:5}
\end{eqnarray}
This is the same as Eq.\ (\ref{Q:17bc}) except for a difference in a {pre}factor.  
Here, the factor $-2\times{_4{\rm C}_2}$ comes from the number of combinations for perturbation expansion 
and the Wick theorem.  Namely, 
\begin{equation}
\frac{1}{3!}\times 2\times 3\times 2\times {_4{\rm C}_3}\times (-1)=-2\times {_4{\rm C}_3},
\label{C:6}
\end{equation}
where the factor $(1/3!)$ comes from the perturbation expansion of $(H-H_{\rm mf})$ to the 3rd order 
in $\Delta$ and $\Delta^{*}$, 
the factor 2 is the  number of ways of choosing $\Delta$ or $\Delta^{*}$ from $(H-H_{\rm mf})$, given by 
Eq.\ (\ref{Q:5}), 
the factor 3 is the  number of ways of choosing two $\Delta^{*}$'s or $\Delta$'s from three products of 
the perturbation terms in $H_{\rm mf}$, given by Eq.\ (\ref{Q:3}), 
another factor 2 is the  number of ways of choosing two $\Delta^{*}$'s from two products of 
the perturbation terms in $H_{\rm mf}$, also given by Eq.\ (\ref{Q:3}), 
the factor $_{4}C_{3}$ is the number of 
combinations how to choose 3 spin states of Green functions connecting a certain pair of $\Delta$ and 
$\Delta^{*}$ from $\alpha$, $\beta$, $\gamma$, and $\delta$, and 
the factor (-1) represents that the number of interchanges 
of Fermion operators is odd in the Wick expansion. Other assignments of 
the spin variables  $\alpha$, $\beta$, $\gamma$, and $\delta$ to Green functions are  
automatically determined by the conservation law of spins.  
Therefore, $D_{2}(T)$ is given in terms of $C_{2}(T)$ as
\begin{equation}
D_{2}(T)=-4C_{2}(T).
\label{C:7}
\end{equation}

The analytical expression $D_{3}(T)$ for the Feynman diagram shown in {Fig.}\ \ref{Fig:3B}(c) is 
identical to that shown in  {Fig.}\ \ref{Fig:3B}(b).  Namely, 
\begin{equation}
D_{3}(T)=D_{2}(T).
\label{C:8}
\end{equation}

Equations (\ref{C:4}), (\ref{C:7}), and (\ref{C:8}) indicate that the contributions of   
Fig.\ \ref{Fig:3B} are twofold those of  Fig.\ \ref{Fig:3A} in size and opposite in sign.  

\begin{figure}[h]
\begin{center}
\rotatebox{0}{\includegraphics[width=0.85\linewidth]{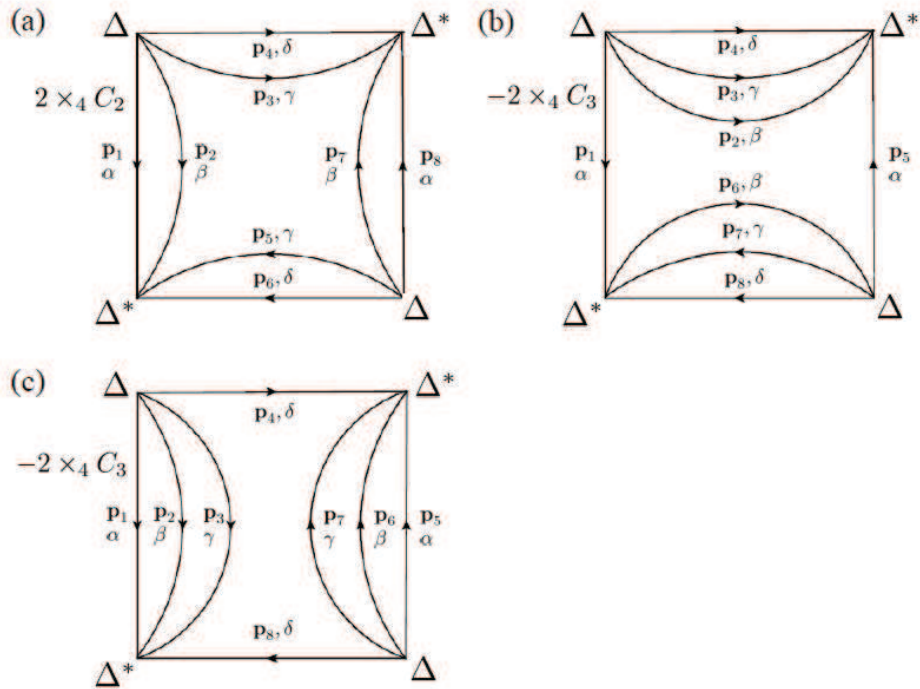}}
\caption{
Feynman diagrams for $\langle H-H_{\rm mf} \rangle_{\rm mf}$ of the quartic terms with 
respect to $\Delta$ and $\Delta^{*}$ {without} the interaction $V$.  
}
\label{Fig:3B}
\end{center}
\end{figure}

\subsection*{Terms with $V$ in $\langle H-H_{\rm mf}\rangle_{\rm mf}$}
The quartic terms including the interaction $V$ in $\langle H-H_{\rm mf} \rangle_{\rm mf}$ are 
given by the Feynman diagrams shown in Fig.\ \ref{Fig:3C}.  

The analytical expression $D_{4}(T)$ for the diagram shown in Fig.\ {\ref{Fig:3C}(a) is 
given as  
\begin{eqnarray}
& &\frac{D_{4}(T)}{V}=\,_4{\rm C}_{2}\,T^6 
\prod_{i=1}^{10}
\displaystyle \displaystyle \frac{1}{N_{L}}\sum_{{\bf p}_i}
\displaystyle \sum \limits _{\epsilon_{n_i}} 
G({\bf p}_{i},{\rm i}\epsilon_{n_i})
\nonumber \\
 & &\qquad\qquad\qquad
\times \delta\left({\bf p}_1+{\bf p}_2+{\bf p}_7+{\bf p}_8\right)  
            \times  \delta_{\epsilon_{n_1}+\epsilon_{n_2}+\epsilon_{n_7}+\epsilon_{n_8},0}
\nonumber \\
 & &\qquad\qquad\qquad
  \times \delta\left({\bf p}_7+{\bf p}_8-{\bf p}_9-{\bf p}_{10}\right)  
            \times  \delta_{\epsilon_{n_7}+\epsilon_{n_8}-\epsilon_{n_9}-\epsilon_{n_{10}},0}
            \nonumber \\
 & &\qquad\qquad\qquad
  \times \delta\left({\bf p}_5+{\bf p}_{6}+{\bf p}_9+{\bf p}_{10}\right)  
            \times  \delta_{\epsilon_{n_5}+\epsilon_{n_{6}}+\epsilon_{n_9}+\epsilon_{n_{10}},0}
            \nonumber \\
 & &\qquad\qquad\qquad
  \times \delta\left({\bf p}_3+{\bf p}_{4}+{\bf p}_5+{\bf p}_6\right)  
            \times  \delta_{\epsilon_{n_3}+\epsilon_{n_{4}}+\epsilon_{n_5}+\epsilon_{n_6},0}.
\label{C:9}
\end{eqnarray}
Here, the combination factor $_4$C$_2$ comes from the number of combinations for perturbation expansion 
and the Wick theorem.  Namely, 
\begin{equation}
\frac{1}{4!}\times {_4{\rm C}_2}\times {_4{\rm C}_2}\times 2\times 2 \times (+1)={_4{\rm C}_2},
\label{C:10}
\end{equation}
where the factor $(1/4!)$ comes from the perturbation expansion of the first term of Eq.\ (\ref{Q:5}) 
to the 4th order in $\Delta$ and $\Delta^{*}$, 
the factor $_{4}{\rm C}_{2}$ is the  number of ways of choosing 2 spin states in the interaction $V$ 
from $\alpha$, $\beta$, $\gamma$, and $\delta$, 
another factor $_4{\rm C}_{2}$ is the  number of ways of choosing two $\Delta$'s from four products of the 
perturbation terms in $H_{\rm mf}$, given by Eq.\ (\ref{Q:3}), 
the factor $2\times 2$ is a product of the  number of ways of choosing two $\Delta$'s from two products of  
the perturbation terms in $H_{\rm mf}$, also given by Eq.\ (\ref{Q:3}), and that of choosing $\Delta^{*}$, 
and the factor (+1) represents that the number of interchanges 
of Fermion operators is even in the Wick expansion. Other assignments of 
the spin variables  $\alpha$, $\beta$, $\gamma$, and $\delta$ to Green functions are  
automatically determined by the conservation law of spins.  
 
By using Eqs.\ (\ref{Q:9L}) and (\ref{Q:10}) and similar ones, the coefficient $D_{4}(T)$ 
is reduced to 
\begin{eqnarray}
& &\frac{D_{4}(T)}{V}= \displaystyle _4{\rm C}_{2}
\prod_{i=1}^{4}\displaystyle \int^\beta _0 d\tau_{i}
\displaystyle \sum_{{\bf r}_i}
\left[\displaystyle \frac{1}{N_{L}}\sum_{{\bf q}_1}
T\sum \limits_{\epsilon_{m_1}} G({\bf q}_1,{\rm i}\epsilon_{m_1}) 
e^{{\rm i}({\bf q}_1\cdot{\bf r}_1-\epsilon_{m_1} \tau_1)}\right]^2 
\nonumber 
\\
& &\qquad\qquad\qquad
\times 
\left[\displaystyle\frac{1}{N_L}\sum_{{\bf q}_2}
T\sum \limits_{\epsilon_{m_2}} G({\bf q}_2,{\rm i}\epsilon_{m_2}) 
e^{{\rm i}({\bf q}_2\cdot({\bf r}_1+{\bf r}_2)-\epsilon_{m_2} (\tau_1+\tau_2))}\right]^2
\nonumber 
\\
& &\qquad\qquad\qquad
\times 
\left[\displaystyle\frac{1}{N_L}\sum_{{\bf q}_3}
T\sum \limits _{\epsilon_{m_3}} 
G({\bf q}_3,{\rm i}\epsilon_{m_3}) e^{{\rm i}({\bf q}_3\cdot (-{\bf r}_2+{\bf r}_3)
-\epsilon_{m_3} (-\tau_2+\tau_3))}\right]^2
\nonumber 
\\
& &\qquad\qquad\qquad
\times 
\left[\displaystyle\frac{1}{N_L}\sum_{{\bf q}_4}
T\sum \limits _{\epsilon_{m_4}}
G({\bf q}_4,{\rm i}\epsilon_{m_4}) e^{{\rm i}({\bf q}_4\cdot ({\bf r}_3+{\bf r}_4)
-\epsilon_{m_4} (\tau_3+\tau_4))}\right]^2
\nonumber 
\\
& &\qquad\qquad\qquad
\times 
\left[\displaystyle\frac{1}{N_L}\sum_{{\bf q}_5}
T\sum \limits _{\epsilon_{m_5}}
G({\bf q}_5,{\rm i}\epsilon_{m_5}) e^{{\rm i}({\bf q}_5\cdot{\bf r}_4-\epsilon_{m_5}\tau_4)}\right]^2.
\label{C:11}
\end{eqnarray}
By calculations similar to those leading to Eq.\ (\ref{Q:18B}) [Eq.\ (\ref{S:11}{)}] 
from Eq.\ (\ref{Q:18}) [Eq.\ (\ref{Q:18bc}{)}], the expression (\ref{C:14}) is reduced to 
\begin{equation}
\frac{D_{4}(T)}{V}={_4{\rm C}_2}{T\over N_L}\sum_{{\bf k}}\sum_{\omega_{n}}
\left[X_{2}(-{\bf k},-{\rm i}\omega_{n})X_{2}({\bf k},{\rm i}\omega_{n})\right]^{3},
\label{Q:11A}
\end{equation} 
where $\omega_{n}$ is the bosonic Matsubara frequency.

The analytical expression $D_{5}(T)$ for the diagram shown in Fig.\ {\ref{Fig:3C}(b) is 
given as  
\begin{eqnarray}
& &\frac{D_{5}(T)}{V}=-2\times{_4{\rm C}_2}\,T^6 
\prod_{i=1}^{10}
\displaystyle \displaystyle \frac{1}{N_{L}}\sum_{{\bf p}_i}
\displaystyle \sum \limits _{\epsilon_{n_i}} 
G({\bf p}_{i},{\rm i}\epsilon_{n_i})
\nonumber \\
 & &\qquad\qquad\qquad
\times \delta\left({\bf p}_1+{\bf p}_5+{\bf p}_6+{\bf p}_9\right)  
            \times  \delta_{\epsilon_{n_1}+\epsilon_{n_5}+\epsilon_{n_6}+\epsilon_{n_9},0}
\nonumber \\
 & &\qquad\qquad\qquad
  \times \delta\left({\bf p}_5+{\bf p}_6-{\bf p}_7-{\bf p}_{8}\right)  
            \times  \delta_{\epsilon_{n_5}+\epsilon_{n_6}-\epsilon_{n_7}-\epsilon_{n_{8}},0}
            \nonumber \\
 & &\qquad\qquad\qquad
  \times \delta\left({\bf p}_7+{\bf p}_{8}+{\bf p}_9+{\bf p}_{10}\right)  
            \times  \delta_{\epsilon_{n_7}+\epsilon_{n_{8}}+\epsilon_{n_9}+\epsilon_{n_{10}},0}
            \nonumber \\
 & &\qquad\qquad\qquad
  \times \delta\left({\bf p}_2+{\bf p}_{3}+{\bf p}_4+{\bf p}_{10}\right)  
            \times  \delta_{\epsilon_{n_2}+\epsilon_{n_{3}}+\epsilon_{n_4}+\epsilon_{n_{10}},0}.
\label{C:12}
\end{eqnarray}
Here, the combination factor $-2\times{_4{\rm C}_2}$ comes from the number of combinations for 
perturbation expansion and the Wick theorem.  Namely, 
\begin{equation}
\frac{1}{4!}\times {_4{\rm C}_2}\times {_4{\rm C}_2}\times 2\times 2\times (-1)=-2\times{_4{\rm C}_2},
\label{C:13}
\end{equation}
where the factor $(1/4!)$ comes from the perturbation expansion of the first term of Eq.\ (\ref{Q:5}) 
to the 4th order in $\Delta$ and $\Delta^{*}$, 
the factor $_{4}{\rm C}_{2}$ is the  number of ways of choosing 2 spin states in the interaction $V$ 
from $\alpha$, $\beta$, $\gamma$, and $\delta$, 
another factor $_4{\rm C}_{2}$ is the  number of ways of choosing two $\Delta$'s from four products of 
the perturbation terms in $H_{\rm mf}$, given by Eq.\ (\ref{Q:3}), 
the factor $2\times 2$ is a product of the  number of ways of choosing two $\Delta$'s from two products of  
the perturbation terms in $H_{\rm mf}$, also given by Eq.\ (\ref{Q:3}), and that of choosing $\Delta^{*}$, 
the factor 2 is the  number of ways of choosing a spin state, $\delta$ or $\gamma$, for the Green function 
on the left side of Fig.\ \ref{Fig:3C}(b), and the factor (-1) represents that the number of interchanges 
of Fermion operators is odd in the Wick expansion. Other assignments of 
the spin variables  $\alpha$, $\beta$, $\gamma$, and $\delta$ to Green functions are  
automatically determined by the conservation law of spins.  
 
By using Eqs.\ (\ref{Q:9L}) and (\ref{Q:10}), the coefficient $D_{5}(T)$ 
is reduced to 
\begin{eqnarray}
& &\frac{D_{5}(T)}{V}= \displaystyle -2\times {_4{\rm C}_2}
\prod_{i=1}^{4}\displaystyle \int^\beta _0 d\tau_{i}
\displaystyle \sum_{{\bf r}_i}
\left[\displaystyle \frac{1}{N_{L}}\sum_{{\bf q}_1}
T\sum \limits_{\epsilon_{m_1}} G({\bf q}_1,{\rm i}\epsilon_{m_1}) 
e^{{\rm i}({\bf q}_1\cdot{\bf r}_1-\epsilon_{m_1} \tau_1)}\right] 
\nonumber 
\\
& &\qquad\qquad\qquad
\times 
\left[\displaystyle\frac{1}{N_L}\sum_{{\bf q}_2}
T\sum \limits_{\epsilon_{m_2}} G({\bf q}_2,{\rm i}\epsilon_{m_2}) 
e^{{\rm i}({\bf q}_2\cdot({\bf r}_1+{\bf r}_2)-\epsilon_{m_2} (\tau_1+\tau_2))}\right]
\nonumber 
\\
& &\qquad\qquad\qquad
\times 
\left[\displaystyle\frac{1}{N_L}\sum_{{\bf q}_3}
T\sum \limits _{\epsilon_{m_3}} 
G({\bf q}_3,{\rm i}\epsilon_{m_3}) e^{{\rm i}({\bf q}_3\cdot ({\bf r}_1+{\bf r}_4)
-\epsilon_{m_3} (\tau_1+\tau_4))}\right]^2
\nonumber 
\\
& &\qquad\qquad\qquad
\times 
\left[\displaystyle\frac{1}{N_L}\sum_{{\bf q}_4}
T\sum \limits _{\epsilon_{m_4}}
G({\bf q}_4,{\rm i}\epsilon_{m_4}) e^{{\rm i}({\bf q}_4\cdot ({\bf r}_2-{\bf r}_4)
-\epsilon_{m_4} (\tau_2-\tau_4))}\right]^2
\nonumber 
\\
& &\qquad\qquad\qquad
\times 
\left[\displaystyle\frac{1}{N_L}\sum_{{\bf q}_5}
T\sum \limits _{\epsilon_{m_5}}
G({\bf q}_5,{\rm i}\epsilon_{m_5}) e^{{\rm i}
({\bf q}_5\cdot({\bf r}_2+{\bf r}_3)-\epsilon_{m_5}(\tau_2+\tau_3))}\right]
\nonumber 
\\
& &\qquad\qquad\qquad
\times 
\left[\displaystyle\frac{1}{N_L}\sum_{{\bf q}_6}
T\sum \limits _{\epsilon_{m_6}}
G({\bf q}_6,{\rm i}\epsilon_{m_6}) e^{{\rm i}({\bf q}_6\cdot{\bf r}_3-\epsilon_{m_6}\tau_3)}\right]^3.
\label{C:14}
\end{eqnarray}
By calculations similar to those leading to Eq.\ (\ref{Q:18B}) [Eq.\ (\ref{S:11}{)}] 
from Eq.\ (\ref{Q:18}) [Eq.\ (\ref{Q:18bc}{)}], the expression (\ref{C:14}) is reduced to 
\begin{equation}
\frac{D_{5}(T)}{V}=-\,2\times{_4{\rm C}_2}{T^{2}\over N_L^{2}}\sum_{{\bf k}_1,{\bf k}_2}
\sum_{\epsilon_{n_{1}},\omega_{n_{2}}}
\left[X_{1}(-{\bf k}_{1},-{\rm i}\epsilon_{n_{1}})X_{2}({\bf k}_{2},{\rm i}\omega_{n_{2}})\right]^{2}
X_{1}({\bf k}_{1}-{\bf k}_{2},{\rm i}\epsilon_{n_{1}}-{\rm i}\omega_{n_{2}})
X_{3}({\bf k}_{1},{\rm i}\epsilon_{n_{1}}), 
\label{C:14A}
\end{equation} 
where $\omega_n$ and $\epsilon_{n}$ are the bosonic and fermionic Matsubara frequencies, as 
mentioned just below Eq.\ (\ref{Q:18B}). 

The analytical expression $D_{6}(T)$ for the diagram shown in Fig.\ {\ref{Fig:3C}(c) is 
given as  
\begin{eqnarray}
& &\frac{D_{6}(T)}{V}=-\,_4{\rm C}_{2}\,
\Biggl[
T^3 \prod_{i=1}^{4}
\displaystyle \displaystyle \frac{1}{N_{L}}\sum_{{\bf p}_i}
\displaystyle \sum \limits _{\epsilon_{n_i}} 
G({\bf p}_{i},{\rm i}\epsilon_{n_i})
\nonumber 
\\
 & &\qquad\qquad\qquad
\times \delta\left({\bf p}_1+{\bf p}_2+{\bf p}_3+{\bf p}_4\right)  
            \times  \delta_{\epsilon_{n_1}+\epsilon_{n_2}+\epsilon_{n_3}+\epsilon_{n_4},0}
{\Biggr]^{2}}.
\label{C:15}
\end{eqnarray}
Here, the combination factor $-_4$C$_2$ comes from the number of combinations for perturbation expansion 
and the Wick theorem.  Namely, 
\begin{equation}
\frac{1}{4!}\times {_4{\rm C}_2}\times {_4{\rm C}_2}\times 4\times 2\times 2\times (-1)
=-{_4{\rm C}_2},
\label{C:16}
\end{equation}
where the factor $(1/4!)$ comes from the perturbation expansion of the first term of Eq.\ (\ref{Q:5}) 
to the 4th order in $\Delta$ and $\Delta^{*}$, 
the factor $_{4}{\rm C}_{2}$ is the  number of ways of choosing 2 spin states in the interaction $V$ 
from $\alpha$, $\beta$, $\gamma$, and $\delta$, 
another factor $_4{\rm C}_{2}$ is the  number of ways of choosing two $\Delta$'s from four products of 
the perturbation terms in $H_{\rm mf}$, given by Eq.\ (\ref{Q:3}), 
the factor $2\times 2$ is a product of the  number of ways of choosing two $\Delta$'s from two products of  
the perturbation terms in $H_{\rm mf}$, also given by Eq.\ (\ref{Q:3}), and that of choosing $\Delta^{*}$, 
and the factor (-1) represents that the number of interchanges 
of Fermion operators is odd in the Wick expansion. Other assignments of 
the spin variables  $\alpha$, $\beta$, $\gamma$, and $\delta$ to Green functions are  
automatically determined by the conservation law of spins.  

Let us define the quantity in the square brackets in Eq.\ (\ref{C:15}) by $Z(T)$. 
 By using Eqs.\ (\ref{Q:9L}) and (\ref{Q:10}), $Z(T)$ is reduced to 
\begin{eqnarray}
& &Z(T)= \displaystyle T^{3}
\displaystyle \int^\beta _0 d\tau
\displaystyle \sum_{{\bf r}}
\left[\displaystyle \frac{1}{N_{L}}\sum_{{\bf q}_1}
T\sum \limits_{\epsilon_{m_1}} G({\bf q}_1,{\rm i}\epsilon_{m_1}) 
e^{{\rm i}({\bf q}_1\cdot{\bf r}-\epsilon_{m_1} \tau)}\right]^3
\nonumber 
\\
& &\qquad\qquad\qquad
\times 
\left[\displaystyle\frac{1}{N_L}\sum_{{\bf q}_2}
T\sum \limits_{\epsilon_{m_2}} [G({\bf q}_2,{\rm i}\epsilon_{m_2})]^{2} 
e^{{\rm i}({\bf q}_2\cdot{\bf r}-\epsilon_{m_2} \tau)}\right]^2.
\label{C:17}
\end{eqnarray}
By calculations similar to those leading to Eq.\ (\ref{Q:18B}) [Eq.\ (\ref{S:11}{)}] 
from Eq.\ (\ref{Q:18}) [Eq.\ (\ref{Q:18bc}{)}], the expression (\ref{C:17}) is reduced to 
\begin{equation}
Z(T)={T\over N_L}\sum_{{\bf k}}\sum_{\epsilon_{n}}
X_{3}(-{\bf k},-{\rm i}\epsilon_{n})\left[X_{1}({\bf k},{\rm i}\epsilon_{n})\right]^{2}.
\label{C:18}
\end{equation} 
Then, $D_6(T)$, given by Eq.\ (\ref{C:15}), is given as 
\begin{equation}
\frac{D_{6}(T)}{V}=-\,_{4}{\rm C}_{2}\left[Z(T)\right]^{2}. 
\label{C:19}
\end{equation}

\begin{figure}[h]
\begin{center}
\rotatebox{0}{\includegraphics[width=0.85\linewidth]{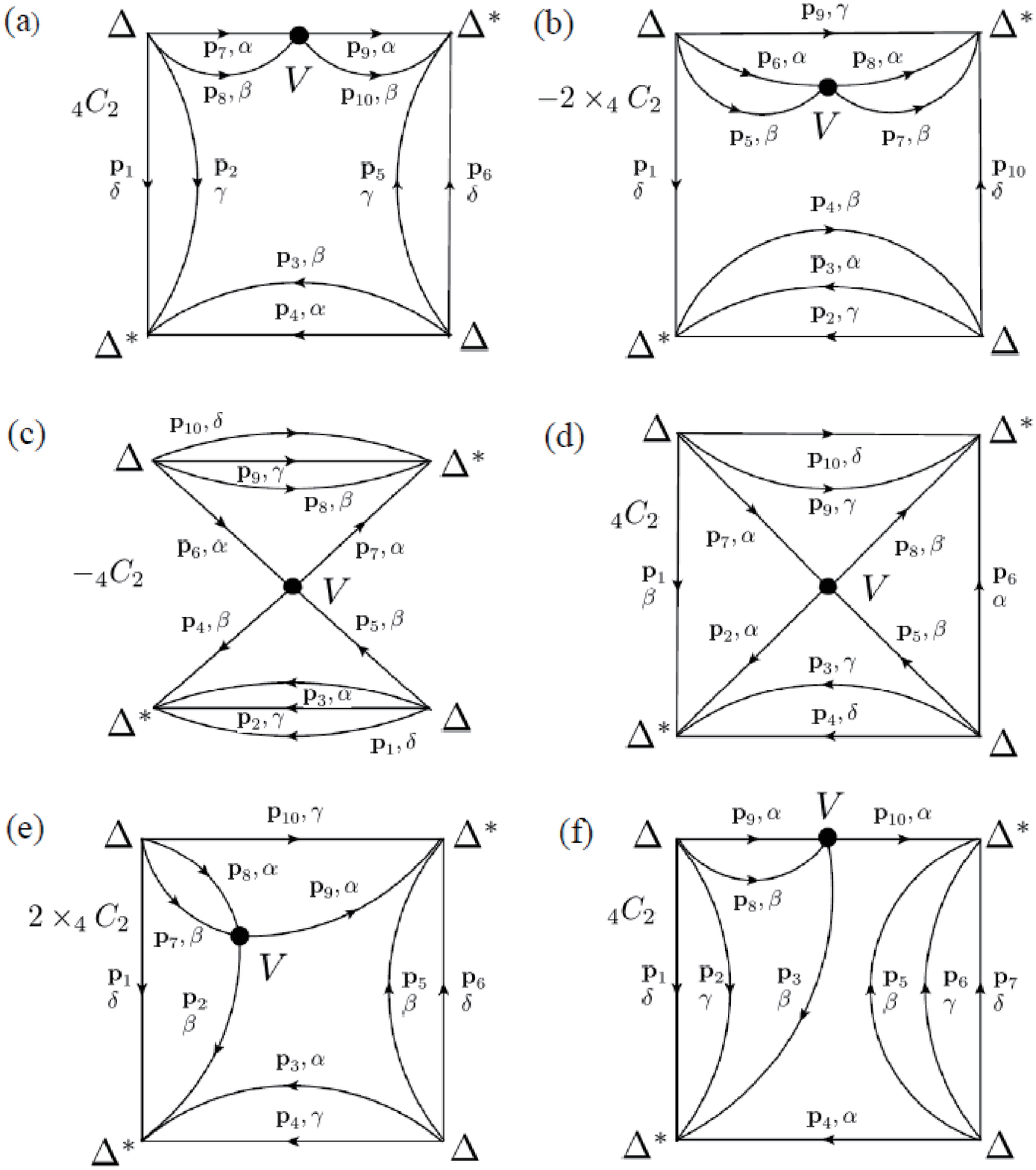}}
\caption{
Feynman diagrams for $\langle H-H_{\rm mf} \rangle_{\rm mf}$ of the quartic terms with 
respect to $\Delta$ and $\Delta^{*}$, which include the interaction $V$.  
}
\label{Fig:3C}
\end{center}
\end{figure}

The analytical expression $D_{7}(T)$ for the diagram shown in Fig.\ {\ref{Fig:3C}(d) is 
given as  
\begin{eqnarray}
& &\frac{D_{7}(T)}{V}={_4{\rm C}_2}\,T^6 
\prod_{i=1}^{10}
\displaystyle \displaystyle \frac{1}{N_{L}}\sum_{{\bf p}_i}
\displaystyle \sum \limits _{\epsilon_{n_i}} 
G({\bf p}_{i},{\rm i}\epsilon_{n_i})
\nonumber \\
 & &\qquad\qquad\qquad
\times \delta\left({\bf p}_1+{\bf p}_7+{\bf p}_9+{\bf p}_{10}\right)  
            \times  \delta_{\epsilon_{n_1}+\epsilon_{n_7}+\epsilon_{n_9}+\epsilon_{n_{10}},0}
\nonumber \\
 & &\qquad\qquad\qquad
  \times \delta\left({\bf p}_2+{\bf p}_8-{\bf p}_5-{\bf p}_{7}\right)  
            \times  \delta_{\epsilon_{n_2}+\epsilon_{n_8}-\epsilon_{n_5}-\epsilon_{n_{7}},0}
            \nonumber \\
 & &\qquad\qquad\qquad
  \times \delta\left({\bf p}_6+{\bf p}_{8}+{\bf p}_9+{\bf p}_{10}\right)  
            \times  \delta_{\epsilon_{n_6}+\epsilon_{n_{8}}+\epsilon_{n_9}+\epsilon_{n_{10}},0}
            \nonumber \\
 & &\qquad\qquad\qquad
  \times \delta\left({\bf p}_3+{\bf p}_{4}+{\bf p}_5+{\bf p}_{6}\right)  
            \times  \delta_{\epsilon_{n_3}+\epsilon_{n_{4}}+\epsilon_{n_5}+\epsilon_{n_{6}},0}.
\label{C:20}
\end{eqnarray}
Here, the combination factor $_4$C$_2$ comes from the number of combinations for perturbation expansion 
and the Wick theorem.  Namely, 
\begin{equation}
\frac{1}{4!}\times _4{\rm C}_{2}\times _4{\rm C}_{2}\times 2\times 2\times (+1)=\,_4{\rm C}_{2},
\label{C:21}
\end{equation}
where the factor $(1/4!)$ comes from the perturbation expansion of the first term of Eq.\ (\ref{Q:5}) 
to the 4th order in $\Delta$ and $\Delta^{*}$, 
the factor $_{4}{\rm C}_{2}$ is the  number of ways of choosing 2 spin states in the interaction $V$ 
from $\alpha$, $\beta$, $\gamma$, and $\delta$, 
another factor $_4{\rm C}_{2}$ is the  number of ways of choosing two $\Delta$'s from four products of 
the perturbation terms in $H_{\rm mf}$, given by Eq.\ (\ref{Q:3}), 
the factor $2\times 2$ is a product of the  number of ways of choosing two $\Delta$'s from two products of  
the perturbation terms in $H_{\rm mf}$, also given by Eq.\ (\ref{Q:3}), and that of choosing $\Delta^{*}$, 
and the factor (+1) represents that the number of interchanges 
of Fermion operators is even in the Wick expansion. Other assignments of 
the spin variables  $\alpha$, $\beta$, $\gamma$, and $\delta$ to Green functions are  
automatically determined by the conservation law of spins.  

By using Eqs.\ (\ref{Q:9L}) and (\ref{Q:10}), the coefficient $D_{7}(T)$ 
is reduced to 
\begin{eqnarray}
& &\frac{D_{7}(T)}{V}= \displaystyle _4{\rm C}_{2}
\prod_{i=1}^{4}\displaystyle \int^\beta _0 d\tau_{i}
\displaystyle \sum_{{\bf r}_i}
\left[\displaystyle \frac{1}{N_{L}}\sum_{{\bf q}_1}
T\sum \limits_{\epsilon_{m_1}} G({\bf q}_1,{\rm i}\epsilon_{m_1}) 
e^{{\rm i}({\bf q}_1\cdot{\bf r}_1-\epsilon_{m_1} \tau_1)}\right]^{2}
\nonumber 
\\
& &\qquad\qquad\qquad
\times 
\left[\displaystyle\frac{1}{N_L}\sum_{{\bf q}_2}
T\sum \limits_{\epsilon_{m_2}} G({\bf q}_2,{\rm i}\epsilon_{m_2}) 
e^{{\rm i}({\bf q}_2\cdot({\bf r}_1+{\bf r}_2)-\epsilon_{m_2} (\tau_1+\tau_2))}\right]
\nonumber 
\\
& &\qquad\qquad\qquad
\times 
\left[\displaystyle\frac{1}{N_L}\sum_{{\bf q}_3}
T\sum \limits _{\epsilon_{m_3}} 
G({\bf q}_3,{\rm i}\epsilon_{m_3}) e^{{\rm i}({\bf q}_3\cdot ({\bf r}_1+{\bf r}_4)
-\epsilon_{m_3} (\tau_1+\tau_4))}\right]
\nonumber 
\\
& &\qquad\qquad\qquad
\times 
\left[\displaystyle\frac{1}{N_L}\sum_{{\bf q}_4}
T\sum \limits _{\epsilon_{m_4}}
G({\bf q}_4,{\rm i}\epsilon_{m_4}) e^{{\rm i}
({\bf q}_4\cdot({\bf r}_2-{\bf r}_4)-\epsilon_{m_4}(\tau_2-\tau_4))}\right]
\nonumber 
\\
& &\qquad\qquad\qquad
\times 
\left[\displaystyle\frac{1}{N_L}\sum_{{\bf q}_5}
T\sum \limits _{\epsilon_{m_5}}
G({\bf q}_5,{\rm i}\epsilon_{m_5}) e^{{\rm i}({\bf q}_5\cdot
({\bf r}_2+{\bf r}_3)-\epsilon_{m_5}(\tau_2+\tau_3))}\right]^2
\nonumber 
\\
& &\qquad\qquad\qquad
\times 
\left[\displaystyle\frac{1}{N_L}\sum_{{\bf q}_6}
T\sum \limits _{\epsilon_{m_6}}
G({\bf q}_6,{\rm i}\epsilon_{m_6}) e^{{\rm i}({\bf q}_6\cdot {\bf r}_3
-\epsilon_{m_6} \tau_3)}\right]
\nonumber 
\\
& &\qquad\qquad\qquad
\times 
\left[\displaystyle\frac{1}{N_L}\sum_{{\bf q}_7}
T\sum \limits _{\epsilon_{m_7}}
G({\bf q}_7,{\rm i}\epsilon_{m_7}) e^{{\rm i}({\bf q}_7\cdot ({\bf r}_3+{\bf r}_4)
-\epsilon_{m_7} (\tau_3+\tau_4))}\right]
\nonumber 
\\
& &\qquad\qquad\qquad
\times 
\left[\displaystyle\frac{1}{N_L}\sum_{{\bf q}_8}
T\sum \limits _{\epsilon_{m_8}}
G({\bf q}_8,{\rm i}\epsilon_{m_8}) e^{{\rm i}({\bf q}_8\cdot (-{\bf r}_4)
-\epsilon_{m_8} (-\tau_4))}\right].
\label{C:22}
\end{eqnarray}
By calculations similar to those leading to Eq.\ (\ref{Q:18B}) [Eq.\ (\ref{S:11}{)}] 
from Eq.\ (\ref{Q:18}) [Eq.\ (\ref{Q:18bc}{)}], the expression (\ref{C:22}) is reduced to 
\begin{eqnarray}
& &\frac{D_{7}(T)}{V}=\,_4{\rm C}_{2}\,{T^{4}\over N_L^{4}}\sum_{{\bf k}_1\sim{\bf k}_4}
\sum_{\epsilon_{n_{1}},\epsilon_{n_{2}}}\sum_{\omega_{n_{3}},\omega_{n_{4}}}
X_{1}(-{\bf k}_{1},-{\rm i}\epsilon_{n_{1}})
X_{1}({\bf k}_{1}-{\bf k}_{3},{\rm i}\epsilon_{n_{1}}-{\rm i}\omega_{n_{3}})
\nonumber
\\
& &\qquad\qquad\qquad\qquad
\times
X_{1}(-{\bf k}_{1}+{\bf k}_{3}-{\bf k}_{4},
-{\rm i}\epsilon_{n_{1}}+{\rm i}\omega_{n_{3}}-{\rm i}\omega_{n_{4}})
X_{1}(-{\bf k}_{2},-{\rm i}\epsilon_{n_{2}})
\nonumber
\\
& &\qquad\qquad\qquad\qquad
\times
X_{1}({\bf k}_{2}-{\bf k}_{3},{\rm i}\epsilon_{n_{2}}-{\rm i}\omega_{n_{3}})
X_{2}({\bf k}_{3},{\rm i}\omega_{n_{3}}). 
\nonumber
\\
& &\qquad\qquad\qquad\qquad
\times
X_{1}(-{\bf k}_{2}+{\bf k}_{3}-{\bf k}_{4},
-{\rm i}\epsilon_{n_{2}}+{\rm i}\omega_{n_{3}}-{\rm i}\omega_{n_{4}})
X_{2}({\bf k}_{4},{\rm i}\omega_{n_{4}}).
\label{C:23}
\end{eqnarray}

The analytical expression $D_{8}(T)$ for the diagram shown in Fig.\ {\ref{Fig:3C}(e) is 
given as  
\begin{eqnarray}
& &\frac{D_{8}(T)}{V}=2\times{_4{\rm C}_2}\,T^6 
\prod_{i=1}^{10}
\displaystyle \displaystyle \frac{1}{N_{L}}\sum_{{\bf p}_i}
\displaystyle \sum \limits _{\epsilon_{n_i}} 
G({\bf p}_{i},{\rm i}\epsilon_{n_i})
\nonumber \\
 & &\qquad\qquad\qquad
\times \delta\left({\bf p}_1+{\bf p}_7+{\bf p}_8+{\bf p}_{10}\right)  
            \times  \delta_{\epsilon_{n_1}+\epsilon_{n_7}+\epsilon_{n_8}+\epsilon_{n_{10}},0}
\nonumber \\
 & &\qquad\qquad\qquad
  \times \delta\left({\bf p}_7+{\bf p}_8-{\bf p}_2-{\bf p}_{9}\right)  
            \times  \delta_{\epsilon_{n_7}+\epsilon_{n_8}-\epsilon_{n_2}-\epsilon_{n_{9}},0}
            \nonumber \\
 & &\qquad\qquad\qquad
  \times \delta\left({\bf p}_5+{\bf p}_{6}+{\bf p}_9+{\bf p}_{10}\right)  
            \times  \delta_{\epsilon_{n_5}+\epsilon_{n_{6}}+\epsilon_{n_9}+\epsilon_{n_{10}},0}
            \nonumber \\
 & &\qquad\qquad\qquad
  \times \delta\left({\bf p}_3+{\bf p}_{4}+{\bf p}_5+{\bf p}_{6}\right)  
            \times  \delta_{\epsilon_{n_3}+\epsilon_{n_{4}}+\epsilon_{n_5}+\epsilon_{n_{6}},0}.
\label{C:24}
\end{eqnarray}
Here, the combination factor $2\times{_4{\rm C}_2}$ comes from the number of combinations for 
perturbation expansion and the Wick theorem.  Namely, 
\begin{equation}
\frac{1}{4!}\times {_4{\rm C}_2}\times {_4{\rm C}_2}\times 2\times 2\times 2\times (+1)
=2\times{_4{\rm C}_2},
\label{C:25}
\end{equation}
where the factor $(1/4!)$ comes from the perturbation expansion of the first term of Eq.\ (\ref{Q:5}) 
to the 4th order in $\Delta$ and $\Delta^{*}$, 
the factor $_{4}{\rm C}_{2}$ is the  number of ways of choosing 2 spin states in the interaction $V$ 
from $\alpha$, $\beta$, $\gamma$, and $\delta$, 
another factor $_4{\rm C}_{2}$ is the  number of ways of choosing two $\Delta$'s from the four products of 
the perturbation terms in $H_{\rm mf}$, given by Eq.\ (\ref{Q:3}), 
the factor $2\times 2$ is a product of the  number of ways of choosing two $\Delta$'s from two products of  
the perturbation terms in $H_{\rm mf}$, also given by Eq.\ (\ref{Q:3}), and that of choosing $\Delta^{*}$, 
the factor 2 is the  number of ways of choosing a spin state, $\delta$ or $\gamma$, for the Green function 
on the left side of Fig.\ \ref{Fig:3C}(e), 
and the factor (+1) represents that the number of interchanges 
of Fermion operators is even in the Wick expansion. Other assignments of 
the spin variables  $\alpha$, $\beta$, $\gamma$, and $\delta$ to Green functions are  
automatically determined by the conservation law of spins.  

By using Eqs.\ (\ref{Q:9L}) and (\ref{Q:10}), the coefficient $D_{8}(T)$ 
is reduced to 
\begin{eqnarray}
& &\frac{D_{8}(T)}{V}= \displaystyle 2\times{_4{\rm C}_2}
\prod_{i=1}^{4}\displaystyle \int^\beta _0 d\tau_{i}
\displaystyle \sum_{{\bf r}_i}
\left[\displaystyle \frac{1}{N_{L}}\sum_{{\bf q}_1}
T\sum \limits_{\epsilon_{m_1}} G({\bf q}_1,{\rm i}\epsilon_{m_1}) 
e^{{\rm i}({\bf q}_1\cdot{\bf r}_1-\epsilon_{m_1} \tau_1)}\right]
\nonumber 
\\
& &\qquad\qquad\qquad
\times 
\left[\displaystyle\frac{1}{N_L}\sum_{{\bf q}_2}
T\sum \limits_{\epsilon_{m_2}} G({\bf q}_2,{\rm i}\epsilon_{m_2}) 
e^{{\rm i}({\bf q}_2\cdot({\bf r}_1+{\bf r}_2)-\epsilon_{m_2} (\tau_1+\tau_2))}\right]
\nonumber 
\\
& &\qquad\qquad\qquad
\times 
\left[\displaystyle\frac{1}{N_L}\sum_{{\bf q}_3}
T\sum \limits _{\epsilon_{m_3}} 
G({\bf q}_3,{\rm i}\epsilon_{m_3}) e^{{\rm i}({\bf q}_3\cdot ({\bf r}_1+{\bf r}_4)
-\epsilon_{m_3} (\tau_1+\tau_4))}\right]^{2}
\nonumber 
\\
& &\qquad\qquad\qquad
\times 
\left[\displaystyle\frac{1}{N_L}\sum_{{\bf q}_4}
T\sum \limits _{\epsilon_{m_4}}
G({\bf q}_4,{\rm i}\epsilon_{m_4}) e^{{\rm i}
({\bf q}_4\cdot({\bf r}_2-{\bf r}_4)-\epsilon_{m_4}(\tau_2-\tau_4))}\right]
\nonumber 
\\
& &\qquad\qquad\qquad
\times 
\left[\displaystyle\frac{1}{N_L}\sum_{{\bf q}_5}
T\sum \limits _{\epsilon_{m_5}}
G({\bf q}_5,{\rm i}\epsilon_{m_5}) e^{{\rm i}({\bf q}_5\cdot
({\bf r}_2+{\bf r}_3)-\epsilon_{m_5}(\tau_2+\tau_3))}\right]^2
\nonumber 
\\
& &\qquad\qquad\qquad
\times 
\left[\displaystyle\frac{1}{N_L}\sum_{{\bf q}_6}
T\sum \limits _{\epsilon_{m_6}}
G({\bf q}_6,{\rm i}\epsilon_{m_6}) e^{{\rm i}({\bf q}_6\cdot {\bf r}_3
-\epsilon_{m_6} \tau_3)}\right]^{2}
\nonumber 
\\
& &\qquad\qquad\qquad
\times 
\left[\displaystyle\frac{1}{N_L}\sum_{{\bf q}_7}
T\sum \limits _{\epsilon_{m_7}}
G({\bf q}_7,{\rm i}\epsilon_{m_7}) e^{{\rm i}({\bf q}_6\cdot (-{\bf r}_4)
-\epsilon_{m_6} (-\tau_4))}\right].
\label{C:26}
\end{eqnarray}
By calculations similar to those leading to Eq.\ (\ref{Q:18B}) [Eq.\ (\ref{S:11}{)}] 
from Eq.\ (\ref{Q:18}) [Eq.\ (\ref{Q:18bc}{)}], the expression (\ref{C:26}) is reduced to 
\begin{eqnarray}
& &\frac{D_{8}(T)}{V}=2\times{_4{\rm C}_2}{T^{3}\over N_L^{3}}\sum_{{\bf k}_1\sim{\bf k}_3}
\sum_{\epsilon_{n_{1}}}\sum_{\omega_{n_{2}},\omega_{n_{3}}}
X_{1}({\bf k}_{1},{\rm i}\epsilon_{n_{1}})X_{1}(-{\bf k}_{1},-{\rm i}\epsilon_{n_{1}})
X_{1}({\bf k}_{1}-{\bf k}_{2},{\rm i}\epsilon_{n_{1}}-{\rm i}\omega_{n_{2}})
\nonumber
\\
& &\qquad\qquad\qquad\qquad\qquad
\times
X_{1}(-{\bf k}_{1}+{\bf k}_{2}-{\bf k}_{3},
-{\rm i}\epsilon_{n_{1}}+{\rm i}\omega_{n_{2}}-{\rm i}\omega_{n_{3}})
X_{2}({\bf k}_{2},{\rm i}\omega_{n_{2}})
\nonumber
\\
& &\qquad\qquad\qquad\qquad\qquad
\times
X_{2}({\bf k}_{3},{\rm i}\omega_{n_{3}})
X_{2}(-{\bf k}_{3},-{\rm i}\omega_{n_{3}}). 
\label{C:27}
\end{eqnarray}

The analytical expression $D_{9}(T)$ for the diagram shown in Fig.\ {\ref{Fig:3C}(f) is 
given as  
\begin{eqnarray}
& &\frac{D_{9}(T)}{V}= {_4{\rm C}_2}\,T^6 
\prod_{i=1}^{10}
\displaystyle \displaystyle \frac{1}{N_{L}}\sum_{{\bf p}_i}
\displaystyle \sum \limits _{\epsilon_{n_i}} 
G({\bf p}_{i},{\rm i}\epsilon_{n_i})
\nonumber \\
 & &\qquad\qquad\qquad
\times \delta\left({\bf p}_1+{\bf p}_2+{\bf p}_3+{\bf p}_4\right)  
            \times  \delta_{\epsilon_{n_1}+\epsilon_{n_2}+\epsilon_{n_3}+\epsilon_{n_4},0}
\nonumber \\
 & &\qquad\qquad\qquad
  \times \delta\left({\bf p}_8+{\bf p}_9-{\bf p}_3-{\bf p}_{10}\right)  
            \times  \delta_{\epsilon_{n_8}+\epsilon_{n_9}-\epsilon_{n_3}-\epsilon_{n_{10}},0}
            \nonumber \\
 & &\qquad\qquad\qquad
  \times \delta\left({\bf p}_1+{\bf p}_{2}+{\bf p}_8+{\bf p}_{9}\right)  
            \times  \delta_{\epsilon_{n_1}+\epsilon_{n_{2}}+\epsilon_{n_8}+\epsilon_{n_{9}},0}
            \nonumber \\
 & &\qquad\qquad\qquad
  \times \delta\left({\bf p}_4+{\bf p}_{5}+{\bf p}_6+{\bf p}_{7}\right)  
            \times  \delta_{\epsilon_{n_4}+\epsilon_{n_{5}}+\epsilon_{n_6}+\epsilon_{n_{7}},0}.
\label{C:28}
\end{eqnarray}
Here, the combination factor $_4$C$_2$ comes from the number of combinations for perturbation expansion 
and the Wick theorem.  Namely, 
\begin{equation}
\frac{1}{4!}\times {_4{\rm C}_2}\times {_4{\rm C}_2}\times 2\times 2\times (+1)= {_4{\rm C}_2},
\label{C:29}
\end{equation}
where the factor $(1/4!)$ comes from the perturbation expansion of the first term of Eq.\ (\ref{Q:5}) 
to the 4th order in $\Delta$ and $\Delta^{*}$, 
the factor $_{4}{\rm C}_{2}$ is the  number of ways of choosing 2 spin states in the interaction $V$ 
from $\alpha$, $\beta$, $\gamma$, and $\delta$, 
another factor $_{4}{\rm C}_{2}$ is the  number of ways of choosing two $\Delta$'s from four products of 
the perturbation terms in $H_{\rm mf}$, given by Eq.\ (\ref{Q:3}), 
the factor $2\times 2$ is a product of the  number of ways of choosing two $\Delta$'s from two products of  
the perturbation terms in $H_{\rm mf}$, also given by Eq.\ (\ref{Q:3}), and that of choosing $\Delta^{*}$, 
and the factor (+1) represents that the number of interchanges 
of Fermion operators is even in the Wick expansion. Other assignments of 
the spin variables  $\alpha$, $\beta$, $\gamma$, and $\delta$ to Green functions are  
automatically determined by the conservation law of spins.  
 
By using Eqs.\ (\ref{Q:9L}) and (\ref{Q:10}), the coefficient $D_{9}(T)$ 
is reduced to 
\begin{eqnarray}
& &\frac{D_{9}(T)}{V}= \displaystyle _4{\rm C}_{2}
\prod_{i=1}^{4}\displaystyle \int^\beta _0 d\tau_{i}
\displaystyle \sum_{{\bf r}_i}
\left[\displaystyle \frac{1}{N_{L}}\sum_{{\bf q}_1}
T\sum \limits_{\epsilon_{m_1}} G({\bf q}_1,{\rm i}\epsilon_{m_1}) 
e^{{\rm i}({\bf q}_1\cdot({\bf r}_1+{\bf r}_2)-\epsilon_{m_1} (\tau_1+\tau_2))}\right]^{2}
\nonumber 
\\
& &\qquad\qquad\qquad
\times 
\left[\displaystyle\frac{1}{N_L}\sum_{{\bf q}_2}
T\sum \limits_{\epsilon_{m_2}} G({\bf q}_2,{\rm i}\epsilon_{m_2}) 
e^{{\rm i}({\bf q}_2\cdot({\bf r}_1+{\bf r}_4)-\epsilon_{m_2} (\tau_1+\tau_4))}\right]^{2}
\nonumber 
\\
& &\qquad\qquad\qquad
\times 
\left[\displaystyle\frac{1}{N_L}\sum_{{\bf q}_3}
T\sum \limits _{\epsilon_{m_3}} 
G({\bf q}_3,{\rm i}\epsilon_{m_3}) e^{{\rm i}({\bf q}_3\cdot ({\bf r}_2-{\bf r}_4)
-\epsilon_{m_3} (\tau_2-\tau_4))}\right]
\nonumber 
\\
& &\qquad\qquad\qquad
\times 
\left[\displaystyle\frac{1}{N_L}\sum_{{\bf q}_4}
T\sum \limits _{\epsilon_{m_4}}
G({\bf q}_4,{\rm i}\epsilon_{m_4}) e^{{\rm i}
({\bf q}_4\cdot({\bf r}_2+{\bf r}_3)-\epsilon_{m_4}(\tau_2+\tau_3))}\right]
\nonumber 
\\
& &\qquad\qquad\qquad
\times 
\left[\displaystyle\frac{1}{N_L}\sum_{{\bf q}_5}
T\sum \limits _{\epsilon_{m_5}}
G({\bf q}_5,{\rm i}\epsilon_{m_5}) e^{{\rm i}({\bf q}_5\cdot{\bf r}_3-\epsilon_{m_5}\tau_3)}\right]^3
\nonumber 
\\
& &\qquad\qquad\qquad
\times 
\left[\displaystyle\frac{1}{N_L}\sum_{{\bf q}_6}
T\sum \limits _{\epsilon_{m_6}}
G({\bf q}_6,{\rm i}\epsilon_{m_6}) e^{{\rm i}({\bf q}_4\cdot (-{\bf r}_4)
-\epsilon_{m_6} (-\tau_4))}\right].
\label{C:30}
\end{eqnarray}
By calculations similar to those leading to Eq.\ (\ref{Q:18B}) [Eq.\ (\ref{S:11}{)}] 
from Eq.\ (\ref{Q:18}) [Eq.\ (\ref{Q:18bc}{)}], the expression (\ref{C:30}) is reduced to 
\begin{eqnarray}
& &\frac{D_{9}(T)}{V}=\,_{4}C_{2}\,{T^{2}\over N_L^{2}}\sum_{{\bf k}_1,{\bf k}_2}
\sum_{\epsilon_{n_{1}},\omega_{n_{2}}}
\left[X_{1}(-{\bf k}_{1},-{\rm i}\epsilon_{n_{1}})\right]^{2}
X_{2}({\bf k}_{2},{\rm i}\omega_{n_{2}})X_{2}(-{\bf k}_{2},-{\rm i}\omega_{n_{2}})
\nonumber
\\
& &\qquad\qquad\qquad\qquad\qquad\qquad\qquad
\times
X_{1}({\bf k}_{1}-{\bf k}_{2},{\rm i}\epsilon_{n_{1}}-{\rm i}\omega_{n_{2}})
X_{3}({\bf k}_{1},{\rm i}\epsilon_{n_{1}}). 
\label{C:31}
\end{eqnarray}

\end{document}